\documentclass[11pt]{article}
\pdfoutput=1 

\usepackage{calc}
\usepackage{rotating}
\usepackage[english]{babel}
\usepackage{graphicx}
\usepackage{subfig}
\usepackage{float}
\usepackage{amsmath}
\usepackage{amssymb}
\usepackage{amsthm}
\usepackage{latexsym}
\usepackage{dcolumn}
\usepackage{geometry}
\usepackage{color}
\usepackage{hyperref}
\usepackage{mciteplus}

\def\gtwid{\mathrel{\raise.3ex\hbox{$>$\kern-.75em\lower1ex\hbox{$\sim
$}}}}
\def\vio{\mathrel{\hbox{$E$\kern-.60em\hbox{$/
$}}}}
\newcommand{\newc}{\newcommand*}

\long\def\begincomment#1\endcomment{%
        \begingroup\sf\baselineskip12pt#1\endgroup}

\newc{\etal}{\textrm{et al.}} 
\newc{\eg}{\textrm{e.g.}} 
\newc{\ie}{\textrm{i.e.}}
\newc{\etc}{\textrm{etc}}
\newc\vs{\textrm{vs.}}
\newc{\cl}{\rm {CL}}

\newc{\ev}{\ensuremath{\,\mathrm{eV}}}
\newc{\kev}{\ensuremath{\,\mathrm{keV}}}
\newc{\mev}{\ensuremath{\,\mathrm{MeV}}}
\newc{\gev}{\ensuremath{\,\mathrm{GeV}}}
\newc{\tev}{\ensuremath{\,\mathrm{TeV}}}
\newc{\MeV}{\mev} 
\newc{\TeV}{\tev}
\newc{\invpb}{\ensuremath{/\text{pb}}}
\newc{\invfb}{\ensuremath{/\text{fb}}}
\newc\nb{\ensuremath{\,\mathrm{nb}}} \newc\pb{\ensuremath{\,\mathrm{pb}}} \newc\fb{\ensuremath{\,\mathrm{fb}}}
\newc\pc{\ensuremath{\,\mathrm{pc}}}
\newc\kpc{\ensuremath{\,\mathrm{kpc}}}
\newc\mpc{\ensuremath{\,\mathrm{Mpc}}}
\newc\ps{\ensuremath{\,\mathrm{ps}}} 
\newc\cmeter{\ensuremath{\,\mathrm{cm}}} 
\newc\meter{\ensuremath{\,\mathrm{m}}} 
\newc\kmeter{\ensuremath{\,\mathrm{km}}}
\newc\second{\ensuremath{\,\mathrm{s}}}
\newc\msecond{\ensuremath{\,\mathrm{ms}}}
\newc\nsecond{\ensuremath{\,\mathrm{ns}}}
\newc\psecond{\ensuremath{\,\mathrm{ps}}}

\newc{\chisqmin}{\ensuremath{\chi^2_{\mathrm{min}}}}
\newc{\Delchisq}{\ensuremath{\Delta\chi^2}}
\newc{\delchisq}{\ensuremath{\delta\chi^2}}
\newc{\chisq}{\ensuremath{\chi^2}}
\newc{\like}{\ensuremath{\mathcal{L}}}

\newc\lsim{\ensuremath{\mathrel{\rlap{\lower4pt\hbox{\hskip1pt$\sim$}}\raise1pt\hbox{$<$}}}}
\newc\gsim{\ensuremath{\mathrel{\rlap{\lower4pt\hbox{\hskip1pt$\sim$}}\raise1pt\hbox{$>$}}}}
\newc{\VEV}[1]{\ensuremath{\langle #1 \rangle}}
\newc{\dl}{\ensuremath{\stackrel{\leftarrow}{D}}}
\newc{\dr}{\ensuremath{\stackrel{\rightarrow}{D}}}

\newc{\bcenter}{\begin{center}}    \newc{\ecenter}{\end{center}}
\newc{\bfl}{\begin{flushleft}}    \newc{\efl}{\end{flushleft}}
\newc{\bfr}{\begin{flushright}}    \newc{\efr}{\end{flushright}}

\newc{\bi}{\begin{itemize}}
\newc{\ei}{\end{itemize}}
\newc{\bed}{\begin{description}}
\newc{\eed}{\end{description}}
\newc{\ben}{\begin{enumerate}}
\newc{\een}{\end{enumerate}}

\newc{\be}{\begin{equation}}
\newc{\ee}{\end{equation}}
\newc{\bea}{\begin{eqnarray}}
\newc{\eea}{\end{eqnarray}}
\newc{\ra}{\rightarrow}

\newc{\alphas}{\ensuremath{\alpha_s}}
\newc{\alphatwo}{\ensuremath{\alpha_2}}
\newc{\alphaone}{\ensuremath{\alpha_1}}
\newc{\alphai}[1]{\ensuremath{\alpha_{#1}}}
\newc{\alphaem}{\ensuremath{\alpha_{\mathrm{em}}}}
\newc{\alphaeff}{\ensuremath{\alpha_{\mathrm{eff}}}}
\newc{\sineff}{\ensuremath{\sin \theta_{\mathrm{eff}}}}
\newc{\sinsqeff}{\ensuremath{\sin^2 \theta_{\mathrm{eff}}}}
\newc{\dalphahad}{\ensuremath{\Delta \alpha_{\mathrm{had}}}}
\newc{\yt}{\ensuremath{h_t}} \newc{\yb}{\ensuremath{h_b}} \newc{\ytau}{\ensuremath{h_{\tau}}}
\newc\mz{\ensuremath{M_Z}} 
\newc\mw{\ensuremath{m_W}}
\newc\mZ{\mz}        \newc\mW{\mw}
\newc\mhsm{\ensuremath{ m_{H_{\mathrm{SM}}}}}
\newc{\mtop}{\ensuremath{ m_t}}               \newc{\mtpole}{\ensuremath{ M_t}}
\newc{\mbottom}{\ensuremath{ m_b}} 
\newc{\mtau}{\ensuremath{ m_{\tau}}}
\newc{\mt}{\mtpole}
\newc{\mb}{\mbottom} 
\newc{\rtwogg}{\ensuremath{R_{h_2}(\gamma\gamma)}}
\newc{\rtwozz}{\ensuremath{R_{h_2}(ZZ)}}
\newc{\ronegg}{\ensuremath{R_{h_1}(\gamma\gamma)}}
\newc{\ronezz}{\ensuremath{R_{h_1}(ZZ)}}
\newc{\rsiggg}{\ensuremath{R_{h_\textrm{sig}}(\gamma\gamma)}}
\newc{\rsigzz}{\ensuremath{R_{h_\textrm{sig}}(ZZ)}}
\newc{\llbar}{\ensuremath{\ell\bar{\ell}}}
\newc{\tauptaum}{\ensuremath{ \tau^+\tau^-}}
\newc{\qqbar}{\ensuremath{ q\bar{q}}} \newc{\ppbar}{\ensuremath{ p\bar{p}}}
\newc{\bbbar}{\ensuremath{ b\bar{b}}} \newc{\ttbar}{\ensuremath{ t\bar{t}}}
\newc{\ffbar}{\ensuremath{ f\bar{f}}} \newc{\tautaubar}{\ensuremath{ \tau\bar{\tau}}}
\newc{\mchi}{\ensuremath{m_\neutone}}
\newc{\squark}{\ensuremath{\tilde{q}}}
\newc{\slepton}{\ensuremath{\tilde{l}}}
\newc{\gluino}{\ensuremath{\tilde{g}}} 
\newc{\mgluino}{\ensuremath{{m_{\gluino}}}}

\newc{\sthw}{\ensuremath{ \sin\theta_W}}              \newc{\cthw}{\ensuremath{\cos\theta_W}}
\newc{\tanthw}{\ensuremath{ \tan\theta_W}}              \newc{\cotthw}{\ensuremath{\cot\theta_W}}
\newc{\ssqthw}{\ensuremath{\sin^2 \theta_W}}
\newc{\msbar}{\ensuremath{\overline{MS}}} \newc{\drbar}{\ensuremath{\overline{DR}}}
\newc{\mtmtsmmsbar}{\ensuremath{ m_t(m_t)^{\msbar}_{{\mathrm{SM}}}}}
\newc{\mtmtsmdrbar}{\ensuremath{ m_t(m_t)^{\drbar}_{{\mathrm{SM}}}}}
\newc{\mtmtmssmdrbar}{\ensuremath{ m_t(m_t)^{\drbar}_{{\mathrm{SUSY}}}}}
\newc{\mbmbmsbar}{\ensuremath{ m_b(m_b)^{\msbar} }}
\newc{\mbmbsmmsbar}{\ensuremath{ m_b(m_b)^{\msbar}_{{\mathrm{SM}}}}}
\newc{\mbmzsmmsbar}{\ensuremath{ m_b(\mz)^{\msbar}_{{\mathrm{SM}}}}}
\newc{\mbmzsmdrbar}{\ensuremath{ m_b(\mz)^{\drbar}_{{\mathrm{SM}}}}}
\newc{\mbmzmssmdrbar}{\ensuremath{ m_b(\mz)^{\drbar}_{{\mathrm{SUSY}}}}}
\newc{\mtaumzsmmsbar}{\ensuremath{ m_{\tau}(\mz)^{\msbar}_{{\mathrm{SM}}}}}
\newc{\mtaumzsmdrbar}{\ensuremath{ m_{\tau}(\mz)^{\drbar}_{{\mathrm{SM}}}}}
\newc{\mtaumzmssmdrbar}{\ensuremath{ m_{\tau}(\mz)^{\drbar}_{{\mathrm{SUSY}}}}}
\newc{\alphasmzms}{\ensuremath{\alpha_s(M_Z)^{\overline{MS}}}}
\newc{\alphaimzms}[1]{\ensuremath{\alpha_{#1}(M_Z)^{\overline{MS}}}}
\newc{\alphaemmz}{\ensuremath{\alpha_{\mathrm{em}}(M_Z)^{\overline{MS}}}}

\newc{\mzero}{\ensuremath{{m_0}}}
\newc{\mhalf}{\ensuremath{ m_{1/2}}}
\newc{\tanb}{\ensuremath{\tan\beta}}
\newc{\azero}{\ensuremath{ A_0}}
\newc{\sgnmu}{\ensuremath{\textrm{sgn}\,\mu}}
\newc{\atau}{\ensuremath{{A_{\tau}}}}
\newc{\mueff}{\ensuremath{\mu_{\rm{eff}}}}
\newc{\lam}{\ensuremath{{\lambda}}}
\newc{\kap}{\ensuremath{{\kappa}}}
\newc{\alam}{\ensuremath{{A_{\lambda}}}}
\newc{\akap}{\ensuremath{{A_{\kappa}}}}
\newc{\hs}{\ensuremath{ H_s}}      
\newc{\mhs}{\ensuremath{ m_{H_s}}} 
\newc{\mgut}{\ensuremath{ M_{\rm GUT}}}
\newc{\mplanck}{\ensuremath{ M_{\rm P}}}      \newc{\mpl}{\ensuremath{ M_{\rm Pl}}}
\newc{\msusy}{\ensuremath{ M_{\rm SUSY}}}      \newc{\ms}{\ensuremath{ M_{\rm S}}}
 \newc{\hu}{\ensuremath{ H_u}}       \newc{\hd}{\ensuremath{ H_d}}
 \newc{\mhu}{\ensuremath{ m_{H_u}}}       \newc{\mhd}{\ensuremath{ m_{H_d}}}
 \newc{\mhuew}{\ensuremath{ m^{\ast}_{H_u}}}       \newc{\mhdew}{\ensuremath{ m^{\ast}_{H_d}}}
 \newc{\mhuewsq}{\ensuremath{ m^{\ast\, 2}_{H_u}}}       \newc{\mhdewsq}{\ensuremath{ m^{\ast\, 2}_{H_d}}}
 \newc{\mhl}{\ensuremath{m_\hl}} 
 \newc{\mhone}{\ensuremath{m_{h_1}}} 
 \newc{\mhtwo}{\ensuremath{m_{h_2}}} 
 \newc{\mglu}{\ensuremath{m_{\tilde g}}} 
 \newc{\mul}{\ensuremath{m_{\tilde{u}_L}}} 
 \newc{\mstopone}{\ensuremath{m_{\tilde{t}_1}}} 
 \newc{\ma}{\ensuremath{m_A}} 
 \newc{\maone}{\ensuremath{m_{a_1}}} 
 \newc{\matwo}{\ensuremath{m_{a_2}}}
 \newc{\hone}{\ensuremath{h_1}}
 \newc{\htwo}{\ensuremath{h_2}}
 \newc{\aone}{\ensuremath{a_1}}
 \newc{\atwo}{\ensuremath{a_2}}

\newc{\sigsip}{\ensuremath{\sigma^{\rm SI}_{p}}}	\newc{\sigsin}{\ensuremath{\sigma^{\rm SI}_{n}}}
\newc{\sigsdp}{\ensuremath{\sigma^{\rm SD}_{p}}}	\newc{\sigsdn}{\ensuremath{\sigma^{\rm SD}_{n}}}
\newc{\sigsi}{\ensuremath{\sigma^{\rm SI}}}	\newc{\sigsd}{\ensuremath{\sigma^{\rm SD}}}
\newc{\abund}{\ensuremath{ \Omega h^2}}
\newc{\omegadm}{\ensuremath{ \Omega_{{\rm DM}}}}     \newc{\abunddm}{\ensuremath{ \Omega_{{\rm DM}} h^2}} 
\newc{\omegam}{\ensuremath{ \Omega_{{\rm m}}}}       \newc{\abundm}{\ensuremath{ \Omega_{{\rm m}} h^2}}
\newc{\omegab}{\ensuremath{ \Omega_{{\rm b}}}}	\newc{\abundb}{\ensuremath{ \Omega_{{\rm b}} h^2}}
\newc{\omegatot}{\ensuremath{ \Omega_{{\rm TOT}}}}
\newc{\omegacdm}{\ensuremath{ \Omega_{{\rm CDM}}}}   \newc{\abundcdm}{\ensuremath{ \Omega_{{\rm CDM}} h^2}}
\newc{\omegalambda}{\ensuremath{ \Omega_{\Lambda}}} \newc{\abundlambda}{\ensuremath{ \Omega_{\Lambda} h^2}}
\newc{\omegarad}{\ensuremath{ \Omega_{{\rm rad}}}}  \newc{\abundrad}{\ensuremath{ \Omega_{{\rm rad}} h^2}}
\newc{\rhocrit}{\ensuremath{ \rho_{\rm crit}}}
\newc{\rhochi}{\ensuremath{ \rho_{\chi}}}
\newc{\abunchi}{\ensuremath{\Omega_\chi h^2}}
\newc{\abundlsp}{\ensuremath{\Omega_{\rm LSP}h^2}}
\newcommand*{\abundchi}{\ensuremath{\Omega_\chi h^2}}

\newc{\amu}{\ensuremath{ a_{\mu}}}        \newc{\amususy}{\ensuremath{ a_{\mu}^{\mathrm{SUSY}}}}
\newc{\amuexpt}{\ensuremath{ a_{\mu}^{\mathrm{expt}}}}        \newc{\amusm}{\ensuremath{ a_{\mu}^{\mathrm{SM}}}}
\newc\deltaamu{\ensuremath{\Delta a_{\mu}}} \newc{\deltaamususy}{\ensuremath{\delta a_{\mu}^{\mathrm{SUSY}}}}
\newc\gmtwo{\ensuremath{ (g-2)_{\mu}}} 
\newc{\deltagmtwomususy}{\ensuremath{\delta\left(g-2\right)_{\mu}^{\mathrm{SUSY}}}}
\newc{\deltagmtwomu}{\ensuremath{\delta\left(g-2\right)_{\mu}}}
\newc\BR{\ensuremath{\textrm{BR}}}
\newc\bsgamma{\ensuremath{ b\rightarrow s \gamma }}
\newc\bxsgamma{\ensuremath{\overline{B}\rightarrow X_{s}\gamma}}
\newc\brbsgamma{\ensuremath{\BR\left(\bsgamma\right)}}
\newc\brbxsgamma{\ensuremath{\BR\left(\bxsgamma\right)}}
\newc\bsmumu{\ensuremath{B_s\to\mu^+\mu^-}}
\newc\brbsmumu{\ensuremath{\BR\left(B_s\to\mu^+\mu^-\right)}}
\newc\bdmmumu{\ensuremath{\overline{B}_d\to\mu^+\mu^-}}
\newc\bbbarmix{\ensuremath{\overline{B}_s\mbox{-}B_s}}      
\newc\delmbs{\ensuremath{\Delta M_{B_s}}}
\newc{\butaunu}{\ensuremath{B_u \rightarrow \tau \nu}}
\newc{\brbutaunu}{\ensuremath{\BR\left(B_u \rightarrow \tau \nu\right)}}

\newcommand*{\reftable}[1]{Table~\ref{#1}}         
       
\newcommand*{\reffig}[1]{Fig.~\ref{#1}}

     \newcommand*{\refsec}[1]{Sec.~\ref{#1}}

\newcommand*{\neutone}{\ensuremath{\chi}}
\newcommand*{\neuttwo}{\ensuremath{{\chi}^0_2}}

\newcommand*{\charone}{\ensuremath{{\chi}^{\pm}_1}}

\newcommand*{\stau}{\ensuremath{\tilde{\tau}}}

\newcommand*{\eight}{\ensuremath{\sqrt{s}=8\tev}}
\newcommand*{\alphaT}{\ensuremath{\alpha_T}}

\newcommand*{\alphaTelefb}{\ensuremath{\cms\ \alphaT\ 11.7\invfb} }

\newcommand*{\cms}{\text{CMS}}

\newcommand*{\xenon}{\text{XENON100}}

\let\oldcite\cite
\renewcommand*{\cite}{~\oldcite}

\newcommand*{\hl}{\ensuremath{h}}

\newcommand*{\mh}{\ensuremath{m_h}}

\restylefloat{figure}

\begin{document}

\title{Dark matter and collider signatures of the MSSM}

\author{Andrew Fowlie,$^1$~Kamila Kowalska,$^2$~Leszek Roszkowski,$^2$\footnote{On leave of absence from the University of Sheffield, U.K.}\\
~Enrico Maria Sessolo,$^2$~and Yue-Lin Sming Tsai$^2$ \\[2ex]
\small\it BayesFITS Group \\ $^1$ Department of Physics and Astronomy, University of
  Sheffield, Sheffield S3 7RH, England \\
 $^2$ National Centre for Nuclear Research,
  Ho{\. z}a 69, 00-681 Warsaw, Poland \\
\bigskip \\
\url{a.fowlie@sheffield.ac.uk, Kamila.Kowalska@fuw.edu.pl}\\ \url{L.Roszkowski@sheffield.ac.uk,Enrico-Maria.Sessolo@fuw.edu.pl},\\
\url{Sming.Tsai@fuw.edu.pl}}


\date{}

\maketitle

\begin{abstract}
  We explore the MSSM with 9 free parameters (p9MSSM) that have
  been selected as a minimum set that allows an investigation of
  neutralino dark matter and collider signatures while maintaining
  consistency with several constraints.  These include measurement
  of the dark matter relic density from PLANCK, main properties of the discovered Higgs boson,
  LHC direct SUSY searches, recent evidence for a Standard Model-like \brbsmumu,
  and the measurement of \deltagmtwomu, plus a number of other
  electroweak and flavor physics constraints.  We perform a
  simulation of two LHC direct SUSY searches at $\sqrt{s}=8\tev$: the
  CMS inclusive $\alpha_T$ search for squarks and gluinos and the CMS
  electroweak production search with $3l+E_T^{\textrm{miss}}$ in the
  final state.  We use the latter to identify the regions of the
  parameter space, consistent at $2\sigma$ with \deltagmtwomu, that are not
  excluded by the direct limits from the electroweak production.  We find
  that they correspond to a neutralino mass in the window $200\gev
  \lesssim\mchi \lesssim 500\gev$.  We also implement the likelihood
  for the XENON100 exclusion bound, in which we consider for the first
  time the impact of a recent determination of the $\Sigma_{\pi N}$
  term from CHAOS data, $\Sigma_{\pi N}=43\pm12\mev$.  We show that in
  light of this measurement, the present statistical impact of the
  XENON100 bound is greatly reduced, although future sensitivities of
  the LUX and XENON1T experiments will have decisive impact on the mixed bino/higgsino
  composition of the neutralino.  We point out some tension
  between the constraints from \deltagmtwomu\ and XENON100.  Finally, we
  present prospects for various indirect
  searches of dark matter, namely $\gamma$-ray fluxes from dSphs and
  the Galactic Center at Fermi-LAT, and the positron flux at AMS02.  We
  also show the 5-year sensitivity on the spin-dependent
  neutralino-proton cross section due to neutrino fluxes from the Sun at
  IceCube.
\end{abstract}
\newpage

\section{\label{sec:intro}Introduction}
The recent discovery of a particle consistent with the Standard Model
(SM)\footnote{The abbreviations used 
  throughout the main text of this paper are summarized in Appendix~A.}  Higgs boson at the
LHC\cite{Chatrchyan:2012ufa,*Aad:2012tfa} provides a strong constraint
on models of low-scale supersymmetry (SUSY).  According to the latest
measurements, the mean value of the particle's mass at CMS is
$125.7\pm0.4$\gev\cite{CMS-PAS-HIG-13-005} and at ATLAS
$125.5\pm0.6$\gev\cite{ATLAS-CONF-2013-014}.  It was shown in many
studies\cite{Hall:2011aa,Baer:2011ab,Heinemeyer:2011aa,Arbey:2011ab,*Arbey:2012dq,
  Carena:2011aa,Cao:2012fz,Christensen:2012ei,Brummer:2012ns} that the
predictions of the Minimal Supersymmetric Standard Model (MSSM) for
the lightest Higgs boson are consistent with the measured value if
large radiative corrections are added to the tree-level mass.  If the
observed new particle is the lightest $CP$-even Higgs $h$ of the MSSM,
$\mhl\simeq 126\gev$ can be achieved, at one loop, either by requiring
the stop masses in the multi-\tev\ range or by requiring maximal stop
mixing, $|X_t|/\msusy\simeq\sqrt{6}$.  Additionally, if one requires
that the present observed value of the relic density is due
exclusively to neutralino dark matter (DM), the fact that only a
limited number of mechanisms can enhance the annihilation cross
section up to the value measured by WMAP or PLANCK puts even more
severe
constraints\cite{Arbey:2012na,Carena:2012gp,Mohanty:2013soa,Han:2013gba}
on the parameter space of models defined at the low-energy SUSY
breaking scale, and even more
so\cite{Ellis:2012aa,Baer:2012uy,Bechtle:2012zk,Balazs:2012qc,Fowlie:2012im,Buchmueller:2012hv,Akula:2012kk,Strege:2012bt,Cabrera:2012vu,Kowalska:2013hha}
in SUSY models constrained by universality conditions at the scale of
grand unification (GUT), like the Constrained MSSM
(CMSSM)\cite{hep-ph/9312272} and the Non-Universal Higgs Mass model
(NUHM)\cite{Berezinsky:1995cj}. In particular, it was
shown, first  in the MSSM\cite{Profumo:2004at} and more recently 
in unified SUSY models\cite{Roszkowski:2009sm,Akula:2012kk,Strege:2012bt,Cabrera:2012vu,Kowalska:2013hha},
that in addition to the previously favored gaugino-like neutralino,
the right relic density can be obtained also with a nearly pure higgsino
lightest SUSY particle (LSP) with mass around 1\tev.

The above picture is in agreement with the recent positive
observation\cite{Aaij:2012nna} of
$\brbsmumu=3.2^{+1.5}_{-1.2}\times10^{-9}$ at LHCb: if this value
converges in future measurements with higher integrated luminosity to
the SM expectation, about
$3.6\times10^{-9}$\cite{Buras:2012ru,Buras:2013uqa}, then it was shown
in\cite{Kowalska:2013hha} that this measurement will strongly disfavor
a substantial part of the CMSSM (specifically, the $A$-funnel (AF)
region), unless the value of the pseudoscalar Higgs mass \ma\ is,
again, in the multi-TeV regime.  The described picture is also in
agreement with the nonobservation of SUSY particles at the LHC so
far. All of this suggests that if low-energy SUSY exists and takes one of the
forms that theorists have hypothesized, it probably comes with masses
larger than the energy scale tested by the most recent LHC runs.
Recent mass limits calculated with simplified model spectra (SMS),
with highly simplified decay chains, which on a case-by-case basis
assume only a few SUSY masses in reach of experimental observation
whereas the others are decoupled, imply that when the neutralino is
very light, the mass of the gluino should be greater than
$\sim1300\gev$\cite{ATLAS-CONF-2012-145,ATLAS-CONF-2012-109}, the
masses of the first- and second-generation squarks should be at least
as heavy\cite{ATLAS-CONF-2012-109}, the masses of third-generation
squarks should exceed
600--700\gev\cite{ATLAS-CONF-2013-037,ATLAS-CONF-2012-165,CMS-PAS-SUS-13-011},
and the mass of the chargino should be greater than $\sim
650\gev$\cite{CMS-PAS-SUS-12-022}.  (We emphasize again that these
limits are, however, obtained under assumptions that are, in general, not
applicable to realistic SUSY models.)

The only piece of information suggesting the sub-TeV scale of SUSY masses is
$\deltagmtwomu=(28.7\pm8.0)\times10^{-10}$\cite{Bennett:2006fi}, 
which is more than $3\sigma$ in excess of zero (axiomatically its SM value),
and which seems to favor light smuons, muon sneutrinos, charginos, and neutralinos, 
if the discrepancy is to be explained within SUSY.
It has been shown in several studies (see, e.g.,\cite{Cabrera:2010xx,Bechtle:2012zk,Fowlie:2012im})
that this cannot be achieved in constrained models, due to the fact that
the soft scalar masses for squarks and sleptons are unified at the GUT scale.
On the other hand, relaxing some of the unification conditions, so to allow the first-two-generation sleptons 
to be light, will easily reduce the tension with the \gmtwo\ constraint, and still maintain consistency 
with the above picture.

Recently, there has been great proliferation of studies investigating the impact of different constraints on 
phenomenological parametrizations of the MSSM. The idea is to treat
the $n$ SUSY-scale parameters of the MSSM that parametrize the sector of the theory to be studied as free,
and constrain the remaining ones by some unifying conditions, or fix them at some decoupled value. 
Most of the papers on these ``p$n$MSSM'' models analyzed the impact of constraints
from direct LHC SUSY searches\cite{CahillRowley:2012gu,AbdusSalam:2012sy,Conley:2010du,CahillRowley:2012cb,AbdusSalam:2012ir,Choudhury:2013jpa,Cahill-Rowley:2013dpa}, 
from the Higgs discovery\cite{Haisch:2012re,Arbey:2012na,Arbey:2012bp,Mahmoudi:2012eh,Mahmoudi:2012ei,Carena:2012he,Carena:2012np,Boehm:2013qva},
from $b$-physics\cite{Arbey:2011aa,Arbey:2012ax,Mahmoudi:2012un,Altmannshofer:2012ks,Boehm:2013qva}, 
and from direct and indirect detection of DM\cite{Arbey:2012na,Mahmoudi:2012ei,Cahill-Rowley:2013dpa,Boehm:2013qva}.
Also, in the past few months there has been a resurgence of interest in scenarios 
that can reconcile the \gmtwo\ anomaly
with the most recent experimental determination from other sectors, 
like the LHC\cite{Endo:2013bba}, DM\cite{Mohanty:2013soa}, the Higgs\cite{Ibe:2013oha}, and lepton-flavor violation\cite{Arana-Catania:2013nha,Zhang:2013hva} 
in SUSY, and also in other beyond-the-SM models\cite{Lee:2013fda}.         

A detailed study of the impact of the experimental 
constraints described above on the
DM sector of a MSSM parametrization with 13 free parameters (p13MSSM) was presented in\cite{Han:2013gba}.
The study focused primarily on neutralino DM in the sub-TeV regime, 
thus only touching the 
parameter space giving higgsino DM at 1~TeV, which has been shown to be an important candidate 
in studies of GUT-constrained model, as explained at the beginning of this section.
We think it would therefore be interesting to also investigate how this sector of the theory 
agrees with the global set of constraints. 
Moreover, it has been demonstrated in many studies (see, e.g.,\cite{Akrami:2010cz}) that a 
proper treatment of the experimental constraints through a likelihood function can lead to significantly different results from  
scans where such constraints are typically implemented in a more simplified boxlike fashion, with
all points accepted when satisfying experimental values within some fixed ranges (e.g., 95\%~C.L.), and otherwise rejected. 

In this paper we perform a statistical analysis of the MSSM at the
SUSY scale with 9 free parameters (p9MSSM), which we identify as a
minimal set of parameters to allow good agreement with constraints
from the relic density, the Higgs mass and decay rates,
\deltagmtwomu, $b$-physics constraints, limits from the XENON100 DM
direct detection (DD) experiment, limits from LHC direct SUSY
searches, and limits from indirect detection (ID) of DM.

To this end we construct approximate but accurate likelihood functions
to incorporate in our analysis limits on SUSY from the
$\sqrt{s}=8\tev$ \alphaT\ inclusive search for squarks and gluinos
with 11.7/fb integrated luminosity\cite{Chatrchyan:2013lya} and the
$\sqrt{s}=8\tev$, 9.2/fb, $3l+E_T^{\textrm{miss}}$ electroweak (EW)
production search at CMS\cite{CMS-PAS-SUS-12-022}.  The latter in
particular gives the strongest limits on chargino-neutralino pair
production to date, so that our detailed implementation can accurately
constrain the sector of the theory consistent with
\deltagmtwomu. Likewise we construct an approximate but accurate
likelihood function to incorporate limits on the spin-independent (SI)
neutralino-nucleon cross section, \sigsip, from
XENON100\cite{Aprile:2012nq}.  We include the theoretical
uncertainties due to the determination of the $\Sigma_{\pi N}$
term\cite{Gasser:1990ce}, in light of the recent determination from
CHAOS data presented in\cite{Stahov:2012ca}.

We will focus, in particular, on the DM sector of the model, by not
only including an accurate implementation of the XENON100 limit, but
also by investigating in detail the impact of ID searches:
$\gamma$-ray fluxes from the Galactic Center (GC) of the Milky Way and
from its dwarf spheroidal satellite galaxies (dSphs) at Fermi-LAT, and
positron fluxes at AMS02.  We will also try to evaluate the reach of
future DM experiments for the parameter space of the model.
   
The article is organized as follows. In \refsec{sec:method} we briefly
revisit the model, highlighting some of its salient features, and detail our methodology, including our statistical
approach.
In \refsec{sec:expcons} we describe the relevant experimental results and their likelihood functions, 
including a detailed discussion of our derivation of the bounds on SUSY from direct searches at the LHC, 
and of the bound from XENON100.
In \refsec{Results} we present the results from our scans
and discuss their novel features. We summarize our findings in
\refsec{Summary}.

\section{\label{sec:method}Model description and numerical methodology}

As mentioned in the Introduction, and shown in many studies (see, e.g., our previous
work\cite{Fowlie:2012im,Kowalska:2012gs,Kowalska:2013hha,Munir:2013wka}),
the discovery of the Higgs mass has made the multi-TeV scale of
$\msusy\approx m_{\tilde{t}}$ difficult to avoid in constrained SUSY
models (although, given the theoretical uncertainties, in the
stau-coannihilation region of the CMSSM the mass of the lightest Higgs
is consistent with the measured value even if $\msusy\simeq 1\tev$,
thanks to maximal
$|X_t|/\msusy$\cite{Fowlie:2012im,Kowalska:2013hha}).  On the other
hand, several groups\cite{Arbey:2012bp,Arbey:2011aa,Cotta:2011ht} have
shown that in the general MSSM, the parameter space is much less
constrained given the large number of free parameters.

In this paper we investigate the impact of the global constraints 
on the p9MSSM, whose 9 free parameters have been chosen on the basis of their relevance 
for the constraints involving DM, the Higgs sector, and other relevant quantities.  
In this section, we will describe our p9MSSM parameter choice.

We make some reasonable and simplifying assumptions, as is usually
done in the MSSM. First, we assume universality for the bino and wino
masses, $M_1=0.5 M_2$, resulting in the absence of wino-like
neutralino as the LSP.  It is known that a wino-like neutralino can
hardly satisfy the relic density constraint, unless taken to be very
heavy ($\mchi\gsim1.6\tev$\cite{Profumo:2004at,ArkaniHamed:2006mb} but
see\cite{Cahill-Rowley:2013dpa} for very recent numerical work on the
issue), and it has been shown to be in potential conflict with ID
experiments due to its large annihilation cross
section\cite{Bertone:2011pq}.  On the other hand, in order to mitigate
possible impacts on the DM sector from LHC multi-jet limits, we treat
the gluino mass $M_3\approx m_{\tilde{g}}$ as a free parameter.  We
scan it in the range $0.7\tev<M_3 <8\tev$ since lower values are now
disfavored by most LHC direct SUSY searches.

The squarks of the first two generations are strongly constrained by direct searches
at the LHC and are basically irrelevant for the constraints that we will employ. 
We therefore fix them at a decoupled scale,
$m_{\tilde{Q}_{1,2}}=m_{\tilde{u}_R^{(1,2)}}=m_{\tilde{d}_R^{(1,2)}}=2.5\tev$.
Instead, we allow wide ranges for the third-generation squark masses, $m_{\tilde{Q}_3}=m_{\tilde{t}_R}=m_{\tilde{b}_R}$,
because they are not constrained by the LHC as strongly
as the first and second generations, and they affect the Higgs sector.

In order to save computer time and make sure we do not generate many points with charged LSP, 
which would then be rejected,
we unify the first- and second-generation sleptons and set them to 
$m_{\tilde{L}_{1,2}}=m_{\tilde{e}_R}=m_{\tilde{\mu}_R}=M_1+50\gev$. 
It will be shown that, when the relic density constraint is taken into account, this choice does not 
compromise good coverage of the sector related to \gmtwo.

On the other hand, we scan over $m_{\tilde{L}_3}=m_{\tilde{\tau}_R}$ 
to investigate scenarios where the relic density is obtained via stau-coannihilation.

We unify all the lepton trilinear couplings
to $A_{e}=A_{\mu}=A_{\tau}$. All up-type squark trilinear couplings
are unified to $A_{t}$ but down-type squark trilinear couplings
are fixed to $A_b=-0.5\tev$, since our likelihood function will not
be sensitive to down-type squark trilinear couplings.
We allow broad ranges for $A_t$ and $A_{\tau}$ to investigate
the Higgs sector and mechanisms for stau-coannihilation, respectively.

We investigate the Higgs sector by scanning over \ma, the Higgs/higgsino mass parameter $\mu$, and the ratio 
of the Higgs doublets' vacuum expectation values (vev), \tanb.

In summary, our multi-dimensional scan is parametrized by 9 free SUSY parameters:
\begin{equation*}
M_2,\,M_3,\,m_{\tilde{Q}_3},\, m_{\tilde{L}_3},\, A_t,\, A_{\tau},\, \ma,\, \mu,\, \tanb\,.
\end{equation*}

In addition, we scan over the bottom quark mass, 
$\mbmbmsbar$, and top quark pole mass, $\mtpole$,
to include SM uncertainties.
In contrast to our previous papers, we do not vary the strong interaction coupling, $\alphas$,
or the electromagnetic coupling $\alphaem$ because they are well constrained.
Moreover, \alphaem\ is used in CMSSM and NUHM scans to set the GUT scale, which
is not an issue in here, since soft SUSY-breaking parameters are defined at the scale 
$\msusy=\sqrt{ m_{\tilde{t}_1} m_{\tilde{t}_2} }$, which is found iteratively by our
RGE code (\texttt{softsusy-3.3.5}\cite{softsusy}).

The set of input parameters we consider in this paper and their scanned ranges are shown in Table~\ref{table:MSSMparams}.

\begin{table}[t]
\begin{center}
\begin{tabular}{|c|c|}
\hline
\hline
Parameter & Range \\
\hline\hline
gluino mass & $0.7<M_3 <8$ \\
wino mass & $0.01 <M_2 <4$ \\
bino mass &  $M_1=0.5 M_2$ \\
stop trilinear coupl. &$-7<A_{t}<7 $\\
$\tau$ trilinear coupl. &$-7<A_{\tau}<7 $\\
sbottom trilinear coupl. &$A_{b}=-0.5 $\\
pseudoscalar mass & $0.2<m_A<4$ \\
$\mu$ parameter & $0.01<\mu<4$ \\
3rd gen. soft squark mass & $0.3<m_{\tilde{Q}_3}<4$ \\
3rd gen. soft slepton mass & $0.1<m_{\tilde{L}_3}<2$ \\
1st/2nd gen. soft slepton mass & $m_{\tilde{L}_{1,2}}=M_1 +50\gev$ \\
1st/2nd gen. soft squark mass & $m_{\tilde{Q}_{1,2}}=2.5$ \\
ratio of Higgs doublet VEVs & $3<\tan\beta <62$ \\
\hline
Nuisance parameter & Central value, error\\
\hline
\hline
Bottom mass \mbmbmsbar\ (GeV) & (4.18, 0.03) \\
Top pole mass $\mtpole$ (GeV)& (173.5, 1.0) \\
\hline
\hline
\end{tabular}
\caption{
Prior ranges for our p9MSSM input parameters, over which we perform our scan. 
All masses and trilinear couplings are in\tev, unless indicated otherwise.}
\label{table:MSSMparams}
\end{center}

\end{table}

We perform our scans by using the package BayesFITS,
which interfaces different publicly available programs.
For sampling, it uses \texttt{MultiNest}\cite{Feroz:2008xx} with 20000 live points,
evidence tolerance factor equal to $10^{-4}$,
and sampling efficiency equal to 0.8.
We compute mass spectra and $\mw$ with \texttt{softsusy-3.3.5}\cite{softsusy}
and pass the spectra via SUSY LesHouches Accord to \texttt{superiso v3.3}\cite{superiso}
to calculate \brbxsgamma, \brbsmumu, \brbutaunu,
and \deltagmtwomususy. DM observables, such as the relic density and
DD and ID observables, are calculated with
\texttt{MicrOMEGAs 2.4.5}\cite{micromegas} and \texttt{DarkSUSY}\cite{darksusy}.
Higgs cross sections,
$\delmbs$, and $\sineff$ are computed with \texttt{FeynHiggs 2.9.4}\cite{feynhiggs:06,feynhiggs:03,feynhiggs:00,feynhiggs:99}.
In addition, we also check the exclusion bounds
obtained from the Higgs searches at LEP and the Tevatron with \texttt{HiggsBounds 3.8.0}\cite{Bechtle:2011sb}.

Note that \texttt{MultiNest} is optimized for Bayesian sampling. The scans are driven 
by the likelihood function, and the input parameters are subject to prior distributions.
We combine six separate scans:
three with log priors in the mass parameters (with the exclusion of $\mu$)
and three with flat priors, so to obtain good coverage of the parameter space.
We take flat priors for the trilinear couplings, for $\mu$, and for $\tanb$ in all six scans.
Moreover, the nuisance parameters are always scanned over with Gaussian prior distributions.

We employ in this work the profile-likelihood approach, which we briefly summarize in the next section,
to draw inferences on the multi-dimensional parameter space of the p9MSSM.
The advantage of the profile-likelihood method with respect to our previous papers\cite{Fowlie:2011mb,Fowlie:2012im,Kowalska:2012gs},
where we were calculating Bayesian inferences based on the posterior probability
density function,
is that we can merge together many chains with different priors to
explore the whole parameter space in good detail,
without worrying about appropriate prior weights.


\section{\label{sec:expcons}Statistical treatment of experimental constraints}

In this section we briefly describe the profile-likelihood method and the experimental constraints
used in this analysis. 
We will later show our results as 68\%~($1\sigma$) and 95\%~($2\sigma$),
or 90\%~($1.65\sigma$)
confidence intervals in the p9MSSM parameter space.

For a theory described by a set of $n$ parameters $m$, one can compare experimental observables $\xi(m)$ with data
$d$ through the likelihood function $\mathcal{L}(m)\equiv p(d|\xi(m))$, which at any point 
$m$ in parameter space gives the probability of the data $d$ given $m$.
   
One can draw inference on a subset of $r\leq n$ specific model
parameters or observables, or a combination of both (collectively
denoted by $\psi_i$), by ``profiling'' the likelihood along the other
directions in the parameter space\cite{Rolke:2004mj,Akrami:2010cz},
\begin{equation}
\mathcal{L}(\psi_{i=1,..,r})=\max_{m\in\mathbb{R}^{n-r}}\mathcal{L}(m)\,.\label{profilike}
\end{equation}

Confidence intervals are calculated from tabulated values
of $\delchisq\equiv-2\ln(\mathcal{L}/\mathcal{L}_{\textrm{max}})$.
For example, in $r=2$ dimensions, 68.3\% confidence regions are given
by $\delta\chi^2=2.30$ and 95.0\% confidence regions by $\delta\chi^2=5.99$.  

\begin{table}[t]\footnotesize
\begin{center}
\begin{tabular}{|l|l|l|l|l|l|}
\hline
Measurement & Mean or range & Error:~exp.,~th. & Distribution & Ref.\\
\hline
\alphaTelefb, \eight & See text. 	& See text. 	& Poisson &\cite{Chatrchyan:2013lya}\\ 
\mhl\ (by CMS) & $125.8\gev$ & $0.6\gev, 3\gev$ & Gaussian &\cite{CMS-PAS-HIG-12-045} \\
\abundchi 			& $0.1199$ 	& $0.0027$,~$10\%$ 		& Gaussian &  \cite{Ade:2013zuv}\\
\brbxsgamma $\times 10^{4}$ 		& $3.43$   	& $0.22$,~$0.21$ 		& Gaussian &  \cite{bsgamma}\\
\brbutaunu $\times 10^{4}$          & $1.66$  	& $0.33$,~$0.38$ 		& Gaussian &  \cite{Amhis:2012bh}\\
$\Delta M_{B_s}$ & $17.719\ps^{-1}$ & $0.043\ps^{-1},~2.400\ps^{-1}$ & Gaussian & \cite{Beringer:1900zz}\\
\sinsqeff 			& $0.23146$     & $0.00012$, $0.00015$             & Gaussian &  \cite{Beringer:1900zz}\\
$M_W$                     	& $80.399\gev$      & $0.023\gev$, $0.015\gev$               & Gaussian &  \cite{Beringer:1900zz}\\
$\brbsmumu\times 10^9$			& 3.2
&  $+1.5, -1.2$, 10\% & Gaussian &  \cite{Aaij:2012nna}\\
\mbmbmsbar\ & 4.18\gev\ & 0.03\gev, 0 & Gaussian & \cite{Beringer:1900zz} \\
\mtpole\ & 173.5\gev & 1.0\gev, 0 & Gaussian & \cite{Beringer:1900zz} \\
\hline
\deltagmtwomususy $\times 10^{10}$ 	& $28.7 $  	& $8.0$,~$1.0$ 		& Gaussian &  \cite{Bennett:2006fi,Miller:2007kk} \\

XENON100 (2012) & See text. 	& See text. 	& Poisson &\cite{Aprile:2012nq}\\ 
\hline
CMS $3l+E_T^{\textrm{miss}}$ 9.2/fb, \eight & See text. 	& See text. 	& Poisson &\cite{CMS-PAS-SUS-12-022}\\
\hline
\end{tabular}

\caption{
The experimental constraints that we include in our likelihood functions to constrain our p9MSSM model.
We denote the first block of constraints as \textbf{basic}. 
} 
\label{tab:exp_constraints}
\end{center}
\end{table}

The constraints that we include in the likelihood function are
listed in Table~\ref{tab:exp_constraints}. For the purpose of presentation, we denote the constraints 
in the first block of the table as \textbf{basic},
as we shall explain in~\refsec{Results}.
As a rule, following the procedure developed
earlier\cite{deAustri:2006pe}, we implement positive measurements through
a Gaussian likelihood, in which the experimental and theoretical uncertainties are added in quadrature.
For the Higgs mass, we use the most recent CMS determination of its central value and experimental uncertainty, 
as it is in good agreement with the determination obtained by ATLAS at the end of the \eight\ run.
The theoretical uncertainty is estimated to be 3\gev\cite{Fowlie:2012im,Heinemeyer:2011aa}.
For the relic density we use the recent determination by PLANCK\cite{Ade:2013zuv}.
 
Additionally, we impose 95\%~C.L. lower limits from direct searches at LEP\cite{Beringer:1900zz}, smeared with 5\% theoretical errors,
for the following particles:

\begin{eqnarray}
\label{eq:LEPlimit}
 \mchi 	  &>& 46  \gev\,,\nonumber\\
 m_{\tilde{e}}   &>& 107 \gev\,,\nonumber\\
 m_{\charone}    &>& 94  \gev \text{~~if~}  m_{\charone}-\mchi    > 3 \gev \text{~and~} \tanb <40\,,\nonumber\\
 m_{\tilde{\mu}} &>& 94  \gev \text{~~if~}  m_{\tilde{\mu}}-\mchi > 10\gev \text{~and~} \tanb <40\,,\nonumber\\
 m_{\stau} 	  &>& 81.9\gev \text{~~if~}  m_{\stau_1}-\mchi      > 15\gev\,,\nonumber\\
 m_{\tilde{b}_1} &>& 89  \gev \text{~~if~}  m_{\tilde{b}_1}-\mchi  > 8 \gev\,,\nonumber\\
 m_{\tilde{t}_1} &>& 95.7\gev \text{~~if~}  m_{\tilde{t}_1}-\mchi  > 10\gev\,.\label{LEP}
\end{eqnarray}

The MSSM predictions for $\brbsmumu$ must be matched with what was measured by the LHCb experiment,
i.e., the time-averaged, flavor-averaged
branching ratio, whose measured value is given in Table~\ref{tab:exp_constraints}. 
This averaging incorporates oscillations between the $B_s$ and $\overline{B}_s$ flavor eigenstates between the
primary vertex and the secondary decay vertex, which occur in the experiment. 
This effect is nontrivial, because the flavor eigenstates have different
decay widths. Therefore, to connect with the experiment, one should multiply the theoretical prediction by a factor
that takes into account this effect\cite{DeBruyn:2012wj,DeBruyn:2012wk}, so that
\begin{equation}
\brbsmumu_{\textrm{time ave}}=
\textrm{corr}\cdot\brbsmumu_{\textrm{theory}}\,,\label{bsmtime}
\end{equation}
where the ``correction" factor is given, for example, in Ref.\cite{Arbey:2012ax}.

\texttt{SuperIso} gives $\brbsmumu^{\textrm{SM}}_{\textrm{time ave}}=3.87\times 10^{-9}$, 
which is higher than
the most recent time-averaged calculation, $\brbsmumu^{\textrm{SM}}_{\textrm{time ave}}=(3.56\pm 0.18)\times 10^{-9}$\cite{Buras:2013uqa}. 
The difference in the two values is due to differences in the chosen value of $f_{B_s}$ and other input parameters 
such as the $B_s$ lifetime, the top mass, and the CKM elements. Other sources of uncertainty 
arise from the chosen renormalization scheme for $\sin\theta_W$ and $M_t$\cite{Mahmoudipriv}.
Since the uncertainties enter multiplicatively in Eq.~(\ref{bsmtime}) as a correction factor 
and we assume a theoretical uncertainty of 10\% 
in our scans, we use the non time-averaged output of \texttt{SuperIso}, which gives in the SM limit
$\brbsmumu^{\textrm{SM}}=3.53\times10^{-9}$, consistent with the determination of\cite{Buras:2013uqa} within $1\sigma$.  
In addition, since the experimental uncertainty in $\brbsmumu$ is asymmetric,
we parametrize the likelihood function as a combination of two
half-Gaussians, one for positive and negative error each.

In Table \ref{tab:exp_constraints}, $\mh$ refers to the mass of the lightest
$CP$-even Higgs boson of the model.
In principle, it is possible for the heavier
$CP$-even Higgs, $H$, to be SM-like with mass $\sim 126\gev$, in the
so-called nondecoupling regime of the model. However, such a
possibility is confined to a very fine-tuned region of the
parameter space, which is essentially ruled out by the direct search limits on
the lightest Higgs $h$ and the pseudo-scalar Higgs $A$ and other experimental constraints from flavor
physics\cite{Christensen:2012ei,Carena:2013qia, Scopel:2013bba, Bhattacherjee:2013vga}. Hence we do not explore such a possibility
here.

The construction of the likelihood for limits from direct SUSY searches and the likelihood for XENON100 deserve a more
detailed explanation, which we give in the following subsections. 

\subsection{Likelihood for LHC direct SUSY searches\label{subsec:LHC}}

In order to implement the impact of direct SUSY search limits on the p9MSSM parameter
space,
we extend the procedure developed previously
in\cite{Fowlie:2012im,Kowalska:2013hha} for the derivation
of approximate but accurate likelihood maps.
For the first time, in this study we apply the SUSY likelihood functions on-the-fly,
point by point in our scan. Note that in our previous studies of constrained
models\cite{Fowlie:2012im,Kowalska:2012gs,Kowalska:2013hha} 
we relied on likelihood maps prepared separately for subsets
of input parameters.  
 
In this paper we apply the results from two different searches: CMS
inclusive search for
SUSY final states with large missing transverse energy and $b$-quark jets
in \eight\ $pp$ collisions
with the variable \alphaT\cite{Chatrchyan:2013lya}; and direct EW
production of charginos and neutralinos
in $pp$ collisions at \eight\cite{CMS-PAS-SUS-12-022}. Each search is
implemented through a step-by-step procedure
that includes generation of a SUSY signal at the scattering level with
\texttt{PYTHIA6.4}\cite{Sjostrand:2006za} and a simulation of the CMS detector response
with \texttt{PGS4}\cite{PGS4} to calculate the efficiency once the kinematic
cuts are applied. We modified the CMS detector card as recommended by the
CMS Collaboration,
and we tuned the algorithm used by \texttt{PGS4} to reproduce
the $b$-tagging efficiency reported by CMS. The obtained signal yields are finally
statistically compared to the publicly available observed and background
yields of the searches,
provided by the CMS Collaboration, as described in\cite{Kowalska:2013hha},
which updated the procedure described in detail
in\cite{Fowlie:2011mb,Fowlie:2012im}.

\paragraph{\underline{CMS \alphaT\ 11.7\invfb, \eight}}\mbox{}\\
The search employs a set of 8 different boxes, with hard jets and missing energy in the
final states,
and different combinations of $b$-tagged jets.
The boxes with zero $b$-tagged jets in the final states are used to impose
limits on
the squarks of the first two generations produced either directly
or in gluino decays.
Increasing numbers of $b$-tagged jets in the final state implies direct
production of
third generation squarks, or gluinos decaying to the latter.
The boxes, together with the number of the observed and background
events provided by the CMS Collaboration, are given in\cite{alphatsite}.

\begin{figure}[t]
\centering
\subfloat[]{%
\label{fig:a}%
\includegraphics[width=0.50\textwidth]{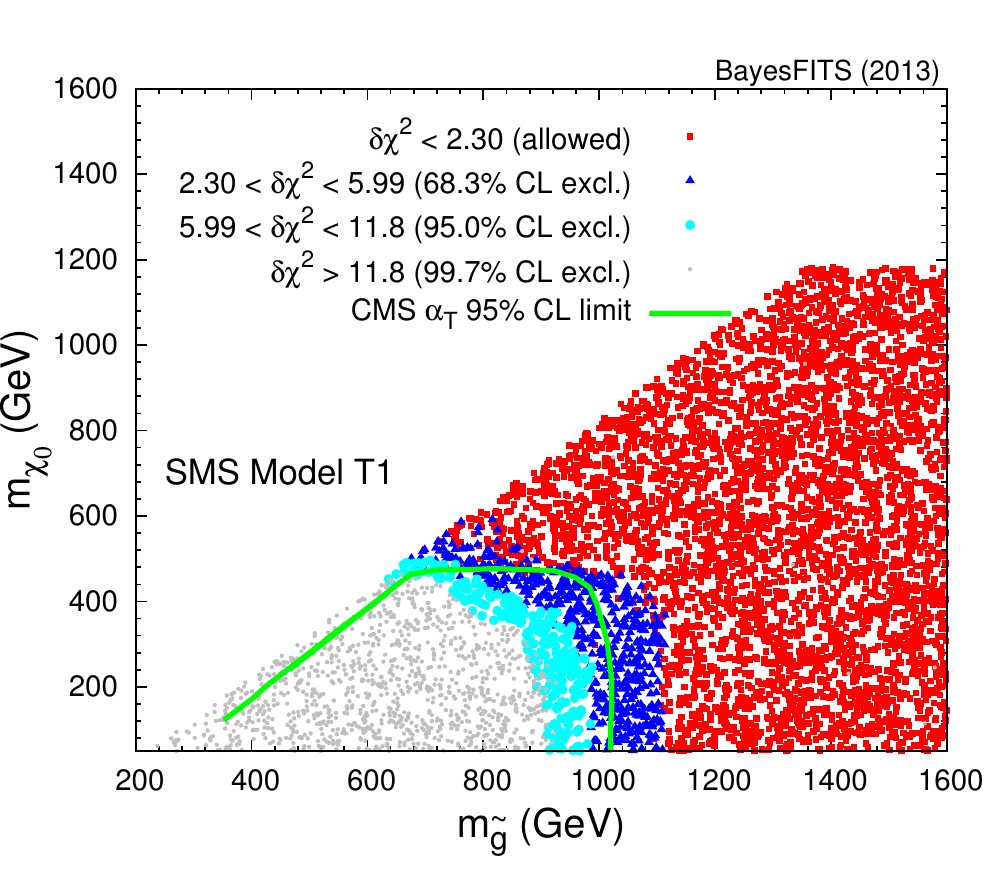}
}%
\subfloat[]{%
\label{fig:b}%
\includegraphics[width=0.50\textwidth]{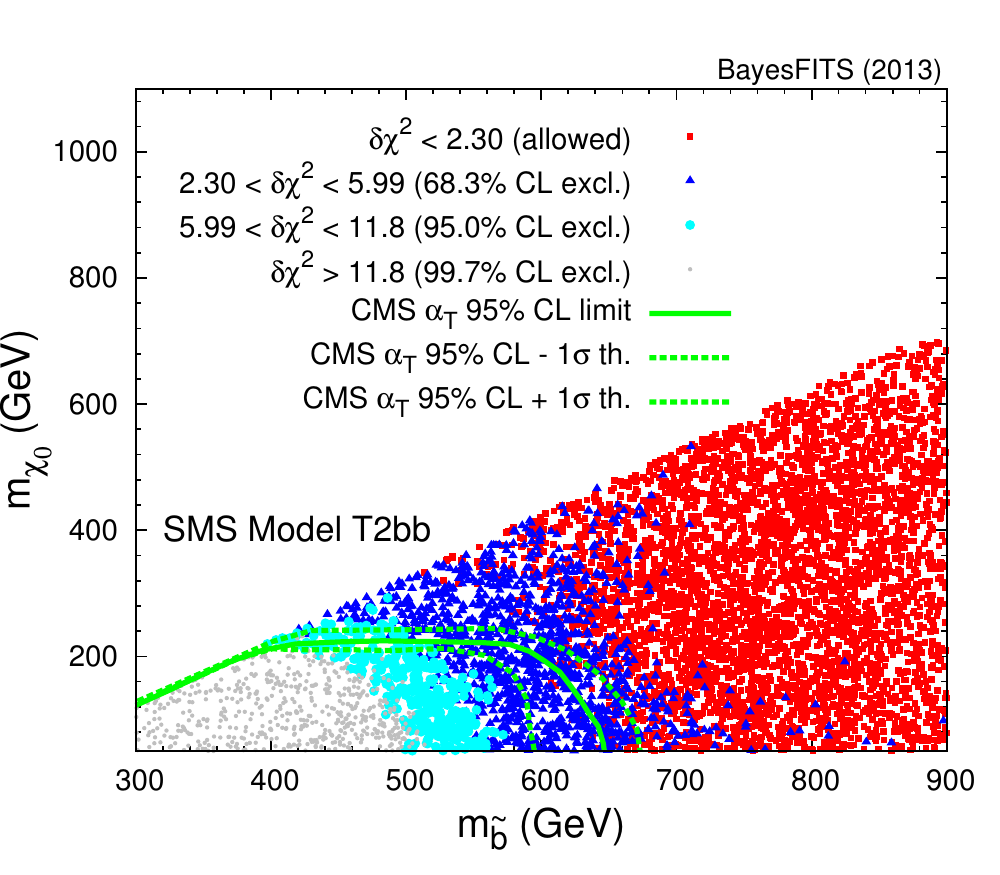}
}%
\caption[]{ Our \alphaT\ likelihood as a function of sparticle masses
  in \subref{fig:a} SMS Model T1 and \subref{fig:b} SMS Model T2bb.
  Points that are excluded at $3\sigma$ are shown as gray dots, at
  $2\sigma$ as cyan circles, and at $1\sigma$ as blue triangles.  The
  95\%~C.L. limit from CMS (solid green line) is also shown to
  facilitate a comparison with our result (boundary of cyan circles and blue triangles).  }
\label{fig:at_8tev_sms}
\end{figure}

We validate our procedure against the official experimental limits in the
framework of SMS.
In \reffig{fig:at_8tev_sms}\subref{fig:a} we present the bounds obtained
from our likelihood in the gluino-LSP plane for model
T1 of pair-produced gluinos decaying into a neutralino and quarks of the first
two generations\cite{Chatrchyan:2013lya}.
In \reffig{fig:at_8tev_sms}\subref{fig:b} we show the bounds
in the sbottom-LSP plane for model T2bb for pair-produced bottom squarks.
Gray dots represent the points excluded by our likelihood function at the
99.7\%~C.L.,
cyan circles those excluded at the 95.0\%~C.L., and blue triangles those
excluded at the 68.3\%~C.L.
The remaining points (red squares) are considered as allowed.
The solid green line shows the official 95\%~C.L. CMS exclusion limit.

As one can see,
we reproduce the experimental exclusion bound in model T1 very well, with
a discrepancy
in the gluino mass limit that stays below 50\gev\ for most of the
parameter space,
although it can be twice as large
for $\mchi\simeq400\gev$. The agreement is slightly worse for model T2bb,
where the discrepancy is about 70\gev. 
The reason is that we use in
our analysis the cross section at the leading order, as provided by \texttt{PYTHIA6.4}, which differs from the cross section implemented in the experimental analysis
by a factor of around 1.8.
Also note that the theoretical sbottom
production cross section is subject to a large theoretical uncertainty (shown in \reffig{fig:at_8tev_sms}\subref{fig:b} as dashed green lines) 
that could affect the determination of the experimental 95\%~C.L. line by around 60\gev.
It is important to keep in mind that such uncertainties
will not have a significant impact on the determination of the 
profile-likelihood confidence regions and the best-fit points, as the other constraints favor
the multi-TeV region for squarks and gluinos. 

We apply the described procedure to calculate the 
\alphaT-given \delchisq\ of all the points of the p9MSSM
sensitive to the
search (about 1\% of the total set).\footnote{With respect
to the SMS used for validation,
the \texttt{PYTHIA} card is changed for the full scan
to include inclusive production.}
One just needs to bear in mind that the \alphaT\ constraints presented in this paper
are conservative. 

Although the \eight\ \alphaT\ search imposes strong limits on the masses
of third-generation squarks
in the p9MSSM, very recently the ATLAS and CMS collaborations updated their results
for direct
top squark pair production\cite{ATLAS-CONF-2013-037,CMS-PAS-SUS-13-011} with the analyses
based on 20.7\invfb\ and 19.5\invfb\ of data, respectively.
We will test at the end the consistency of our best-fit point with
the most updated ATLAS and CMS third-generation searches.

\begin{figure}[t]
\centering
\label{fig:a}%
\includegraphics[width=0.50\textwidth]{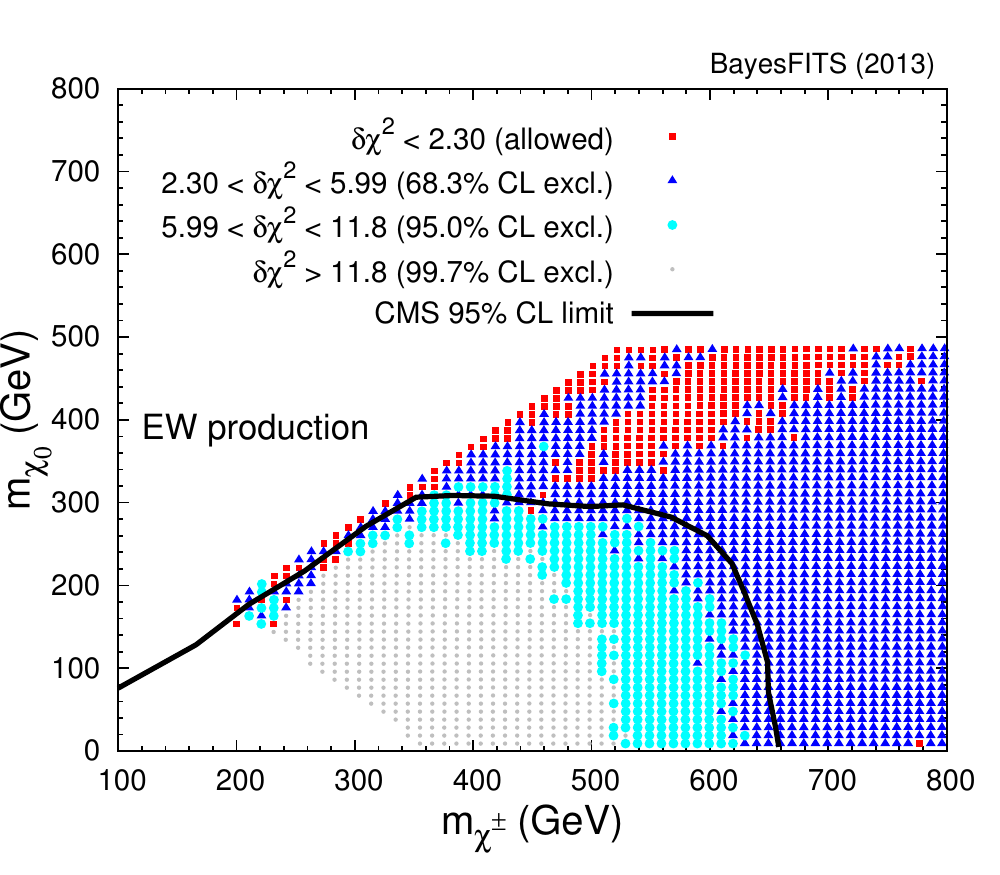}
\caption[]{
Our EW-production likelihood as a function of neutralino and chargino masses. 
Points that are excluded at $3\sigma$ are shown as gray dots, at
$2\sigma$ as cyan circles, and at $1\sigma$ as blue triangles. 
The 95\%~C.L. limit from CMS (solid black line) 
is also shown to facilitate a comparison with our result (boundary of cyan circles and blue triangles).
}
\label{fig:EWprod}
\end{figure}

\paragraph{\underline{CMS $3l+E_T^{\textrm{miss}}$ 9.2\invfb, \eight\
(EW production)}}\mbox{}\\
We follow the same procedure to construct the likelihood function for
chargino-neutralino pair production (EW production). We use the CMS search\cite{CMS-PAS-SUS-12-022},
which currently gives the strongest direct limits.

We consider the
channels with three leptons in the final state: an
opposite-sign-same-flavor lepton pair,
$ee$ or $\mu\mu$, and a third lepton being either an electron or a muon.
We have checked on a few
test scans that these channels yield the highest sensitivity. The observed and background events
are given in Table~1 of\cite{CMS-PAS-SUS-12-022}.
We validate our procedure against the official CMS 95\%~C.L. exclusion bound
for a SMS
with $m_{\tilde{l}}=0.5\mchi+0.5m_{\chi_1^{\pm}}$\cite{CMS-PAS-SUS-12-022}.
The result is shown in \reffig{fig:EWprod}, where the color code is the
same as
in \reffig{fig:at_8tev_sms}.
As one can see the discrepancy in the chargino mass bound is less than
50\gev\
for the neutralino mass range.

Given our parameter choice and ranges, the likelihood for EW production of
neutralinos and charginos
can rule out a large fraction of the total number of
scanned points. Even by preliminarily
considering only the points that can satisfy all other constraints, the
number of points
in our scans that can be potentially affected by the EW-production
likelihood is about 400,000 (out of 1.8 millions in total).
For this reason, including this search in the global
likelihood function has proven to be a numerically unmanageable task.
Thus, the contribution to the \chisq\ of this search is calculated 
on a randomly chosen sample
of approximately 40,000 points from our chains.
This will be enough to draw general conclusions. 
We will use this information to check the consistency of this search with the other
constraints
(and particularly \deltagmtwomu), and also we will test the best-fit points 
from the global likelihood against EW production
to make sure that they 
are not excluded.

\subsection{XENON100 likelihood\label{sec:X100like}}

A proper treatment of the XENON100 90\%~C.L. bound in the (\mchi, \sigsip) 
plane\cite{Aprile:2011dd,Aprile:2011hx,arXiv:1104.2549} is not straightforward because,
as it has been long known, the limits from DD of DM experiments on SUSY parameter space 
are marred by large
nuclear physics uncertainties\cite{Jungman:1995df,Ellis:2008hf}.
The astrophysics uncertainties resulting from the DM local density and velocity distribution, on the other hand, affect the elastic scattering 
cross section by only around 50\% in the mass range considered in this paper\cite{Catena:2011kv}.

Nuclear physics uncertainties enter the picture 
through the calculation of
the cross section of DM-quark elastic scattering.
To connect this prediction with experiment one has to estimate a nucleon mass matrix,
$\langle N| \bar{q}q |N\rangle$,
to transform the cross section from the quark level to the nucleon
level. The nucleon mass matrix calculation is subject to uncertainties
on the quark masses $m_{d,c,b,t}$,
on the ratios $m_u/m_d$ and $m_s/m_d$, and on the hadronic quantities 
related to the change
in the nucleon mass due to nonzero quark masses,
$\sigma_0$ and $\Sigma_{\pi N}$:

\begin{eqnarray}
\label{eq:sigmaterm}
 \sigma_{0}&=&\frac{m_u+m_d}{2}\langle N| \bar{u}u+\bar{d}{d}-2\bar{s}s |N\rangle,  \nonumber \\
 \Sigma_{\pi N}&=&\frac{m_u+m_d}{2}\langle N| \bar{u}u+\bar{d}{d} |N\rangle\,.\label{sigmaterms}
\end{eqnarray}

$\Sigma_{\pi N}$ is generally derived by extrapolating information 
from experimental input, generally $\pi$--$N$ elastic scattering cross sections.
It has been long known that these hadronic uncertainties
can be much larger than astrophysical uncertainties (see\cite{Roszkowski:2012uf} and references therein). 

In a recent paper\cite{Stahov:2012ca}, the differential elastic $\pi$--$N$ scattering cross sections\cite{Denz:2005jq} 
measured with the CHAOS detector at TRIUMF\cite{Raywood:1994qf} were employed to derive a new determination of the $\Sigma_{\pi N}$
term, $\Sigma_{\pi N}=43\pm12\mev$, where the error bar is mostly due to the experimental uncertainties.
This value is substantially lower than the values previously 
calculated using phase-shift analyses from the GWU/SAID database\cite{PhysRevC.86.035202}, 
or using chiral perturbation theory\cite{Alarcon:2011zs,*Alarcon:2012nr},
and it can have substantial implications,
as we shall see, when deriving limits on SUSY from DD experiments.  
 
\begin{figure}[t]
\centering
\label{fig:a}%
\includegraphics[width=0.60\textwidth]{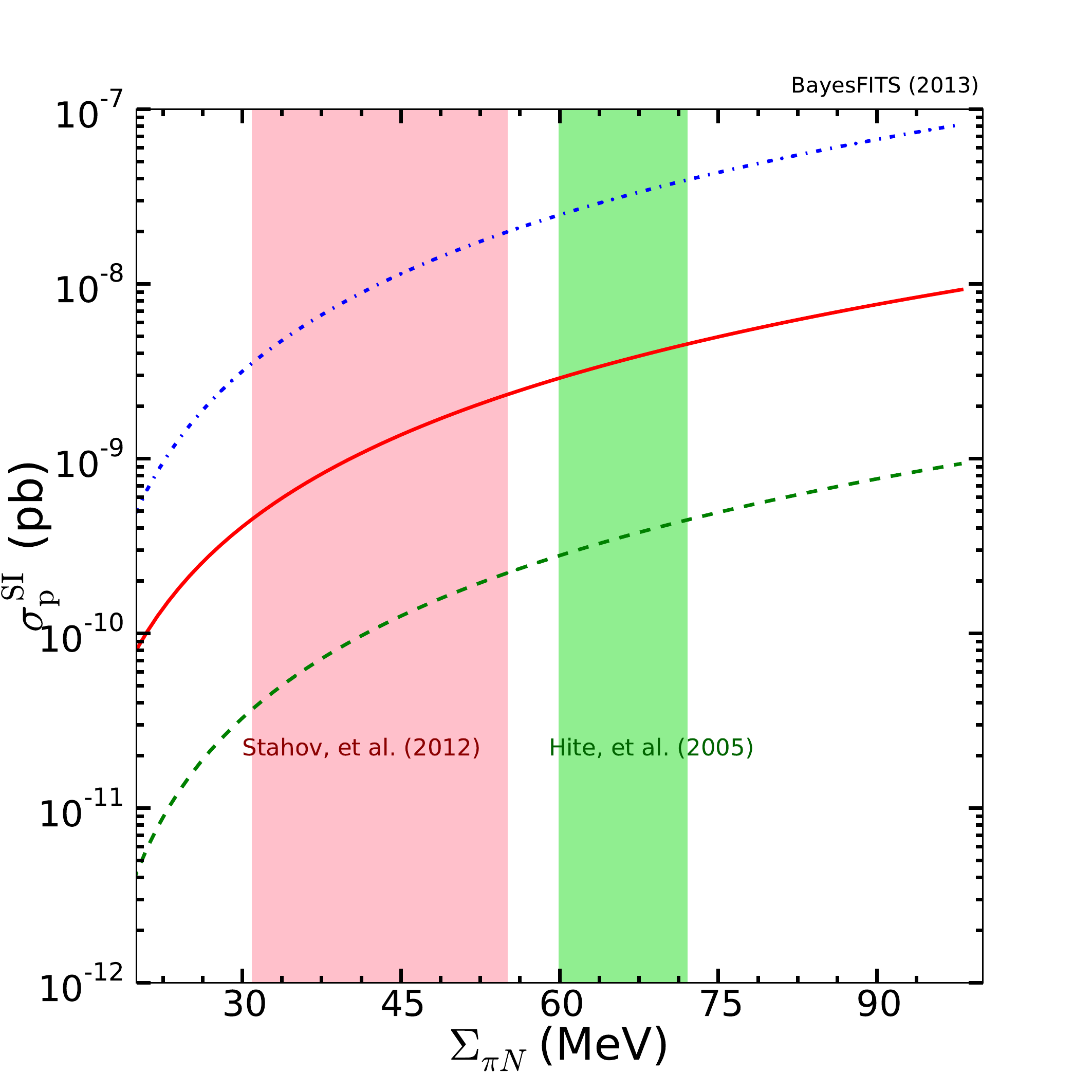}
\caption[]{
The spin-independent neutralino-proton scattering cross section versus 
the pion-nucleon $\Sigma$ term for three p9MSSM points (shown in \reftable{tab:susypts}), 
characterised by their neutralino composition: the dash-dotted blue line shows a point corresponding to mixed neutralino, 
the solid red line to a higgsino-like neutralino, and the dashed green line to a gaugino-like neutralino. 
$1\sigma$ confidence intervals for the pion-nucleon $\Sigma$ term from\cite{Stahov:2012ca} 
(light red, left) and\cite{Hite:2005tg} (light green, right) are shown by vertical shaded blocks. 
}
\label{fig:sigmapin}
\end{figure} 
 
To illustrate this point, we show in \reffig{fig:sigmapin} the dependence of the SI
cross section on the $\Sigma_{\pi N}$ term for three different neutralino masses and 
gaugino/higgsino fractions.
One can see that \sigsip\ can vary by more than one order of magnitude over the plotted
range of $\Sigma_{\pi N}$, and by a factor of five over the $1\sigma$ range of\cite{Stahov:2012ca} (light red band on the left).
Thus, in this study we include the most recent $\Sigma_{\pi N}$ determination of\cite{Stahov:2012ca} (with its uncertainties)
in the likelihood function for XENON100.

The likelihood function for XENON100 is given by the product of an experimental and a theoretical
part. We build the experimental, model-independent part following the procedure described in detail in Sec.~IIIB of Ref.\cite{Cheung:2012xb}.
We assume that number of observed events follows a Poisson 
distribution about the number of ``signal+background" events. 
The systematic uncertainties are parametrized by marginalizing the background prediction with a Gaussian distribution
of mean $b=1$ and standard deviation $\delta b=0.2$, as given by the XENON Collaboration\cite{Aprile:2012nq}.
An ``exclusion signal," $s^{\ast}_{90}$, is thus calculated,
\begin{equation}
0.1=\frac{\int_{s^{\ast}_{90}}^\infty\mathcal{P}(s'+b|o)ds'}{\int_0^\infty\mathcal{P}(s'+b|o)ds'}\,,\label{conflevel}
\end{equation}
where the probability distribution is given by
\begin{equation}
\mathcal{P}(s+b|o)= \int_0^\infty\frac{e^{-(s + b')}\left(s + b' \right)^{o}}{o !}
\exp \left[ -\frac{(b'-b)^2}{2\delta b^2}\right]db'\,,\label{bgmarg}
\end{equation}
and $o=2$ is the number of observed events\cite{Aprile:2012nq}. 

For each pair (\mchi, $\sigma_{p,90}^{\textrm{SI}}$) lying on the 90\%~C.L. exclusion line,
a signal $s_\textrm{mo}$ is then calculated with \texttt{micrOMEGAs}, in the nuclear
recoil energy range of 6.6--30.5 $\rm{keV}_{\rm{nr}}$.
We shall use the default setting of DM velocity distribution
(the truncated Maxwell distribution). One can thus derive
experimental ``efficiencies,"
\begin{equation}
\varepsilon(\mchi,\sigma_{p,90}^{\textrm{SI}})=\frac{s^{\ast}_{90}}{s_\textrm{mo}(\mchi, \sigma_{p,90}^{\textrm{SI}})}\,,\label{efficiencies}
\end{equation}
which account for the experimental acceptances.

We finally account for nuclear physics uncertainties by profiling over $\Sigma_{\pi N}$
with Gaussian distribution, 
\begin{equation}
\mathcal{L}[\mchi,\sigsip(\Sigma_{\pi N})]=\max_{\Sigma_{\pi N}'}\,\mathcal{P}[\varepsilon s_{\textrm{mo}}(\mchi,\sigsip(\Sigma_{\pi N}'))+b|o]
\cdot\exp\left[ -\frac{(\Sigma_{\pi N}'-\Sigma_{\pi N})^2}{2\sigma_{\Sigma_{\pi N}}^2}\right]\,,\label{profilelike}
\end{equation}
where the probability $\mathcal{P}$ is given in Eq.~(\ref{bgmarg}).
Here, we only vary $\Sigma_{\pi N}$ as the largest source of uncertainty
but fix other mass matrix parameters, such as
$m_u/m_d =0.553$, $m_s/m_d=18.9$ and $\sigma_{0}=35.5\mev$.

\begin{figure}[t]
\centering
\subfloat[]{%
\label{fig:X100_mx_sigsipa}%
\includegraphics[width=0.50\textwidth]{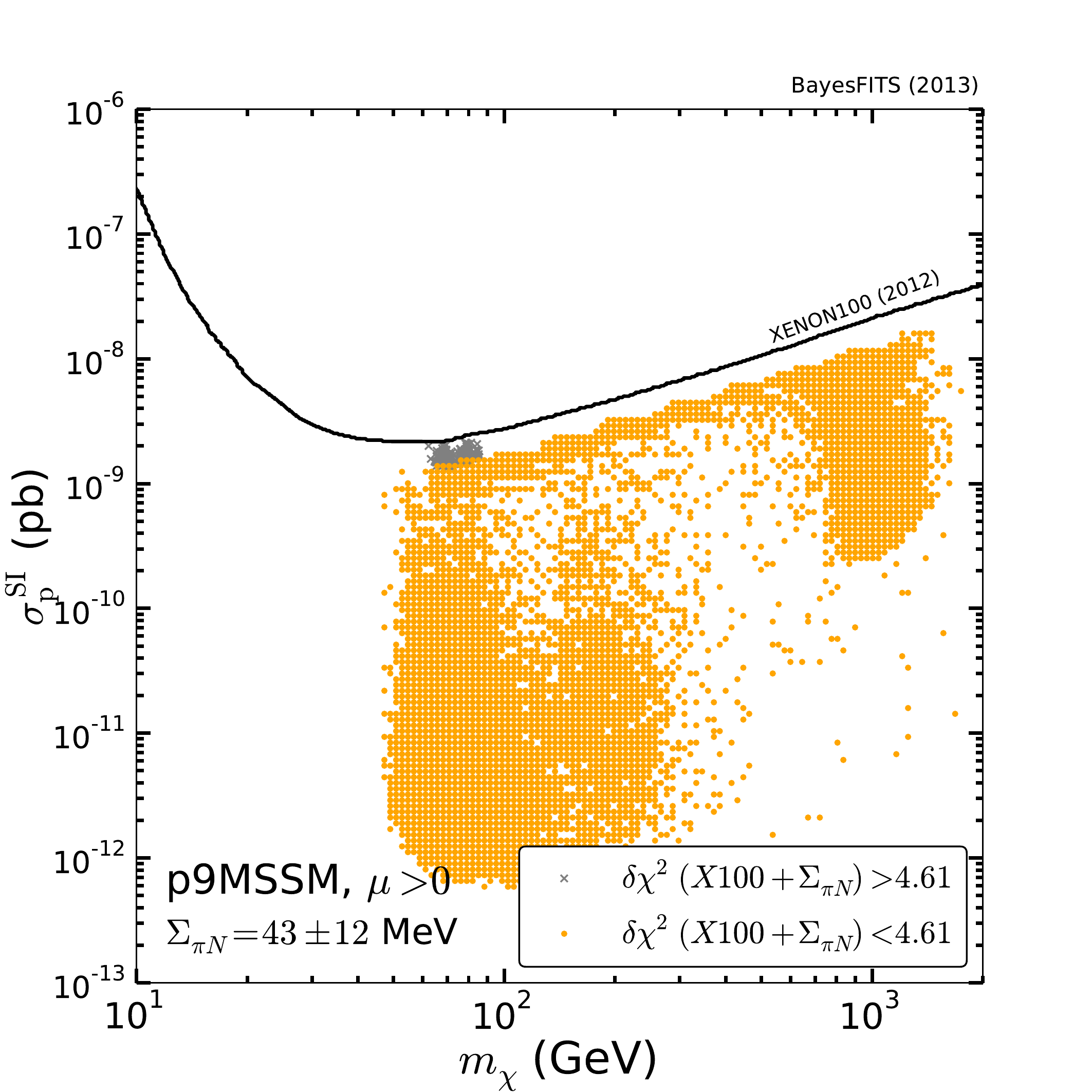}
}%
\subfloat[]{%
\label{fig:X100_mx_sigsipb}%
\includegraphics[width=0.50\textwidth]{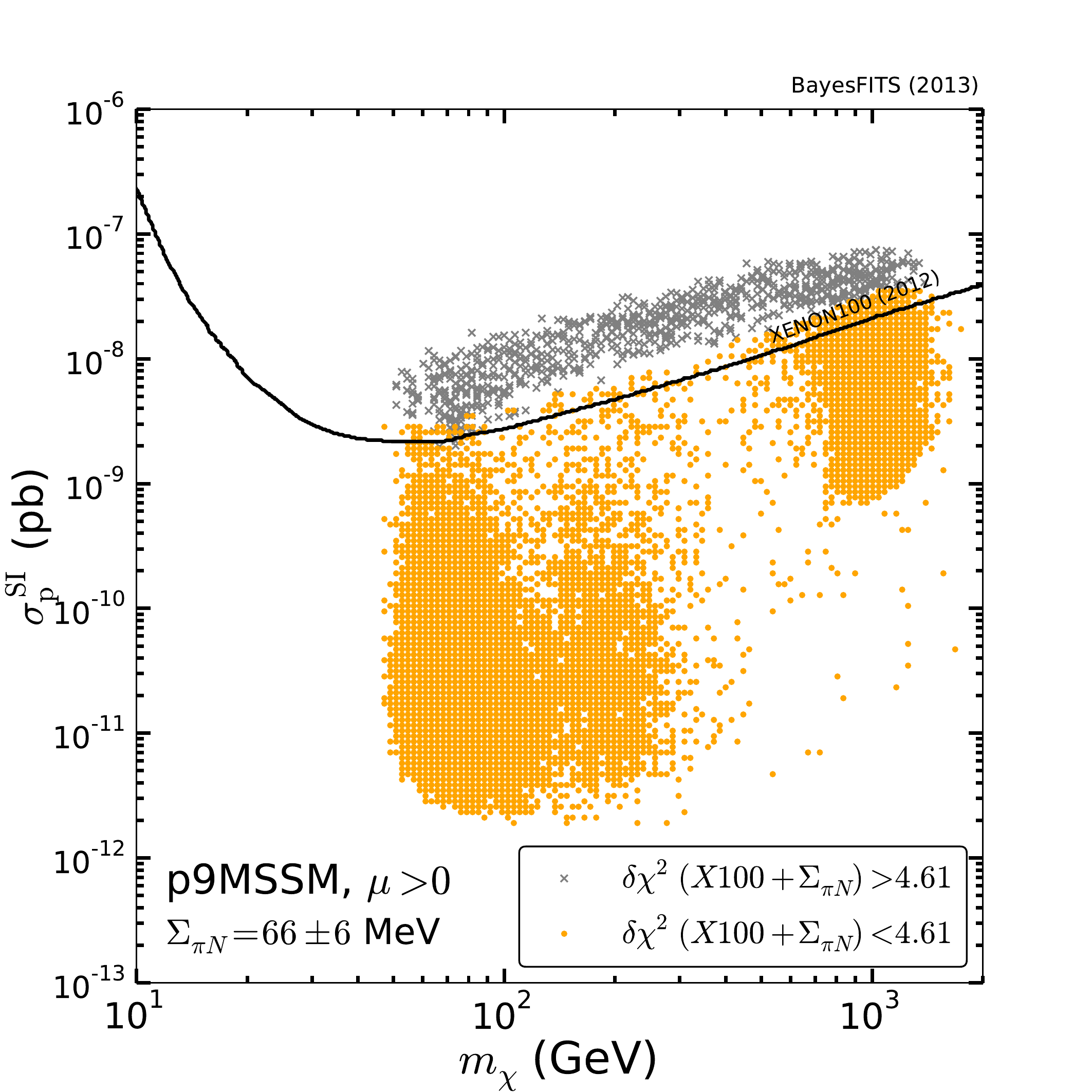}
}
\caption[]{ p9MSSM points scattered in the ($\mchi$, $\sigsip$) plane
  assuming \subref{fig:X100_mx_sigsipa} $\Sigma_{\pi N}=43\pm 12\mev$
  and \subref{fig:X100_mx_sigsipb} $\Sigma_{\pi N}=66\pm 6\mev$. The
  XENON100 90\%~C.L. exclusion contour is shown in solid black.
  Points excluded through the likelihood of Eq.~(\ref{profilelike}) at
  the 90\%~C.L. ($\delta \chi^2 > 4.61$) are shown as gray crosses.
  All the points in the plots satisfy the constraints from LEP,
  \alphaT, Higgs mass, PLANCK, and flavor physics.  }
\label{fig:cf_mx_sigsip}
\end{figure} 

In \reffig{fig:cf_mx_sigsip}\subref{fig:X100_mx_sigsipa}, 
we show the impact of incorporating the $\Sigma_{\pi N}$ uncertainties into the
likelihood. 
All the points in the plot 
satisfy the constraints from LEP, $\alpha_T$ limits, Higgs mass, PLANCK, and flavor physics. 
The points excluded by the likelihood function of Eq.~(\ref{profilelike}) at the 90\%~C.L. ($\delchisq=4.61$) are shown as 
gray crosses and the rest of the points as yellow circles.
Note that, since we profile on the theoretical uncertainty, the position of the points in the
plane can float according to the value of $\Sigma_{\pi N}$ producing the largest likelihood,
see Eq.~(\ref{profilelike}). One can see that, given the new determination and uncertainties on 
$\Sigma_{\pi N}$, the impact of the XENON100 constraint on the parameter space 
of the p9MSSM is almost negligible.

In contrast, we repeated the same procedure by considering a recent determination 
$\Sigma_{\pi N}=66\pm6\mev$\cite{Hite:2005tg,Alvarez-Ruso:2013fza} obtained 
from the most recent GWU pion-nucleon phase-shift analysis\cite{Arndt:2003if}.  
One can see in \reffig{fig:cf_mx_sigsip}\subref{fig:X100_mx_sigsipb} 
that in this case a substantial fraction of points is excluded by XENON100.
 
In our results we will show the effect of applying in turn one or the other determination, 
and discuss the ensuing implications for the p9MSSM.

\section{\label{Results}Results}

We collected a total of about $1.8\times10^6$ points through several scans 
of the p9MSSM parameter space, as
defined in the previous sections.

We identify three different sets of constraints, which are shown 
separately in Table~\ref{tab:exp_constraints}. 
The upper box encapsulates what we define as the \textbf{basic} set of constraints,
which are taken into account in all of the plots presented below.  
On the other hand, the \gmtwo\ and XENON100 constraints are included in the global likelihood 
when discussing the impact of these specific constraints, and 
we will indicate explicitly in the text and figures when this is the case.  
Finally, the EW-production constraint is treated separately, as this contribution 
is added to the likelihood only for a randomly chosen thinned selection of points, 
as explained in Sec.~\ref{subsec:LHC}. 

\begin{figure}[t]
\centering
\subfloat[]{%
\label{fig:a}%
\includegraphics[width=0.50\textwidth]{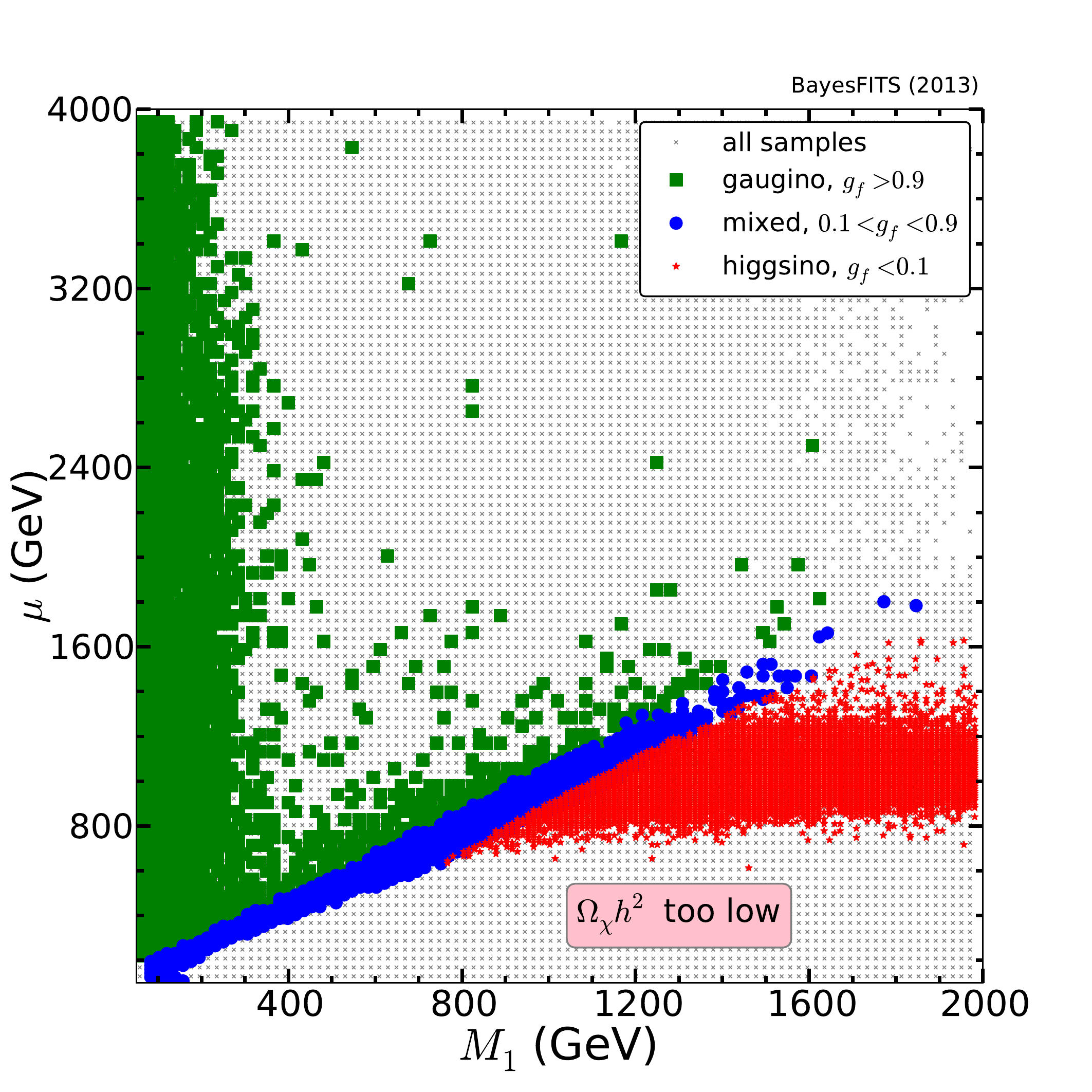}
}%
\subfloat[]{%
\label{fig:b}%
\includegraphics[width=0.50\textwidth]{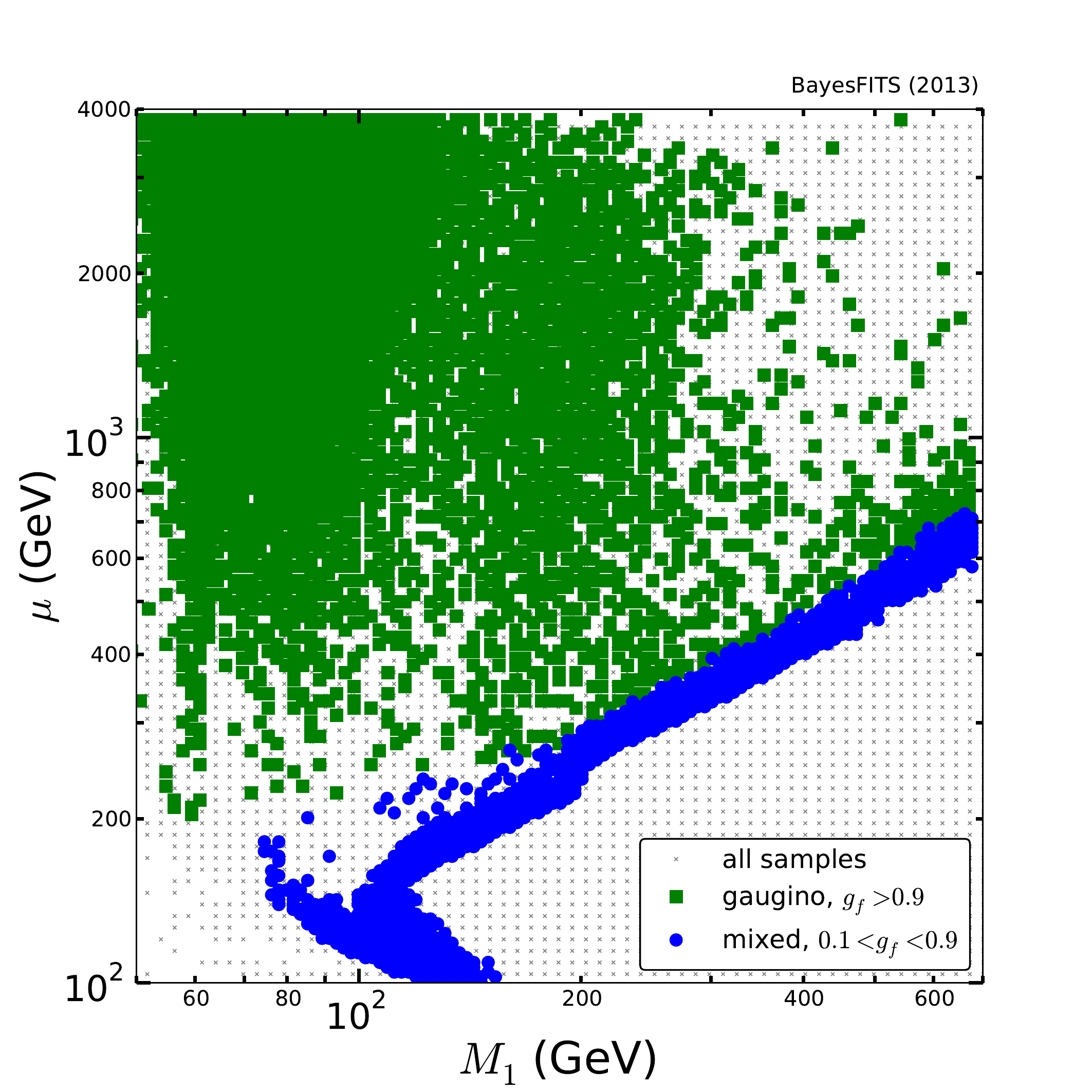}
}%
\caption[]{ \subref{fig:a} Scatter plot of p9MSSM points in the
  ($M_1$, $\mu$) plane. Gray crosses show the points excluded by the
  \textbf{basic} likelihood at the 95\%~C.L. Allowed points are
  divided by the composition of the neutralino: gaugino-like (green
  squares), mixed (blue circles), or higgsino-like (red stars).
  \subref{fig:b} Magnified plot of the low-mass region of
  \subref{fig:a} with logarithmic scales.  }
\label{fig:mssm_m2_mu}
\end{figure}

In \reffig{fig:mssm_m2_mu}\subref{fig:a}, we show the distribution of our points in the 
($M_1$, $\mu$) plane.
The gray dots are excluded at the 95\%~C.L. based on the profile-likelihood method, 
after applying the \textbf{basic} set of constraints ($\delchisq_{\textrm{basic}}>5.99$). 
In the remainder of this section, we will not show the gray dots
again, i.e., we will present our results as 95\%~confidence
regions in two-dimensional (2D) projections
based on the profile-likelihood method. However, before we move on, we want to point out 
a couple of features of the gray dots. 
The first is that their distribution appears in the plot to lie on a grid. 
The reason is that, although all of our $>10^6$ points are generated randomly, we binned
the data excluded at the 95\%~C.L. in a $100\times 100$-step grid to reduce the size of the picture. 
The second is that some of the points are missing from the upper right corner of    
\reffig{fig:mssm_m2_mu}\subref{fig:a}. The reason has to do with our choice of parameter ranges
(Table~\ref{table:MSSMparams}): those points are characterized by a neutralino mass around 2000\gev, and the upper limit on the stau mass is also
$\sim2000\gev$. Thus, in that region \texttt{MultiNest} is likely to generate points with stau LSP, which are automatically rejected.  

The relic density is a strong constraint, since it is a positive 
measurement with a rather small experimental uncertainty.
Therefore, the shape and size of the 2D 95\% confidence regions presented in this section
will be determined predominantly by the relic density as measured by PLANCK.

The color code in the 95\% confidence region of \reffig{fig:mssm_m2_mu}\subref{fig:a}
shows the composition of the lightest neutralino, which is the lightest mass eigenstate of
mixed bino, wino, up-type higgsino and down-type higgsino gauge eigenstates,
\begin{equation}
\chi\equiv\chi^0_{1}=Z_1 \tilde{B} +Z_2 \tilde{W}+
Z_3 \tilde{H_u} + Z_4 \tilde{H_d}\,,\label{netralino}
\end{equation}
where the coefficients $Z_{i}$ ($i\leq4$) are
determined by diagonalizing the neutralino mass matrix.
To describe the neutralino compositions, it is convenient to introduce
a gaugino fraction, $g_f =Z^{2}_1+Z^{2}_2$.
When $g_f$ is close to 1, gauginos dominate the neutralino; on the
other hand the neutralino will be higgsino-like if $g_f\simeq 0$.
The points for which the neutralino is a nearly pure 
gaugino are presented as green squares, the points for which the neutralino is higgsino-like  
are marked as red stars, and the points for which the neutralino composition
is some mixture of gaugino and higgsino states are shown as blue circles. 

A bino-like LSP is obtained when $\mchi\approx M_1<\mu$. One can see in \reffig{fig:mssm_m2_mu}\subref{fig:a}
that, when $M_1\lesssim 400\gev$, gaugino-like neutralino DM satisfies the relic abundance for a broad range of $\mu$ values 
(green squares on the left). In \reffig{fig:mssm_m2_mu}\subref{fig:b} we 
show a zoomed-in view of this region of the parameter space.
One can identify two separate gaugino-like branches. On the left, for $M_1\lesssim 65\gev$
the correct value of the relic abundance is obtained by efficient annihilation to $b$-quarks 
in the early universe through $s$-channel diagram exchange of the lightest Higgs boson
(note that for this to occur the neutralino must have a nonzero higgsino component) and, for slightly larger $M_1\lesssim 100\gev$, 
through ``bulk" annihilation to leptons through $t$-channel slepton exchange. We will refer to this region as the $h$-resonance/bulk region\cite{Ellis:1989pg} (HR/bulk). 
To the right of the HR/bulk region, for $100-150\gev \lesssim M_1\lesssim 300\gev$, a second area for gaugino DM can be observed, 
where the correct relic density is obtained through 
coannihilation with sleptons of the three generations\cite{Ellis:1998kh} (selectrons and smuons for $100\gev \lesssim M_1\lesssim 200\gev$, staus for 
$200\gev \lesssim M_1\lesssim 300\gev$). We will refer to this region as the slepton-coannihilation (SC) region.

By decreasing $\mu$ down to $\mu\approx M_1$ ($<M_2$), the higgsino
fraction in the neutralino increases (mixed composition, blue
circles). Along the blue strip that extends to $\mu\approx M_1\lesssim
800\gev$ in \reffig{fig:mssm_m2_mu}\subref{fig:a}, the relic density
constraint is satisfied thanks to $\chi\chi$ annihilation into gauge
bosons, through $t$-channel exchange of higgsino-like $\chi^\pm_1$
and/or $\chi^0_2$.  This is the p9MSSM equivalent of the focus
point/hyperbolic branch (FP/HB) region of the
CMSSM\cite{Chan:1997bi,Feng:1999zg}, and we will loosely use the same
acronym to describe this region of the p9MSSM in what follows.  The
``hook" feature in points with mixed composition at
$\mu\lesssim150\gev$ results from a $WW$ threshold.  In only that
region, the neutralino has a bino/higgsino composition and, at the
same time, $\mchi<M_W$ while $m_{\charone}>94\gev$.  Because the
$\chi\chi\rightarrow W^+W^-$ annihilation is suppressed by a threshold
due to chargino exchange, the relic density is not too small.  Note that
the green strip of gaugino-like DM, adjacent and above the FP/HB
region up to $M_1\simeq1.2-1.6\tev$ in
Fig.~\ref{fig:mssm_m2_mu}\subref{fig:a} is the AF
region\cite{Drees:1992am}.

As one considers ever larger $\mu$ along the FP/HB region, the
neutralino becomes almost purely higgsino-like, and its mass
stabilizes at $\mchi\approx\mu\simeq 1\tev$. We call this the 1TH
region and it is indicated with red stars in
\reffig{fig:mssm_m2_mu}\subref{fig:a}.  Here $\chi$ and $\chi^0_2$ are
either both higgsino-like or one of them is higgsino- and the other
bino-like, respectively, while $\chi^\pm_1$ is always higgsino-like.
The relic density constraint is satisfied for broad ranges of $M_1$,
partially through LSP co-annihilation with the second lightest
neutralino, $\chi^0_2$, and/or the lightest chargino $\chi^\pm_1$.

Note, finally, the lower density of gray dots in the lower left 
corner of \reffig{fig:mssm_m2_mu}\subref{fig:b}. In fact, in that region of the parameter
space the LEP constraints on the chargino mass, Eq.~(\ref{LEP}), become much harder to satisfy.


\subsection{Impact of the XENON100 limit}

\begin{figure}[t]
\centering
\subfloat[]{%
\label{fig:a}%
\includegraphics[width=0.50\textwidth]{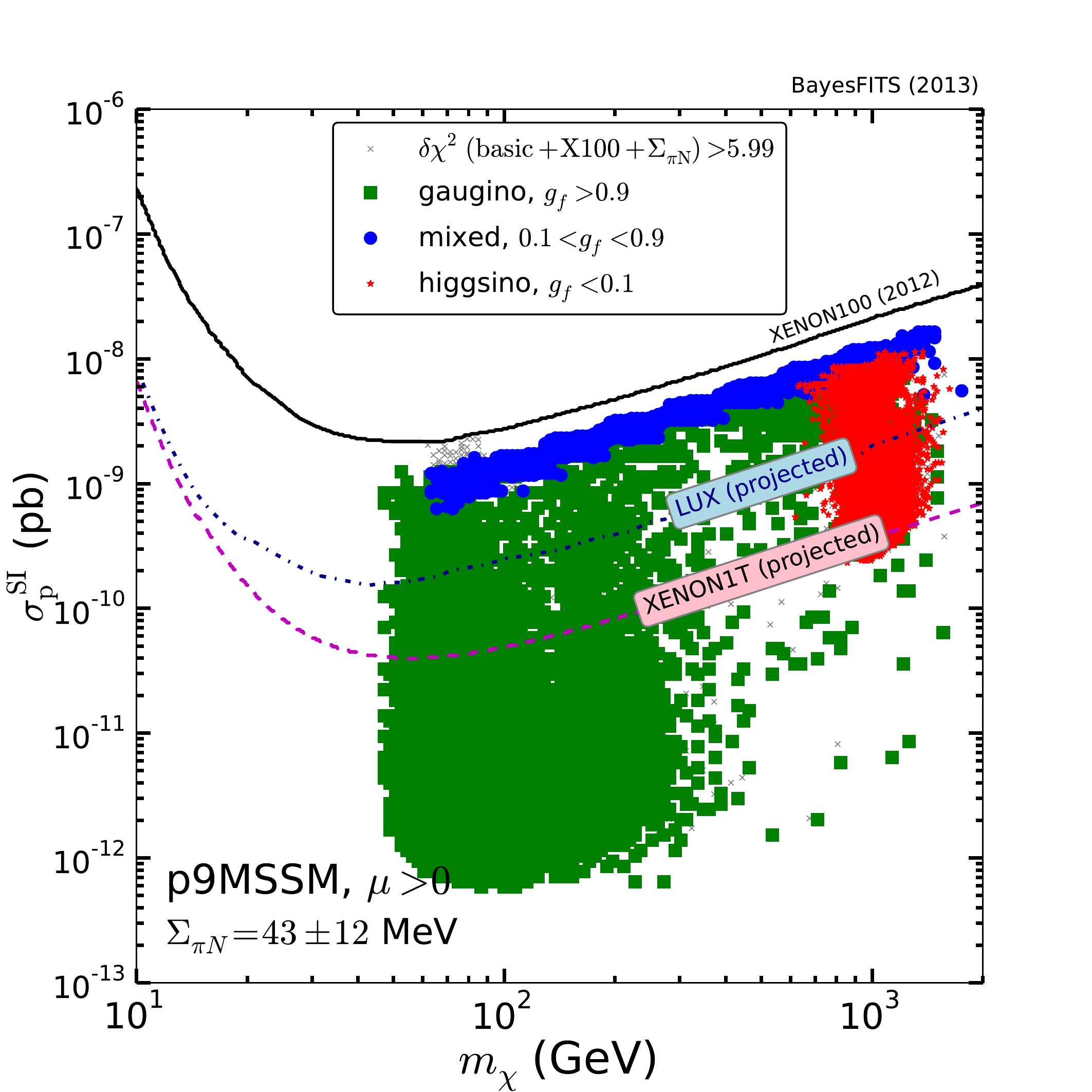}
}%
\subfloat[]{%
\label{fig:b}%
\includegraphics[width=0.50\textwidth]{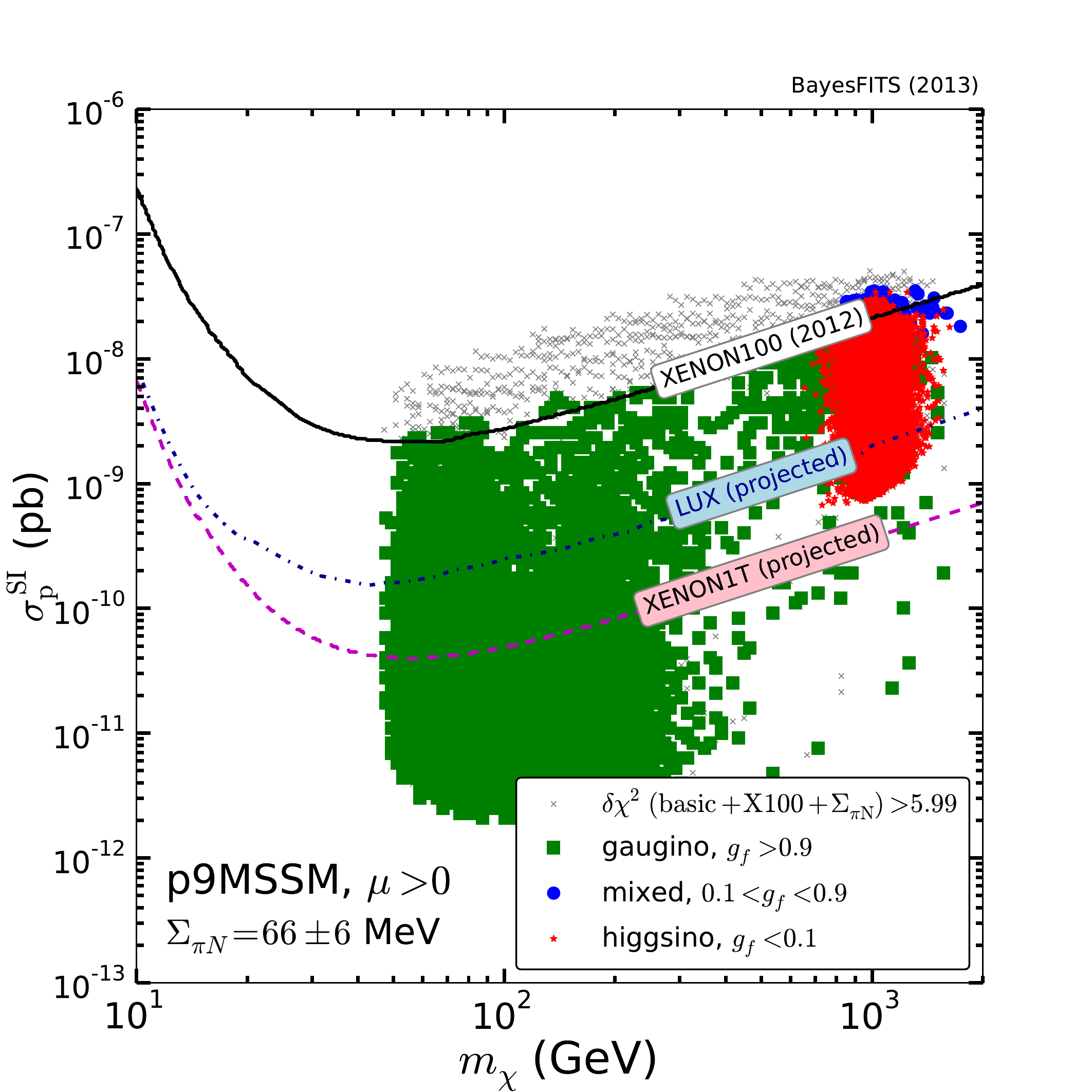}
}%
\caption[]{p9MSSM points that are allowed at 2$\sigma$ by the \textbf{basic}
constraints in the ($\mchi$, $\sigsip$) plane.  
The points consistent at 2$\sigma$ with the \textbf{basic}
\textit{and} XENON100 constraints are divided by the composition of
the neutralino:  
gaugino-like (green squares), mixed (blue circles), or higgsino-like
(red stars). 
Points excluded at the 95\%~C.L. by \textbf{basic}+XENON100 are shown
as gray crosses. 
\subref{fig:a} $\Sigma_{\pi N}= 43\pm12$ MeV, \subref{fig:b} $\Sigma_{\pi N}= 66\pm6$ MeV.

}
\label{fig:X100_mx}
\end{figure}

In this subsection we analyze the impact of the XENON100 90\%~C.L. upper bound 
on the parameter space of the p9MSSM.
We emphasize that the bound is applied through the likelihood function 
given by Eq.~(\ref{profilelike}).

In Figs.~\ref{fig:X100_mx}\subref{fig:a} and \ref{fig:X100_mx}\subref{fig:b}, 
we show the difference between the \textbf{basic} 95\% confidence region (all points)
and the 95\% confidence region
obtained by adding the likelihood of Eq.~(\ref{profilelike}) (all points
except gray crosses)
in the (\mchi, \sigsip) plane.
The color code describes the gaugino fraction of the LSP, 
and it is the same as in \reffig{fig:mssm_m2_mu}. 

In \reffig{fig:X100_mx}\subref{fig:a}, we show the case with nuclear physics uncertainties
parametrized around $\Sigma_{\pi N}=43\pm 12\mev$,    
as described in Sec.~\ref{sec:X100like}. One can see that a small fraction of points characterized by mixed gaugino-higgsino composition 
and $\mchi\simeq 60-90\gev$ is excluded at the 95\%~C.L. by the global likelihood. 
For these points, in fact, the lightest Higgs boson exchange in the $t$ channel 
due to the non-negligible higgsino fraction of the neutralino can enhance \sigsip.
At the tree level, there are only two Feynman diagrams contributing to
$\sigsip$\cite{Drees:1993bu}; $t$-channel diagram Higgs
exchange and $s$-channel squark resonance. While the squark resonance
is suppressed by the fact that LHC limits now imply heavy squarks,
one can always tune the gaugino fraction
to increase or decrease the contribution from the Higgs exchange mode.
Also, the mass values for the excluded points correspond to the region of greater 
sensitivity for XENON100. 

Note that, as pointed out in Sec.~\ref{sec:X100like}, the large theoretical uncertainties drastically reduce the 
impact of the XENON100 constraint on the parameter space. 
This point is emphasized in \reffig{fig:X100_mx}\subref{fig:b}, where we
parametrize the theoretical uncertainties around $\Sigma_{\pi N}=66\pm 6\mev$,
following the determination of\cite{Hite:2005tg,Alvarez-Ruso:2013fza}. 
One can see that, in this case, the mixed gaugino/higgsino region of the parameter 
space becomes almost entirely excluded at the 95\%~C.L.   

In Figs.~\ref{fig:X100_mx}\subref{fig:a} and \ref{fig:X100_mx}\subref{fig:b}
we also plot the expected reaches of LUX\cite{Akerib:2012ys} and XENON1T\cite{Aprile:2012zx}.
One can see that, even when substantial theoretical uncertainties
are taken into account, those experiments have the potential to bite into a large fraction of the 
mixed gaugino/higgsino region,
particularly if the determination of $\Sigma_{\pi N}$ stabilizes in the future around 
the larger value.


\subsection{Impact of \deltagmtwomu\ and limits from the LHC\label{sec:g2}}

\begin{figure}[t]
\centering
\subfloat[]{%
\label{fig:a}%
\includegraphics[width=0.50\textwidth]{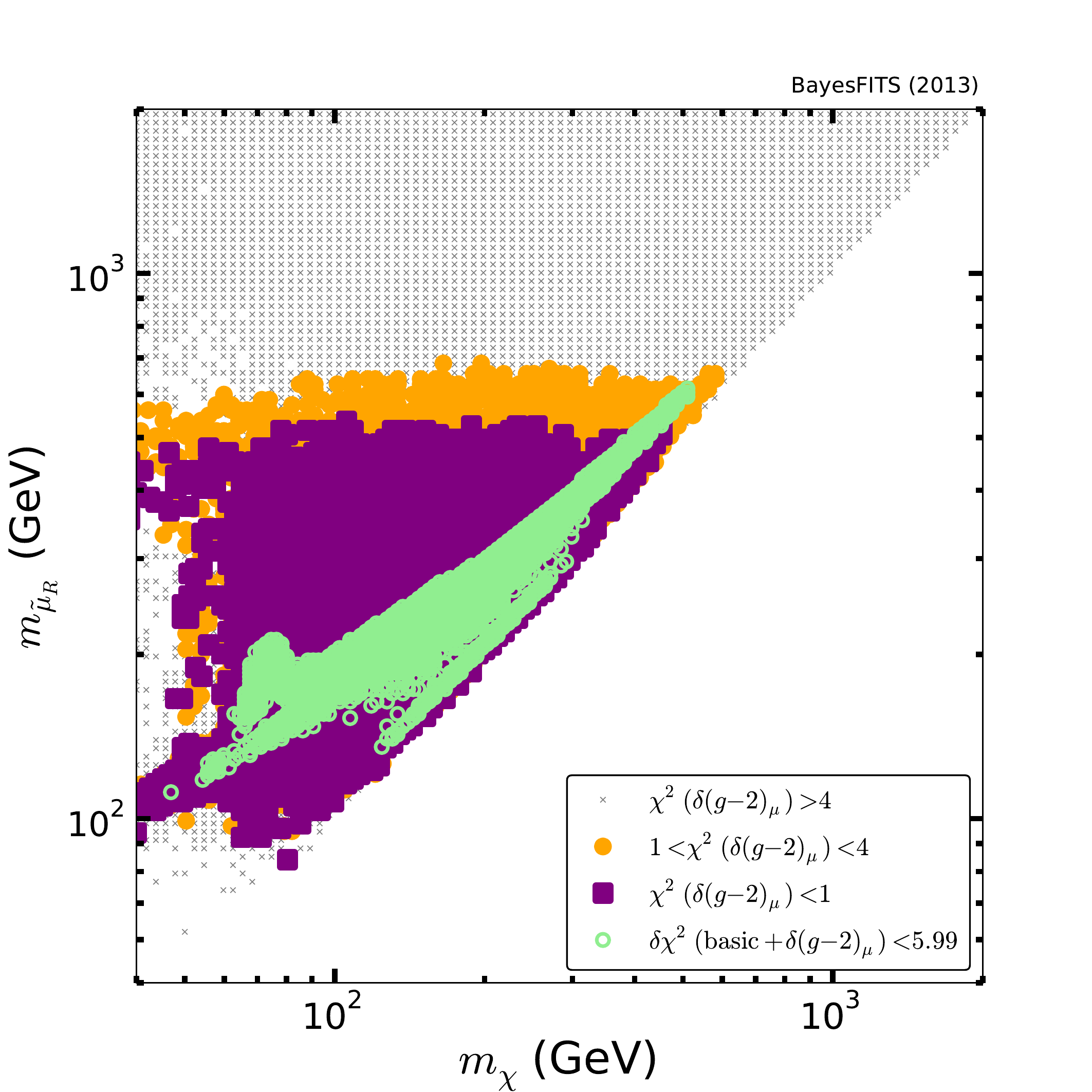}
}%
\subfloat[]{%
\label{fig:b}%
\includegraphics[width=0.50\textwidth]{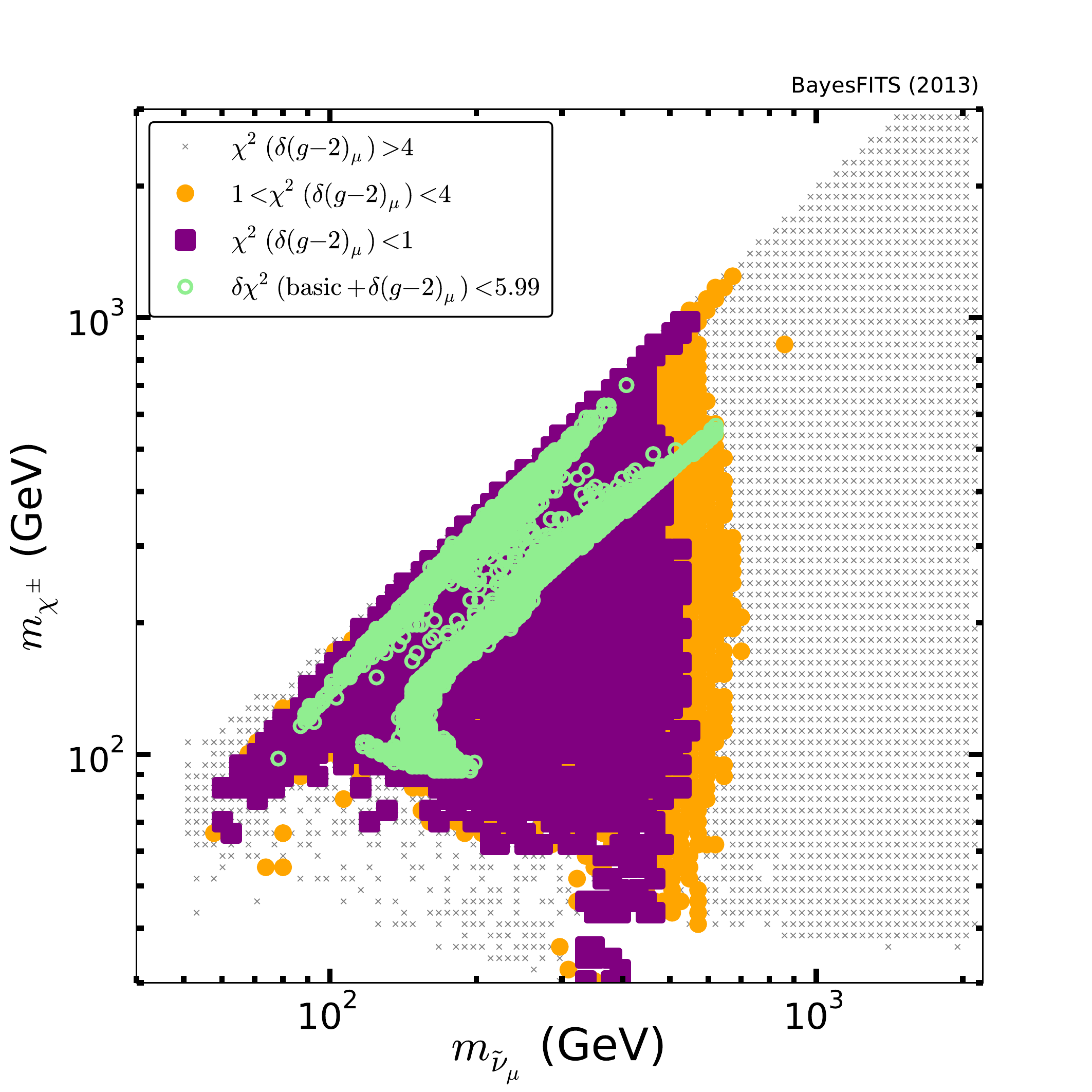}
}%
\caption[]{
\subref{fig:a} p9MSSM points scattered in the ($\mchi$, $m_{\tilde{\mu}}$) plane. 
Points with a \chisq\ from the $\gmtwo$ constraint of greater than 4 are shown with gray crosses. 
Points with $\chisq<1$ and $1<\chisq<4$ are shown with purple squares and orange circles, respectively. 
Points compatible with both $\gmtwo$ and \textbf{basic} at $2\sigma$ are shown with light green circles.
\subref{fig:b} p9MSSM points scattered in the ($m_{\tilde{\nu}_\mu}$, $m_{\charone}$) plane. The color
code is the same as in \subref{fig:a}.
}
\label{fig:mssm_g2}
\end{figure} 

\begin{figure}[t]
\centering
\includegraphics[width=0.60\textwidth]{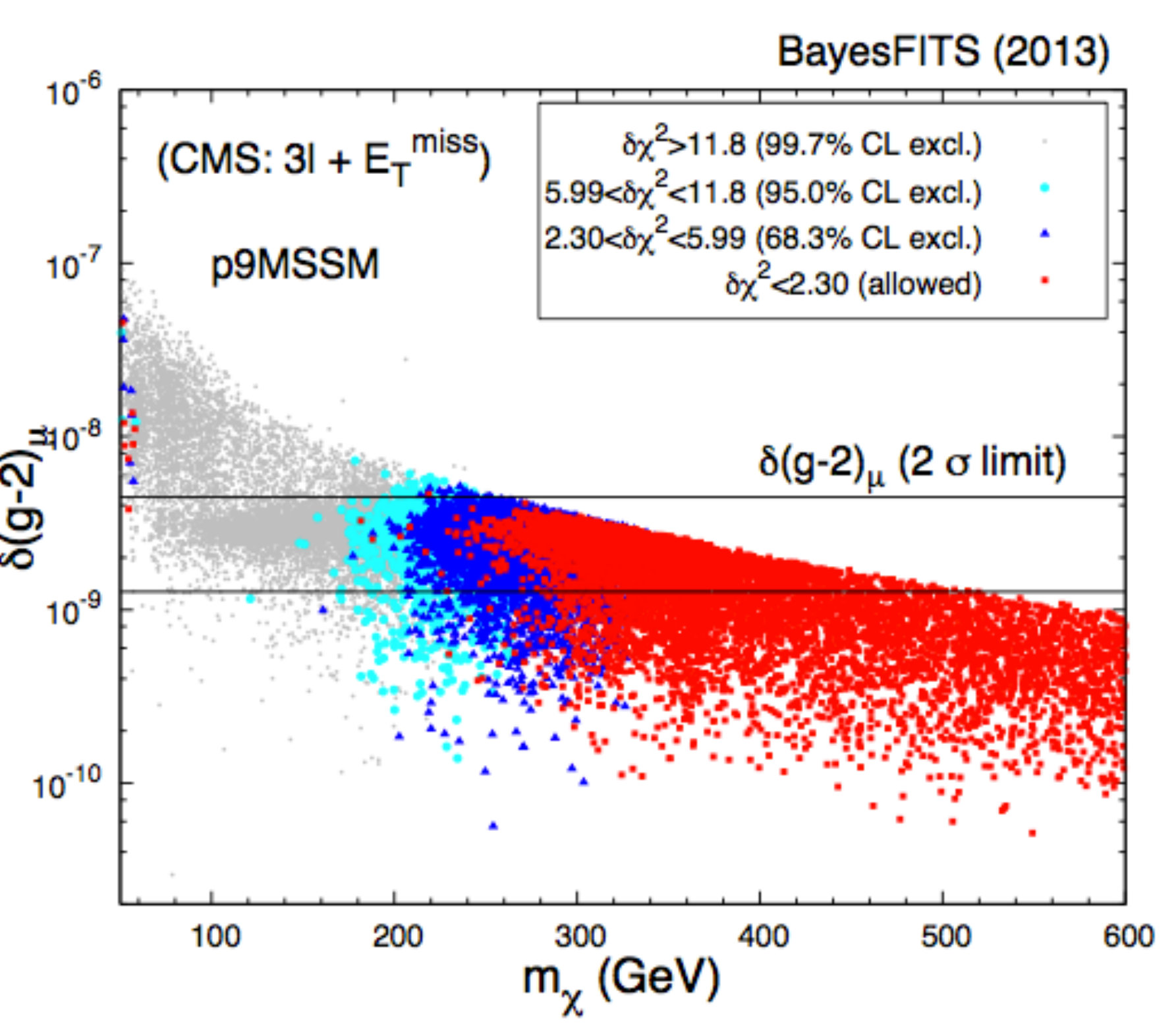}
\caption[]{
Our chargino-neutralino pair production (EW) likelihood in the (\mchi, \deltagmtwomu) plane for a thinned sample 
of p9MSSM points consistent at $2\sigma$ with the \textbf{basic} constraints. 
The allowed $2\sigma$ interval for \gmtwo\ is shown
with horizontal solid lines. 
The \delchisq\ from EW production is indicated by the different colors: grey dots, 
cyan circles and blue triangles are excluded at the $99.7\%$, $95\%$ and $68.3\%$~C.L., respectively. 
}
\label{fig:g2_ew}
\end{figure} 

As mentioned in Sec.~\ref{sec:method}, we imposed a GUT-inspired universality
condition, $M_1=0.5 M_2$, on our parameter space. We additionally assumed
$m_{\tilde{L}_{1,2}}=M_1+50\gev$, in order to enhance \gmtwo.

The anomalous magnetic moment of the muon, \gmtwo, is subject to SUSY
contributions that can enhance its SM value for $\mu>0$. 
At the leading order, the dominant terms
are given by a chargino-sneutrino (of the second generation) loop and by a neutralino-smuon loop (Ref.\cite{Endo:2013bba} and references therein).

In \reffig{fig:mssm_g2}\subref{fig:a} we show the distribution of scan
points in the (\mchi, $m_{\tilde{\mu}_R}$) plane and in
\reffig{fig:mssm_g2}\subref{fig:b} the one in the
($m_{\tilde{\nu}_{\mu}}$, $m_{\chi_1^{\pm}}$) plane.  The purple
(orange) squares (circles) indicate the points for which
\deltagmtwomu\ is satisfied at $1\sigma$ ($2\sigma$), i.e.,
$\chisq_{g-2}<1$ ($\chisq_{g-2}<4$).

For $\mu\ll M_2=2 M_1$, $\chi$ and $\chi_1^{\pm}$ are higgsino-like, or $\chi$ is a 
bino/higgsino mixed state and $\chi_1^{\pm}$ is higgsino-like, and their masses are comparable. They are both 
lighter than the sleptons, so \deltagmtwomu\ gives the bound
$m_{\tilde{\mu}}\simeq m_{\tilde{\nu}_{\mu}}\lesssim 600\gev$, which is independent 
of our parametrization.
On the other hand, for $\mu>M_2$, the neutralino is bino-like and the (wino-like) chargino mass presents an upper limit 
$m_{\chi_1^{\pm}}\approx2\mchi$.
Thus, by placing an upper bound on the mass of the neutralino, $\mchi\lesssim500\gev$,
the \gmtwo\ constraint indirectly places a limit on the chargino mass $m_{\chi_1^{\pm}}\lesssim1\tev$.
Besides, since we have set $m_{\tilde{L}_{1,2}}=M_1+50\gev$, when the neutralino is bino-like an upper bound on its mass translates 
on an indirect upper bound on the smuon mass. 

So, the two parameters of relevance to the \gmtwo\ constraint are $\mu$ and $M_2$.
On the other hand, when the relic density constraint is included,
the region where the neutralino is almost purely higgsino-like is excluded 
at the 95\%~C.L. for $\mchi\ll 1\tev$,
as shown  in Figs.~\ref{fig:mssm_m2_mu} and \ref{fig:X100_mx}.
In Figs.~\ref{fig:mssm_g2}\subref{fig:a} and \ref{fig:mssm_g2}\subref{fig:b}, the points for which the \textbf{basic} set 
of constraints is satisfied together with \deltagmtwomu\ at $2\sigma$ are shown as green empty circles. 
One can identify two main branches: on the diagonal, close to the edge of the parameter space, $\mu>M_2$;
elsewhere $\mu\leq M_2$.
 
Inclusive searches for SUSY particles, like the \alphaT\ search that we included in the likelihood function, 
have little sensitivity to models in which the squarks are heavy and the sleptons are light.
On the other hand, searches for EW production of pair-produced charginos
with multiple leptons and missing energy in the final state are designed to probe the 
parameter space of the theory that overlaps
with the \gmtwo\ sector. As described in Sec.~\ref{subsec:LHC}, 
we calculated the likelihood for the CMS EW-production search at 
$\sqrt{s}=8\tev$ and $\mathcal{L}=9.2$/fb, which is the one giving the strongest limits. 
Due to the great number of points in our scan that can be affected by this search, we 
apply the numerical procedure to calculate the likelihood to a 
randomly selected sample of approximately 40,000 points, all of which 
satisfy the \textbf{basic} set of constraint at the 95\%~C.L.

We show the exclusion due to EW production in \reffig{fig:g2_ew},
where we plot the 95\% confidence region for the \textbf{basic}
constraints applied to the thinned chain in the (\mchi, \deltagmtwomu)
plane (note that after the other constraints are taken into account,
in the p9MSSM \deltagmtwomu\ is parametrized only by \mchi).  The
color code is the same as in Figs.~\ref{fig:at_8tev_sms} and
\ref{fig:EWprod}. The cyan circles are excluded at the 95.0\%~C.L.,
and the gray dots at the 99.7\%~C.L.  One can see again that the
\gmtwo-- $2\sigma$ region requires $\mchi\lesssim 500\gev$, whereas
the strongest EW-production search at the LHC requires $\mchi\gsim
200-250 \gev$, the range depending on the parameters and on the
chargino and neutralino compositions.  Thus, there is a window of
availability in the parameter space for $200\gev \lesssim\mchi
\lesssim 500\gev$, of points in good agreement with all constraints.

As a side note, on the left of \reffig{fig:g2_ew} one can notice the presence of some
points at $\mchi\simeq 50\gev$ not excluded by the LHC EW production search.
Those points are characterized by $|m_{\neuttwo(\charone)}-m_{\tilde{l}}|\lesssim 1\gev$ (the neutralino is bino-like and we chose 
$m_{\tilde{L}_{1,2}}=M_1+50\gev$ and $M_1=0.5M_2$)
and the intermediate sleptons are considered on-shell by \texttt{PYTHIA}. As a consequence,
two of the final state leptons are soft and the search loses sensitivity.

\subsection{Higgs mass and $h\rightarrow \gamma\gamma$ signal rate }

\begin{figure}[t]
\centering
\subfloat[]{%
\label{fig:a}%
\includegraphics[width=0.57\textwidth]{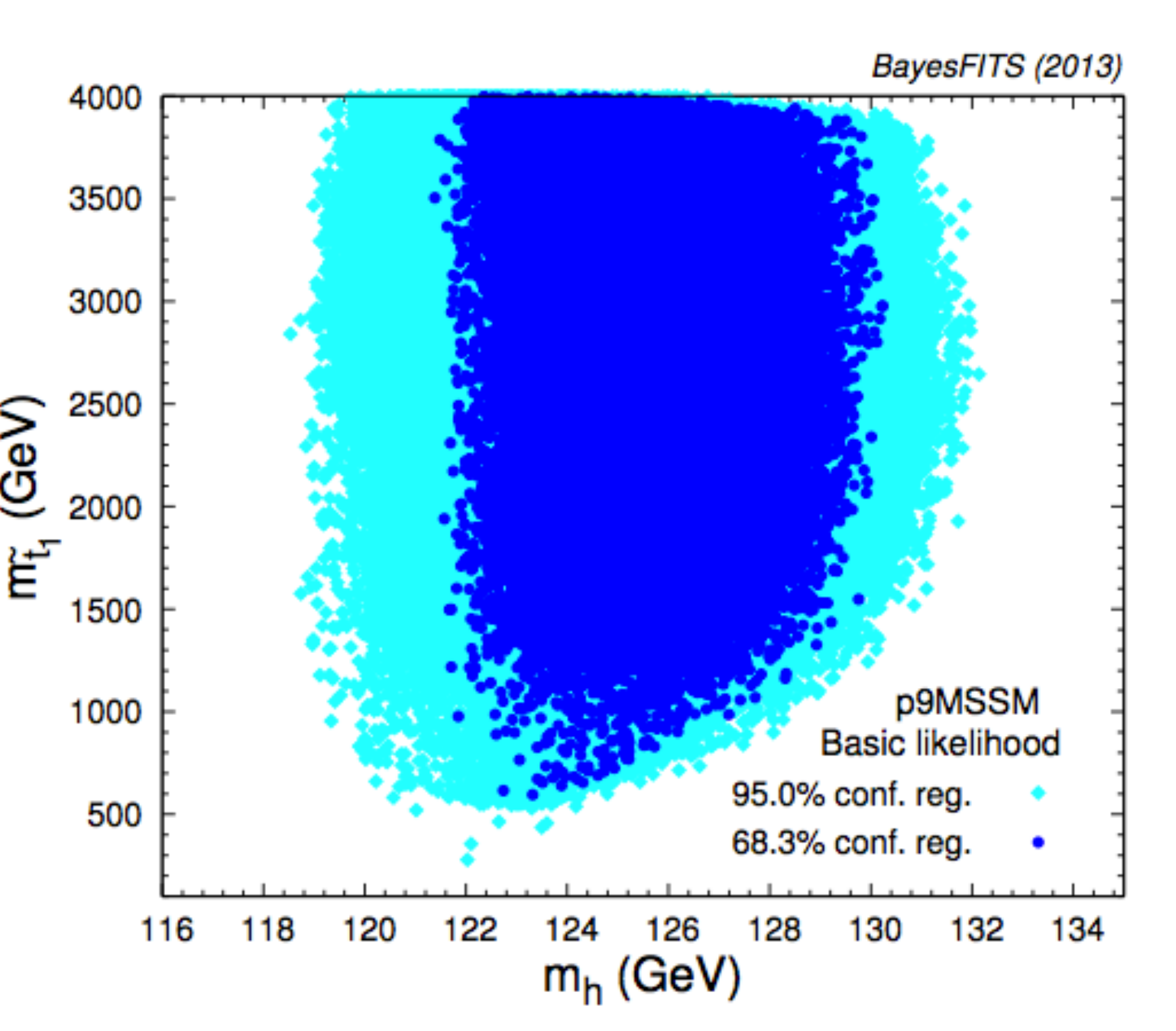}
}%
\subfloat[]{%
\label{fig:b}%
\includegraphics[width=0.50\textwidth]{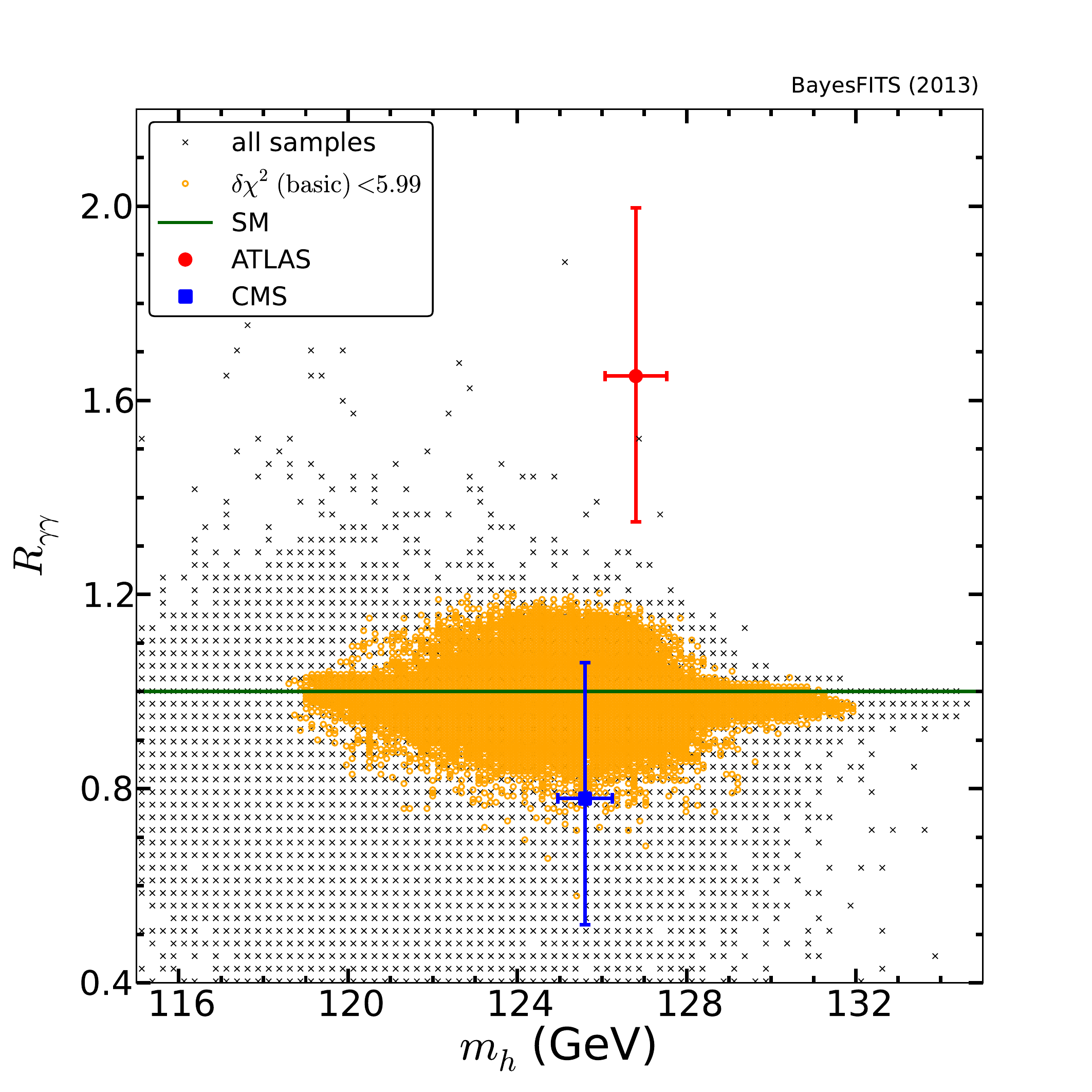}
}%
\caption[]{ \subref{fig:a} 68.3\% (blue points) and 95.0\% (cyan
  diamonds) confidence regions for the \textbf{basic} likelihood in
  the (\mhl, \mstopone) plane.  \subref{fig:b} Scatter plot of p9MSSM
  points in the (\mhl, $R_{\gamma\gamma}$) plane.  The points
  consistent with \textbf{basic} constraints at $2\sigma$ are shown as
  orange circles.  The Standard Model value $R_{\gamma\gamma}=1$ is
  marked with a horizontal solid line.  The ATLAS and CMS measurements
  are marked with red and blue cross hairs respectively.}
\label{fig:diphot}
\end{figure} 

In \reffig{fig:diphot}\subref{fig:a}, we present the $1\sigma$ (blue
points) and $2\sigma$ (cyan diamonds) confidence regions of the
\textbf{basic} likelihood in the (\mhl, \mstopone) plane.  One can see
that \mhl\ presents a normal distribution around the central value, as
expected from the Gaussian likelihood.  A Higgs mass consistent at
$1\sigma$ with all constraints including the measured value of the
Higgs mass can be obtained for stop masses as small as $\sim600\gev$,
thanks to maximal stop mixing, $|X_t|/\msusy\simeq\sqrt{6}$. The
$2\sigma$ region allows for stop masses as small as $\sim280\gev$.
For $\mchi\lesssim 200\gev$ the \alphaT\ constraint excludes
$\mstopone\lesssim350\gev$ at the 95\%~C.L. (we remind the reader that
our implementation of the \alphaT\ search is very conservative for the
third generation squarks, as was discussed in \refsec{subsec:LHC})
unless the SUSY spectrum is compressed ($\mchi\gsim\mstopone-M_t$), in
which case the search becomes sensitive to initial state radiation,
and a reliable bound cannot be produced. This is the case for
the two points at $\mstopone\lesssim350\gev$ shown in the plot, which
are characterized by neutralino masses of order
$\mchi\sim220-230\gev$.

In this subsection we also check to what extent the MSSM 
lightest Higgs boson $h$ complies with the LHC observations in the $\gamma\gamma$ channel.
In order to do so, we calculate its reduced cross section, defined in literature (see, e.g.,\cite{Gunion:2012gc})
as
\be
R_{h}(\gamma\gamma) =  \frac{\sigma(pp\rightarrow h)} {\sigma(pp\rightarrow
  h_{\textrm{SM}})}\times \frac{BR(h \rightarrow \gamma\gamma)}{BR(h_{\textrm{SM}} \rightarrow
  \gamma\gamma)}\,.
\label{eq:RX1}
\ee

The branching ratios in Eq.~(\ref{eq:RX1}) are calculated using \texttt{FeynHiggs 2.9.4}
\cite{feynhiggs:06,feynhiggs:03,feynhiggs:00,feynhiggs:99} both for the MSSM $h$ and the SM Higgs, 
$h_{\textrm{SM}}$, with the same mass. 
The total Higgs production cross sections are computed from the parton level production cross sections as follows:
\be
\frac{\sigma(pp\rightarrow h)}{\sigma(pp\rightarrow h_{\textrm{SM}})}=
\sum_{Y\in\textrm{ prod}}\frac{\sigma(pp\rightarrow Y\rightarrow h_{\textrm{SM}})}{\sigma(pp\rightarrow h_{\textrm{SM}})}
\times\frac{\sigma(Y\rightarrow h)}{\sigma(Y\rightarrow h_{\textrm{SM}})}\,,\label{eq:RX2}
\ee
where $Y$ spans over the different production channels: 
gluon-fusion, vector boson-fusion, Higgs-strahlung off a $W/Z$ boson,
and associated Higgs production with top
quarks. The parton level cross sections $\sigma(Y\rightarrow h)$ and $\sigma(Y\rightarrow h_{\textrm{SM}})$
are computed for $h$ and $h_{\textrm{SM}}$ using \texttt{FeynHiggs 2.9.4}. 
For each production mode $Y$, we then obtain a coefficient $\sigma(pp\rightarrow Y\rightarrow h_{\textrm{SM}})/\sigma(pp\rightarrow h_{\textrm{SM}})$
to transform the  
calculation from the parton to the hadron level 
using the public tables provided by the LHC Higgs Cross Section Working
Group\cite{HXSWG:2011ti,Dittmaier:2012vm} for $\sqrt{s}=8\tev$.

In \reffig{fig:diphot}\subref{fig:b} we show the p9MSSM distribution
of points in the (\mhl, $R_{h}(\gamma\gamma)$) plane. The complete
sample is shown in gray crosses, while the points that satisfy the
\textbf{basic} constraints at $2\sigma$ are shown as orange
circles. We also superimpose the CMS and ATLAS central values and
experimental errors.  One can notice the the $\gamma\gamma$ rate can
be enhanced in the p9MSSM by 20\% with respect to the SM, when all
constraints are taken into account.  The value of
$R_{h}(\gamma\gamma)$ can, however, also be as low as 0.6.  Different
mechanism of di-photon rate enhancement were discussed in the
literature, including the effects of the light
staus\cite{Carena:2012gp} or light charginos\cite{Casas:2013pta}.  In
our scan, di-photon rate enhancement is in general a combination of
different mechanisms.


\subsection{Indirect detection of DM\label{sec:IDdetect}}

Having tested the compatibility of our model with the limits from XENON100, 
\gmtwo\ and the LHC SUSY searches, we now proceed to examine the implications 
from ID of DM experiments on the allowed regions of the parameter space.  
We derive constraints from Fermi $\gamma$-ray data from the 
GC of the Milky Way as well as from its dSphs
and from IceCube data on neutrinos from the Sun.   

\begin{figure}[t]
\centering
\subfloat[]{%
\label{fig:a}%
\includegraphics[width=0.50\textwidth]{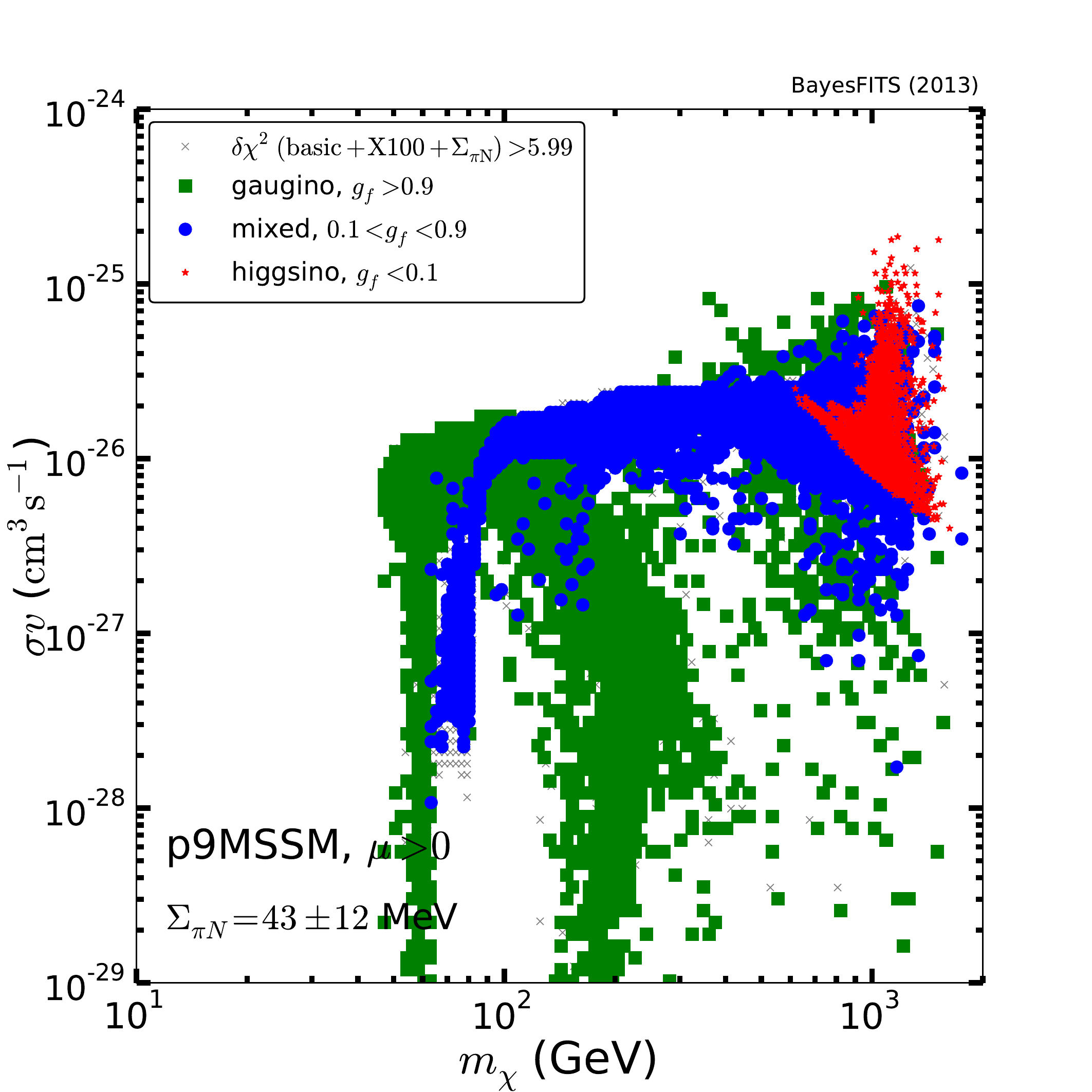}
}%
\subfloat[]{%
\label{fig:b}%
\includegraphics[width=0.50\textwidth]{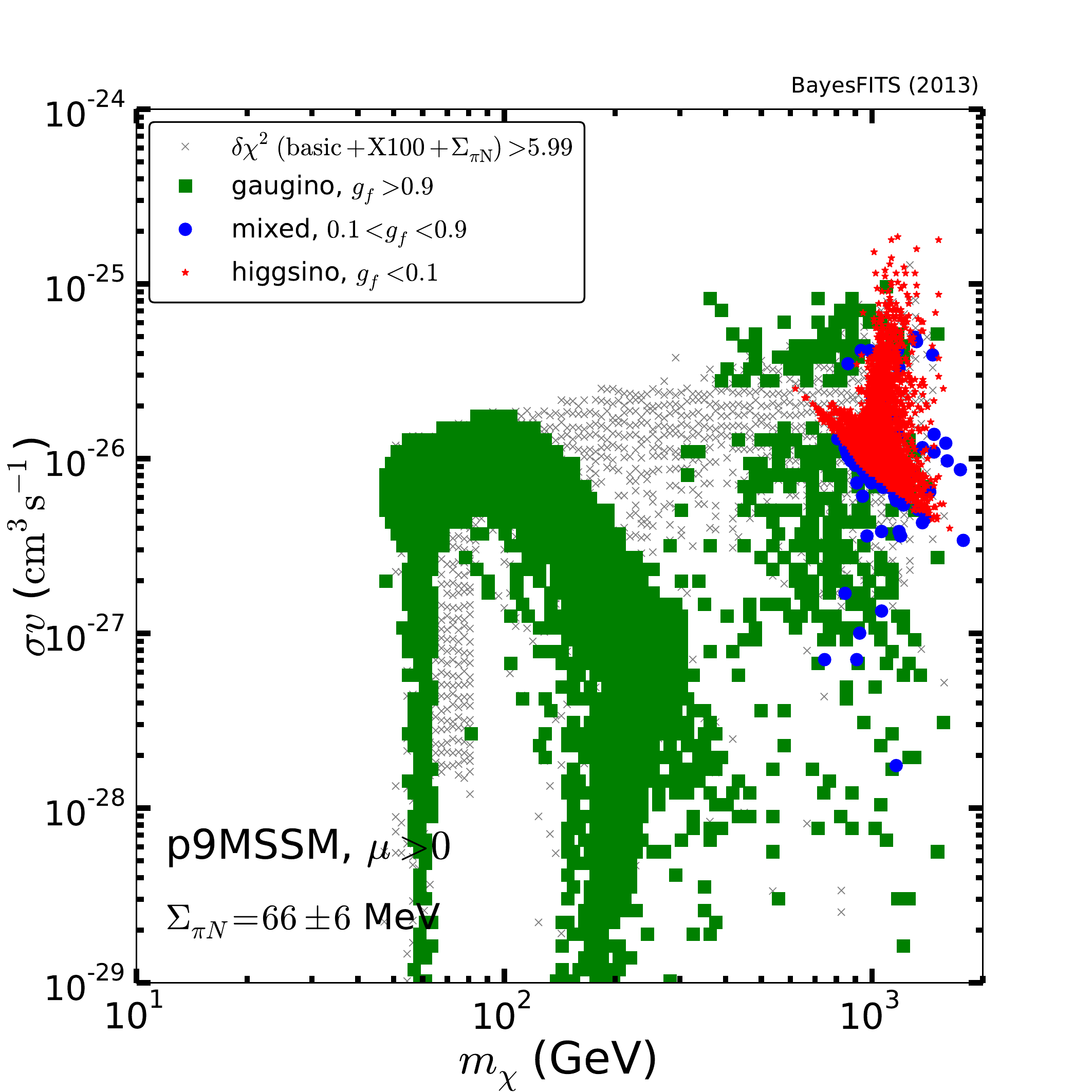}
}%
\caption[]{ p9MSSM points allowed at 2$\sigma$ by the \textbf{basic}
  constraints in the ($\mchi$, $\sigma v$) plane.  The points
  consistent at 2$\sigma$ with the \textbf{basic} \textit{and}
  XENON100 constraints are shown for different composition of the
  neutralino: gaugino-like (green squares), mixed (blue circles), or
  higgsino-like (red stars).  \subref{fig:a} $\Sigma_{\pi N}= 43\pm12$
  MeV, \subref{fig:b} $\Sigma_{\pi N}= 66\pm6$ MeV.  }
\label{fig:gf}
\end{figure}

The quantity relevant for indirect DM detection
searches is the neutralino annihilation
cross section in the limit of small momenta, $\sigma v\equiv\sigma v|_{p\rightarrow 0}$.
In \reffig{fig:gf} we present the 95\% confidence regions obtained  by adding the XENON100 likelihood to the \textbf{basic} 
set of constraints, projected onto the ($\mchi$, $\sigma v$)
plane. As was done in \reffig{fig:X100_mx}, we show the case with $\Sigma_{\pi N}=43\pm 12\mev$ in \reffig{fig:gf}\subref{fig:a} and 
the one with $\Sigma_{\pi N}=66\pm 6\mev$ in \reffig{fig:gf}\subref{fig:b}.
The color code describing the gaugino fraction of the LSP is the same 
as in the previous figures. 

Different mechanisms of generating the correct value of the relic
density, associated with different regions of the ($M_1$, $\mu$) plane
in \reffig{fig:mssm_m2_mu}, can be also identified in
\reffig{fig:gf}\subref{fig:a}.  The first vertical branch on the left,
characterized by gaugino-like neutralinos at $\mchi\simeq 60\gev$,
corresponds to the HR region of the ($M_1$, $\mu$) plane, while the
adjacent gaugino region at $\sigma v\simeq 10^{-26}$~cm$^3$~s$^{-1}$
corresponds to the bulk region.

The second vertical branch at $\mchi\simeq 80\gev$, with mixed
gaugino/higgsino composition, becomes horizontal for larger masses and
extends to $\sim 800\gev$. As we pointed out while discussing
\reffig{fig:mssm_m2_mu}, this is the MSSM counterpart of the FP/HB
region in GUT-constrained models. One can see that the opening of the
$WW$ annihilation channel at $\mchi\gsim 80\gev$ increases the value
of $\sigma v$. Also note that the only points excluded at the
95\%~C.L. by adding XENON100 in the case $\Sigma_{\pi N}=43\pm 12\mev$
lie in this region (gray crosses).

More to the right, there is an additional vertical branch, at $\mchi\simeq 100-300\gev$, characterized by 
a large gaugino fraction. This is the SC region. The AF region can instead be identified with the widespread area of gaugino-like LSP 
at $400\gev\lesssim\mchi\lesssim 1.2\tev$.

In the 1TH region, for $800\gev\lesssim\mchi\lesssim 1.2\tev$, about 50\% of the contribution
to the relic density reduction comes from $\chi_{1}\chi^\pm$ and $\chi_{1}\chi_{2}$ coannihilations.
Most of the remaining half of the total contribution in this branch is due to
$\chi\chi$ annihilation to fermions.

We remind the reader that, although the value of $\sigma v$ is related
to the value of the relic abundance of neutralinos at present, the correlation between the two quantities is not straightforward, 
since the annihilation (and coannihilation)
cross sections used to determine the relic density are thermally averaged in the early Universe.
This is the reason why many of the points in \reffig{fig:gf} present cross sections in excess of or below
the ``default'' value of $3\times 10^{-26}\textrm{cm}^3/\textrm{s}$ for the relic density.
We give a short review of this issue in Appendix~B.

\subsubsection{$\gamma$ rays from dSphs}

\begin{figure}[t]
\centering
\subfloat[]{%
\label{fig:dSphsBF}%
\includegraphics[width=0.50\textwidth]{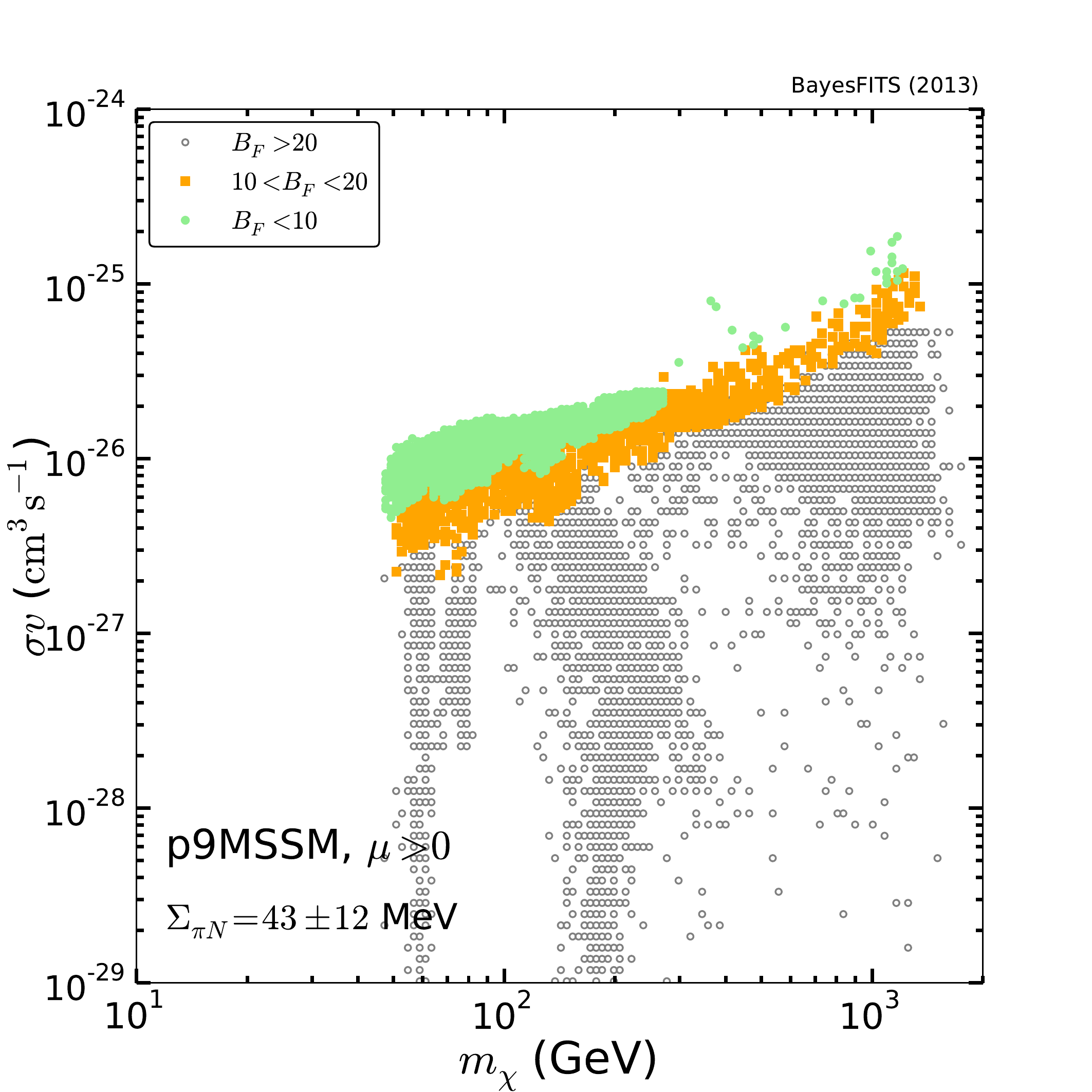}
}%
\subfloat[]{%
\label{fig:gammaGC}%
\includegraphics[width=0.50\textwidth]{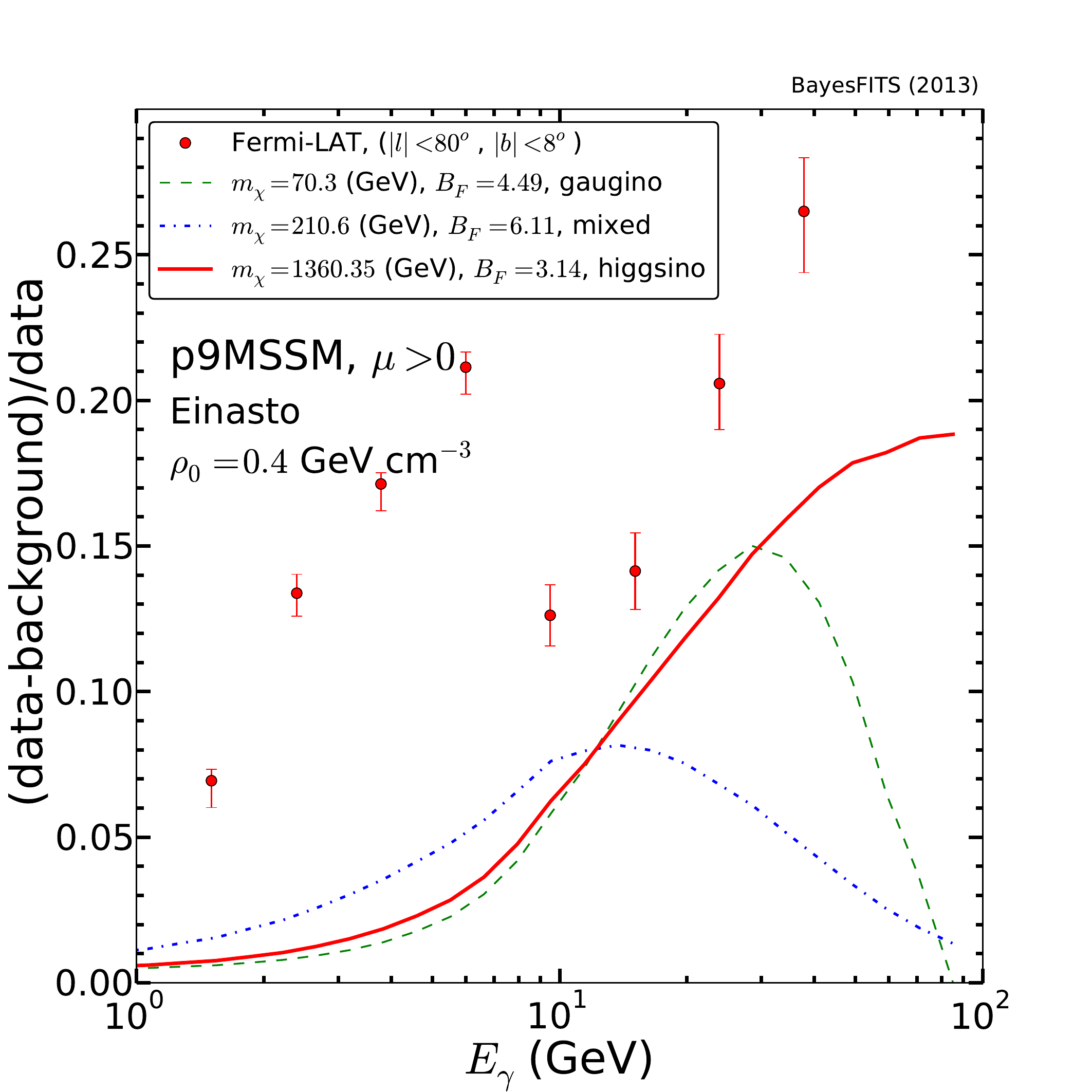}
}%
\caption[]{ \subref{fig:dSphsBF} p9MSSM points consistent with the
  \textbf{basic} and XENON100 constraints at $2\sigma$ in the (\mchi,
  $\sigma v$) plane.  The $\gamma$-ray flux boost factor $B_F$ is
  indicated by the different colors: points with $B_F>20$ are shown
  with gray circles, $10<B_F<20$ with orange squares, and $B_F<10$
  with green diamonds.  \subref{fig:gammaGC} The expected $\gamma$-ray
  flux distribution above the background from the GC for three p9MSSM
  points (shown in \reftable{tab:susypts}) with small $B_F$, but with
  different neutralino composition: a point with gaugino-like
  neutralinos is shown with a dashed green line, mixed neutralinos
  with a dash-dotted blue line, and higgsino-like neutralinos with a
  solid red line.  The Einasto halo profile and a local density of
  $\rho_0=0.4\gev \textrm{cm}^{-3}$ have been assumed.  The measured
  $\gamma$-ray flux distribution from Fermi-LAT is shown with red
  circles.  }
\label{fig:FermiGa}
\end{figure} 

In order to quantify the impact of the Fermi-LAT exclusion bounds
from dSphs in the ($\mchi$, $\sigma v$) plane of the p9MSSM
we use a publicly available code written by one of us\cite{Tsai:2012cs}, 
which calculates the likelihood function for each model point given the 
4-year data made available by the experimental collaboration\cite{Fermitools}.
The program, described in detail in\cite{Tsai:2012cs}, takes the 
$\gamma$-ray yield spectrum for $\chi\chi$ annihilation at each model point,
$dN_{\gamma}/dE_{\gamma}$, as an input.

The likelihood does not exclude any of the points in our scan.
One should bear in mind, though, that the values of $dN_{\gamma}/dE_{\gamma}$ given by different simulation
codes such as \texttt{PYTHIA6}\cite{Sjostrand:2006za}, \texttt{PYTHIA8}\cite{Sjostrand:2007gs}, 
or \texttt{Herwig++}\cite{Bahr:2008pv} can differ by a significant factor\cite{Cembranos:2013cfa},
hence the predictions are subject to a certain amount of theoretical uncertainty.
We incorporated possible uncertainties in our results by multiplying 
the SUSY-predicted $\gamma$-ray flux with a phenomenological boost factor, $B_F$,
adjusted point by point to maximize the likelihood.
We show in \reffig{fig:FermiGa}\subref{fig:dSphsBF} 
the points of the 95\% confidence region in the ($\mchi$, $\sigma v$) 
plane from the combined \textbf{basic}+XENON100 profile-likelihood (we assume
$\Sigma_{\pi N}=43\pm 12\mev$). 
The color code indicates the boost factor required at each point to maximize the dSphs
likelihood\cite{Tsai:2012cs}. For the green diamonds, $B_F<10$, for the orange squares $10<B_F<20$, 
and for the gray circles $B_F>20$. 
One can see that the dSphs 4-year data can be used to test 
some of the points (green points) not tested by XENON100 (see \reffig{fig:gf}\subref{fig:a} for comparison), 
provided a small boost factor is 
assumed.

\subsubsection{$\gamma$ rays from the Galactic Center}

\begin{table} [t]
\begin{center}
\begin{tabular}{|c|c|c|c|}
\hline\hline
Parameter& gaugino & mixed& higgsino  \\
\hline\hline
$\tanb$ & 7.94 & 52.63 & 4.76 \\
\hline
$M_{2}$ (GeV)& 148.56 & 457.82 & 3810.98 \\
\hline
$M_{3}$ (GeV)& 1847.66 & 2785.12 & 2281.29 \\
\hline
$\mu$ (GeV)& 620.95 & 275.72 & 1345.21 \\
\hline
$\ma$ (GeV)& 1454.49 & 3648.84 & 2716.89 \\
\hline
$A_{t}$ (GeV)& 2086.52 & 3607.58 & 5277.32 \\
\hline
$A_{\tau}$ (GeV)& $-2786.17$ & 5256.12 & $-4519.05$ \\
\hline
$m_{\tilde{Q}_3}$ (GeV)& 3335.38 & 2330.03 & 3577.35 \\
\hline
$m_{\tilde{L}_3}$ (GeV)& 144.74 & 1613.40 & 1500.82 \\
\hline\hline
$m_{\chi}$ (GeV)& 70.29 & 210.61 & 1360.36 \\
\hline
$\sigma v$ (cm$^{3}$s$^{-1}$)& $1.33\times 10^{-26} $ & $2.64\times 10^{-26} $ & $3.75\times 10^{-25} $\\
\hline
$B_F$ & $4.49 $ & $6.11 $ & $3.14 $\\
\hline\hline
\end{tabular}
\end{center} 
\caption{
Our three benchmark p9MSSM DM candidate points, which are denoted by their neutralino compositions.
All the three points are consistent with LEP, \alphaT, PLANCK, flavor physics, 
measurements of nuisance parameters, Higgs mass measurements and XENON100 limit at 2$\sigma$.
}
\label{tab:susypts}
\end{table}

Of the points shown in green in
\reffig{fig:FermiGa}\subref{fig:dSphsBF}, we take the three with the
smallest $B_F$ and we simulate numerically the diffused $\gamma$-ray
fluxes from the GC with the \texttt{GALPROP}
package\cite{Strong:2009xj}.  The p9MSSM parameters of the three
selected points are given in Table~\ref{tab:susypts}.  We assume the
Einasto halo profile\cite{Einasto} and a local density
$\rho_0=0.4\gev\cmeter^{-3}$\cite{Salucci:2010qr}.  The $\gamma$ rays
from the GC can be produced via $\pi^0$ decay, as well as inverse
Compton scattering and bremsstrahlung of the electrons and positrons
along the propagation line in the interstellar medium. For each point,
we insert $dN_{\gamma}/dE_{\gamma}$ into \texttt{GALPROP} to calculate
$\gamma$-ray fluxes from $\pi^0$ decay. For inverse Compton scattering
and bremsstrahlung, we insert $dN_{e^{\pm}}/dE_{e^{\pm}}$ into
\texttt{GALPROP} to obtain $\gamma$-ray fluxes due to the propagation
of the $e^{\pm}$ from the GC to the Earth.  In our analysis, we use
the best-fit propagation model\cite{Trotta:2010mx}.  In
\reffig{fig:FermiGa}\subref{fig:gammaGC}, we show the comparison of
the residual ``data$-$background/data'' ratio, calculated for the
three selected points in the Galactic coordinates $|l|<80^o$ and
$|b|<8^o$, with the experimental data published
in\cite{FermiLAT:2012aa}. Interestingly, the points that seem to be
closer to reproducing the data (with $B_F<5$) are the ones
characterized either by heavy higgsino-like neutralinos, because of
the enhancement in $\sigma v$ and of a substantial integrated flux, or
by a light gaugino-like $\chi$, because the flux suppression due to
$\mchi^2$ is reduced.  A more refined analysis of the DM halo and
astrophysical background sources will, however, be required to draw
more quantitative conclusions.

\subsubsection{Sensitivity of IceCube}

As mentioned earlier, data from the IceCube neutrino telescope can
also be used to test the parameter space of the model.  
In particular, neutrinos from the Sun provide a means to test the
spin-dependent cross section of neutralino-nucleon elastic scattering, 
\sigsdp,
to which DD experiments are insensitive\cite{Jungman:1995df}. 

We use \texttt{DarkSUSY} to compute the 5-year, 95\%~C.L. sensitivity
of the IceCube's 86 string configuration.  We adopt the default values
of the effective area, of the atmospheric neutrino background, and of
the quark spin content of the nucleon.  The maximum opening angle is
set to $20^o$ from the center of the Sun.

\begin{figure}[t]
\centering
\subfloat[]{%
\label{fig:a}%
\includegraphics[width=0.44\textwidth]{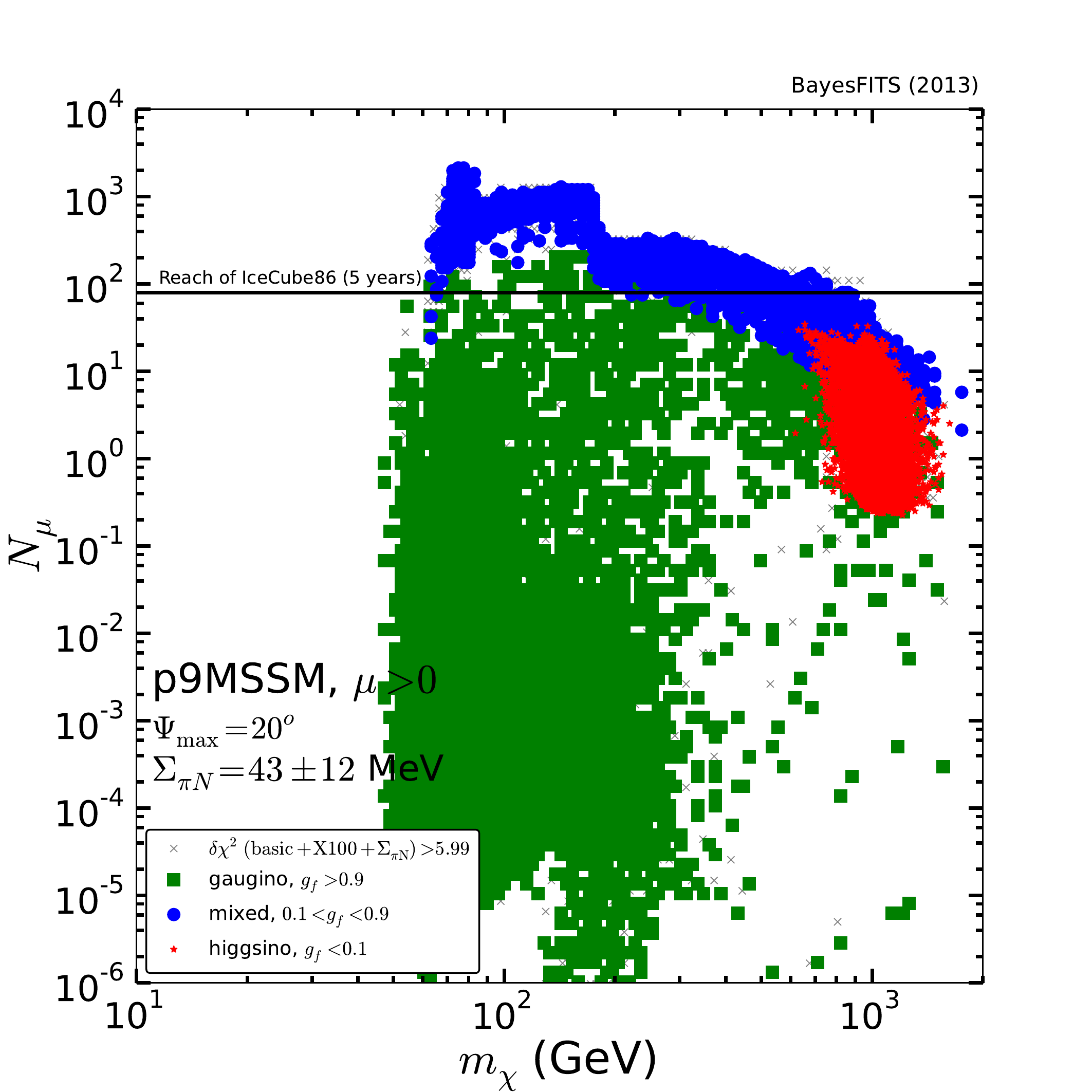}
}%
\subfloat[]{%
\label{fig:b}%
\includegraphics[width=0.44\textwidth]{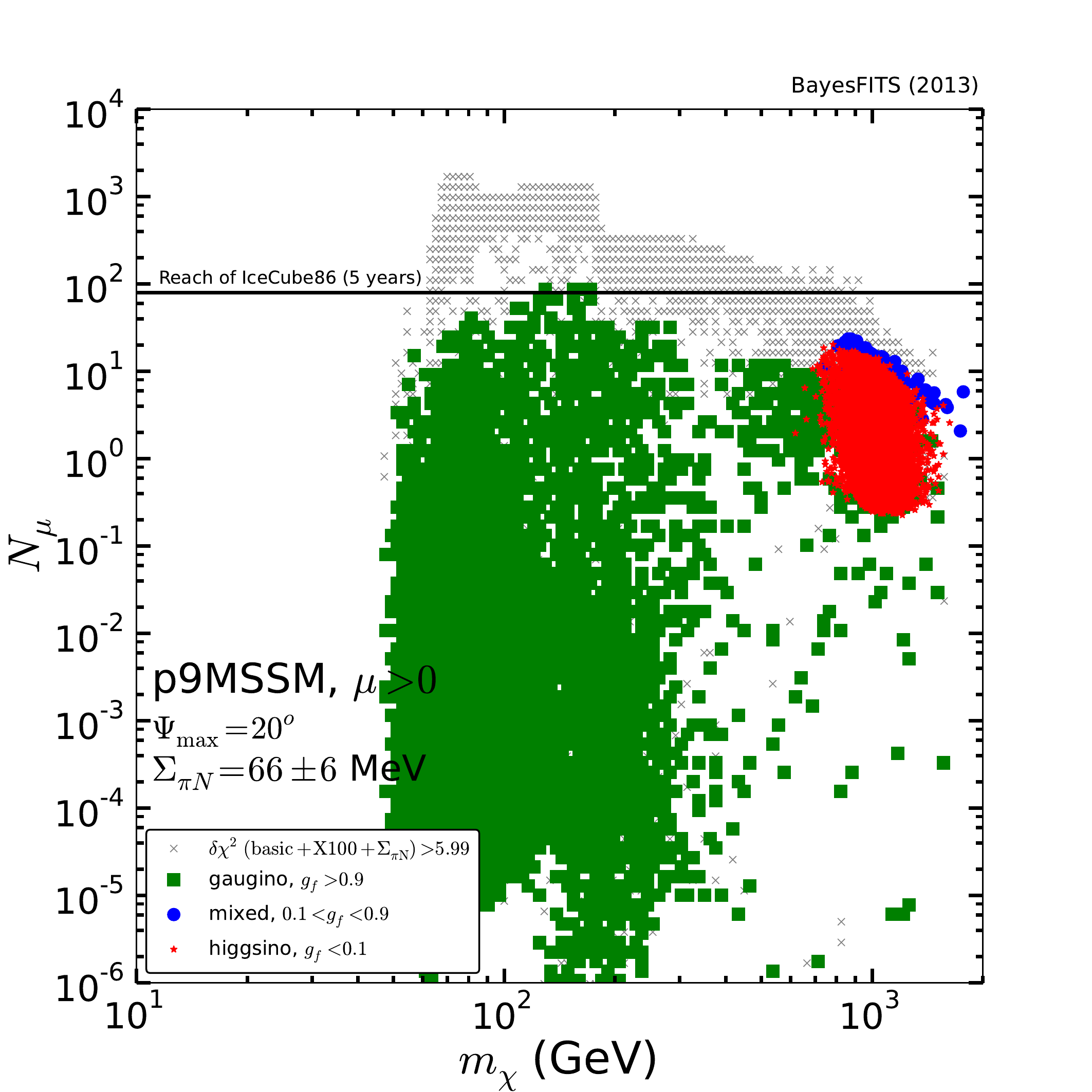}
}\\
\subfloat[]{%
\label{fig:c}%
\includegraphics[width=0.44\textwidth]{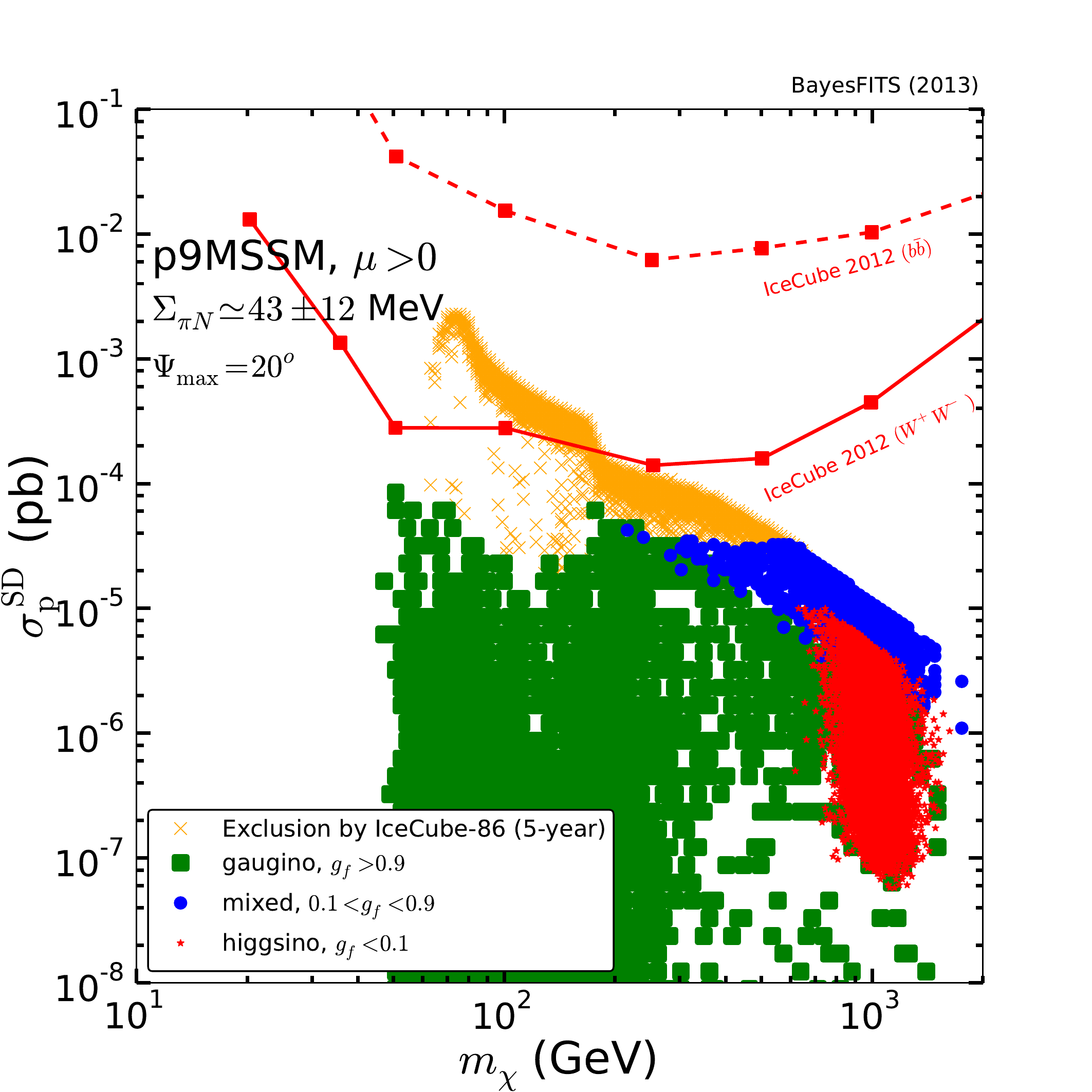}
}%
\subfloat[]{%
\label{fig:d}%
\includegraphics[width=0.44\textwidth]{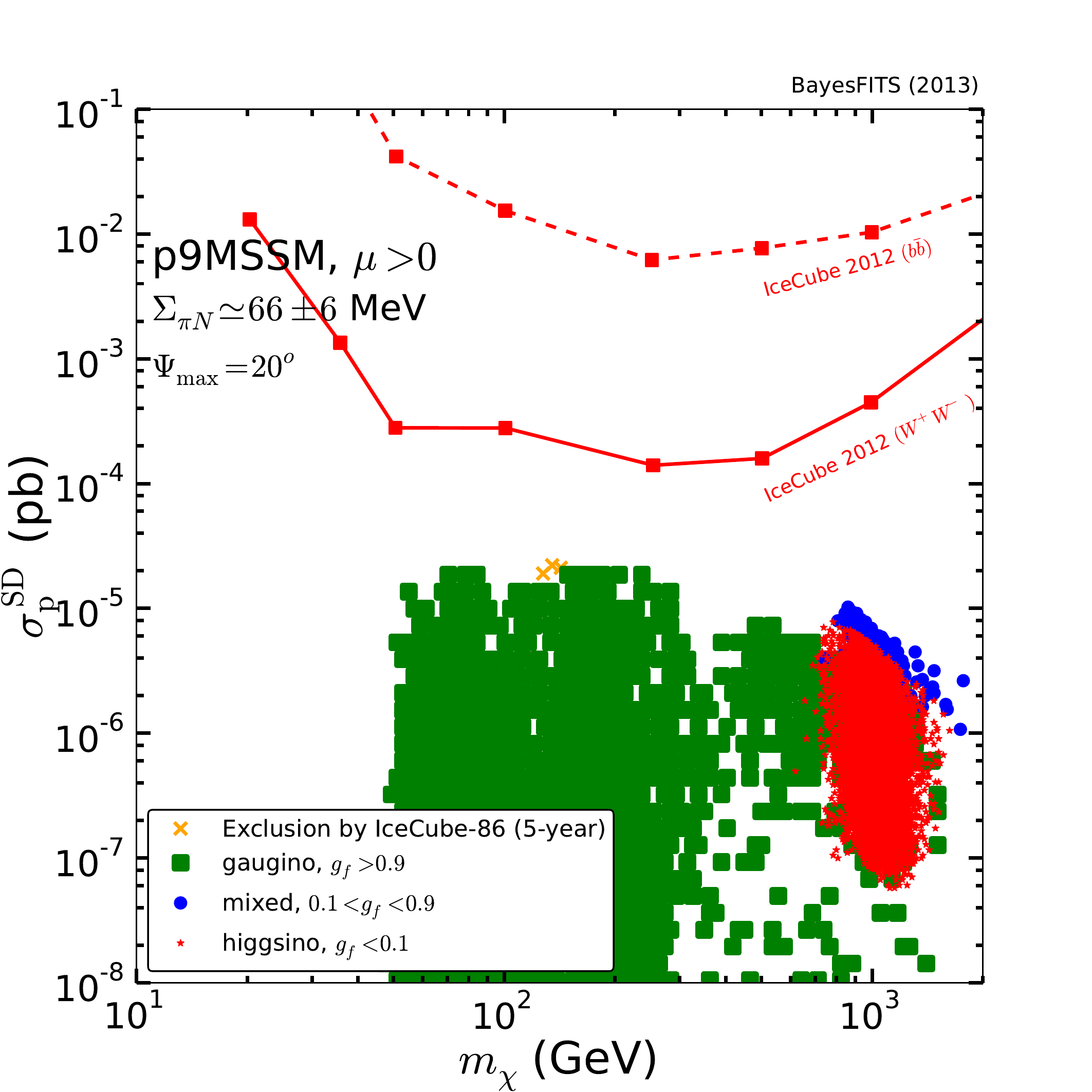}
}%
\caption[]{
\subref{fig:a} p9MSSM points allowed at 2$\sigma$ by the \textbf{basic} constraints in the (\mchi, $N_\mu$) plane. 
The points consistent at 2$\sigma$ with the \textbf{basic} \textit{and} XENON100 ($\Sigma_{\pi N}=43\pm12\mev$) 
constraints are characterized by the composition of the neutralino: 
gaugino-like (green squares), mixed (blue circles), or higgsino-like (red stars). 
The expected 95\%~C.L. sensitivity of IceCube-86 with 5 years of data is shown with a horizontal solid line.
\subref{fig:b} Same as \subref{fig:a}, but with $\Sigma_{\pi N}=66\pm6$ MeV. 
\subref{fig:c} p9MSSM points consistent at 2$\sigma$ with the \textbf{basic} and XENON100 
($\Sigma_{\pi N}=43\pm12\mev$) constraints in the (\mchi, \sigsdp) plane. The points to be excluded by 
IceCube-86 (5-year sensitivity) are shown as orange crosses.
Achieved 95\%~C.L. limits from IceCube-79\cite{Aartsen:2012kia} for neutralinos annihilating into $b\bar{b}$ 
and $W^+W^-$ channels are shown as dashed red and solid red lines, respectively.
Note, however, that for $\mchi< m_W$, the solid red line shows
the limit for neutralinos annihilating into $\tau^+\tau^-$.
\subref{fig:d} Same as \subref{fig:c}, but with $\Sigma_{\pi N}=66\pm6$ MeV.
}
\label{fig:IC86}
\end{figure} 

In \reffig{fig:IC86}\subref{fig:a}, we show the 
95\% confidence region from the \textbf{basic}+XENON100
profile-likelihood ($\Sigma_{\pi N}=43\pm 12\mev$) in the (\mchi, $N_{\mu}$) plane, where 
$N_{\mu}$ is the number of muon events at the detector in 5 years.
The color code is the same as in \reffig{fig:X100_mx}. 
In \reffig{fig:IC86}\subref{fig:b} we show the case with 
$\Sigma_{\pi N}=66\pm 6\mev$. One can see from \reffig{fig:IC86}\subref{fig:a} that many of the points
at present not excluded by the XENON100 likelihood with the most
conservative choice of theoretical uncertainty, 
are in reach of the 5-year sensitivity at IceCube.

In \reffig{fig:IC86}\subref{fig:c} we show the 95\% confidence region for the \textbf{basic}+XENON100 
($\Sigma_{\pi N}=43\pm12\mev$) likelihood in the (\mchi, \sigsdp) plane. The points excluded by 
the IceCube-86, 5-year sensitivity are shown as orange crosses. The case with $\Sigma_{\pi N}=66\pm6\mev$
is shown in \reffig{fig:IC86}\subref{fig:d}.

 
\subsubsection{Positron flux} 
 
\begin{figure}[t!]
\centering
\includegraphics[width=0.60\textwidth]{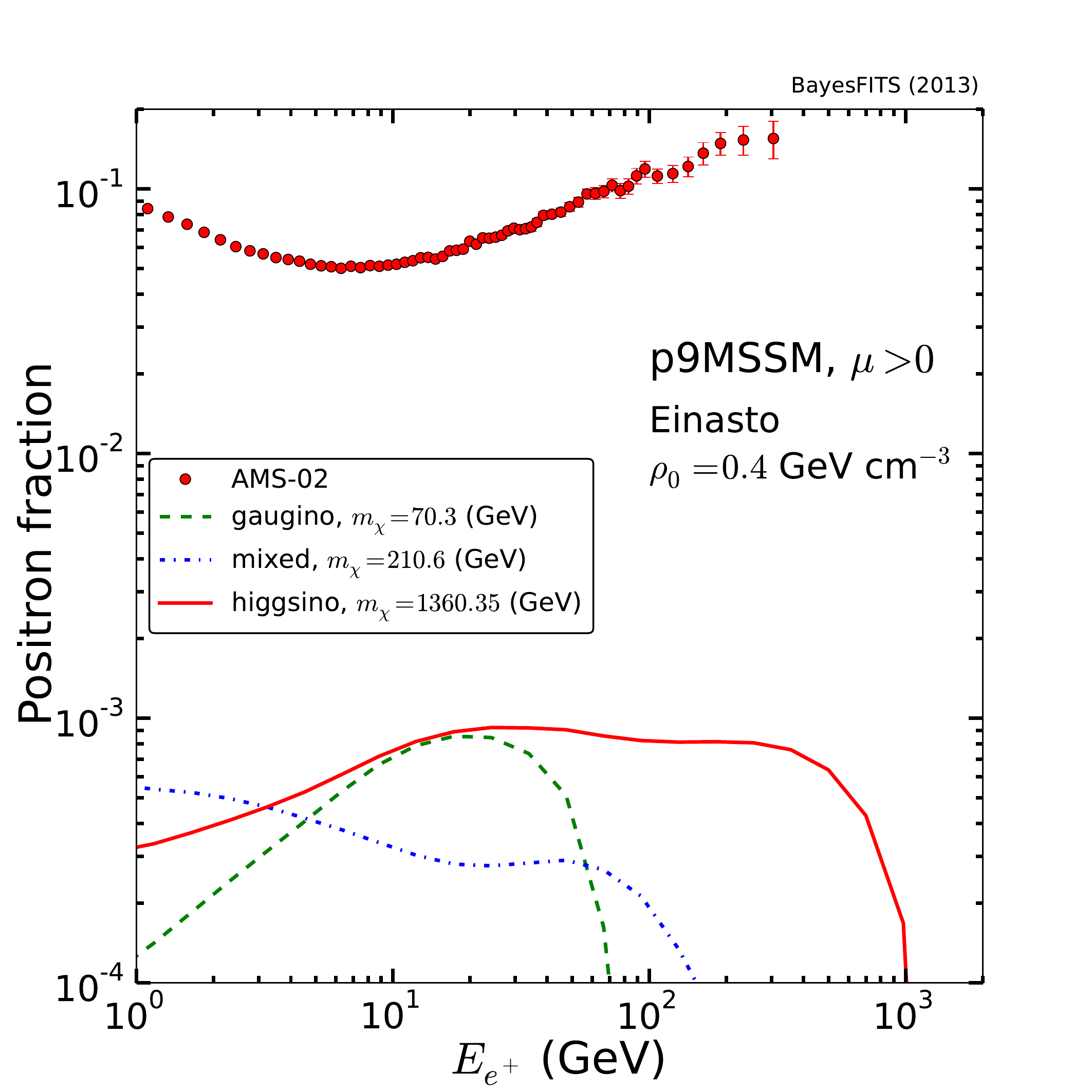}
\caption{\label{fig:ep}
Positron flux fraction from dark matter annihilation in the Galactic halo versus
the positron energy $E_{e^+}$, for three p9MSSM points with different neutralino composition
(shown in \reftable{tab:susypts}): the point with gaugino-like neutralino is shown with a dashed green line, 
mixed neutralino with a dash-dotted blue line and higgsino-like neutralino with a solid red line. 
We assume a boost factor of $B_F=1$, a local density of $\rho=0.4\gev\textrm{cm}^{-3}$ and the Einasto profile. 
The AMS-02 data is shown with red circles. The highest part of the positron fraction distributions 
would reach the AMS-02 data if the boost factors were $B_F\ge65$, 120, and 60 for gaugino-like, mixed, and higgsino-like neutralino, respectively.
}
\end{figure}

Finally we present the model's predictions for the positron flux from
neutralino annihilation in the local halo.
Annihilation of DM particles
may give rise to a large enough signal of positrons
that can be identified by several antimatter search experiments.
PAMELA\cite{Adriani:2010rc} first observed a rise of
the cosmic ray positron fraction for positron energies greater than
10 GeV. This anomalous enhancement was subsequently confirmed by
Fermi-LAT\cite{Abdo:2009zk} and by the recently released AMS02 first result\cite{Aguilar:2013qda}.
In the recent AMS02 result, this continuing rise is extended up to
positron energy $\sim 350\gev$.
Explanation of these anomalies by neutralino DM is not easy,
since the described enhancement requires a large boost factor 
for the annihilation cross section $\sigma v$\cite{Roszkowski:2009sm}.
On the other hand, several 
groups\cite{Cirelli:2008pk,Kopp:2013eka,DeSimone:2013fia,Yuan:2013eja,Cholis:2013psa,Yuan:2013eba}
have fitted the positron results by including AMS02 in a model independent
way, and they found that only heavy DM mass, from $\mchi\simeq 800\gev$ to a few \tev\ can
peak at the right position of positron spectrum.

One must bear in mind that in the p9MSSM analyzed in this study, the
positron flux would be suppressed by such large masses.  However, one
can also consider the case in which such enhancement could come from
the combined effect of nearby pulsars and DM annihilation, so that
$\mchi$ can be lowered down to a few hundred GeV\cite{Yuan:2013eja}.

Similar to what we did for the $\gamma$-ray flux, we employ the Einasto profile
with the caution that the halo uncertainties can affect the $\gamma$-ray 
flux by as much as a factor of 10\cite{Cumberbatch:2010ii},
but they do not affect the positron flux. This is because positrons will
lose energy during the propagation and, since most high-energy positrons
originate from a local neighborhood the size of a few kpc\cite{Baltz:1998xv,Lavalle:2006vb},
their flux is less dependent on the halo.
However, for the purpose of this paper here we ignore the uncertainties 
and we use the best-fit
propagation model from the GALPROP group\cite{Trotta:2010mx},
in which they fitted a number of isotopic abundance ratios such as B/C, 
$\rm{Be}^{10}/\rm{Be}^9$ and others.
Again, we set the local density to $\rho_0=0.4\gev\cdot\rm{cm}^{-3}$.
We modified \texttt{DarkSUSY} to
generate the positron-flux spectrum $dN_{e^+}/dE_{e^+}$.
The source term for solving the diffusion equation to obtain
the positron spectrum is given in terms of the 
DM density profile $\rho$ as a function of the distance from 
the GC, \textbf{r},
\begin{equation}
Q_{e^+}(\textbf{r},E_{e^+}) \sim\frac{\rho(\textbf{r})^2}{\mchi^2}\times\sigma v\times\frac{dN_{e^+}}{dE_{e^+}}
\,.
\end{equation}
This source term is then fed into \texttt{GALPROP} with the running
parameters of the best-fit model\cite{Trotta:2010mx}.

Since we are interested in the positron fraction at large energy,
we do not include the solar modulation effect.
In \reffig{fig:ep}, we present the positron flux fraction
produced by DM annihilation in the Galactic halo
against the positron energy $E_{e^+}$ for a point with gaugino-like neutralino (blue dot-dashed),
mixed composition (green dashed) and higgsino-like neutralino (red solid).
We find that the neutralino contribution to the positron fraction
does not match the observed data. In order to do so, one would require a boost factor
of order 65, 120, and 60 for the three cases, respectively.

It is a known fact that $\sigma v$ for $1\tev$ higgsino-like (or
wino-like, for that matter) neutralinos can be boosted by a Sommerfeld
enhancement factor of order
3--6\cite{Cirelli:2007xd,Hisano:2004ds,Hryczuk:2010zi} (for multi-TeV
neutralinos the $\sigma v$ boost factor could even be of several
orders of magnitude).  However, for $\mchi\lesssim1\tev$ the
Sommerfeld enhancement would still not be enough to obtain the boost
factors cited above.  On the other hand, the boost factor can also
come from astrophysical sources like mini-spikes\cite{Brun:2007tn} or
subhalos\cite{Lavalle:1900wn}.  Such astrophysical sources can give an
energy-dependent boost factor for positron fluxes.  (For example,
Fig.~5 of Ref.\cite{Brun:2007tn} shows enhancements from mini-spikes
of order $10^4$ at $E_{e^+}\simeq 300\gev$ for $\sim1\tev$ DM.)

\subsection{Best fits and mass spectrum}

\begin{figure}[t]
\centering
\subfloat[][Basic likelihood only]{\label{fig:chisq:i}
\includegraphics[width=0.49\linewidth, clip=true, trim=3cm 3cm 0cm 0cm]{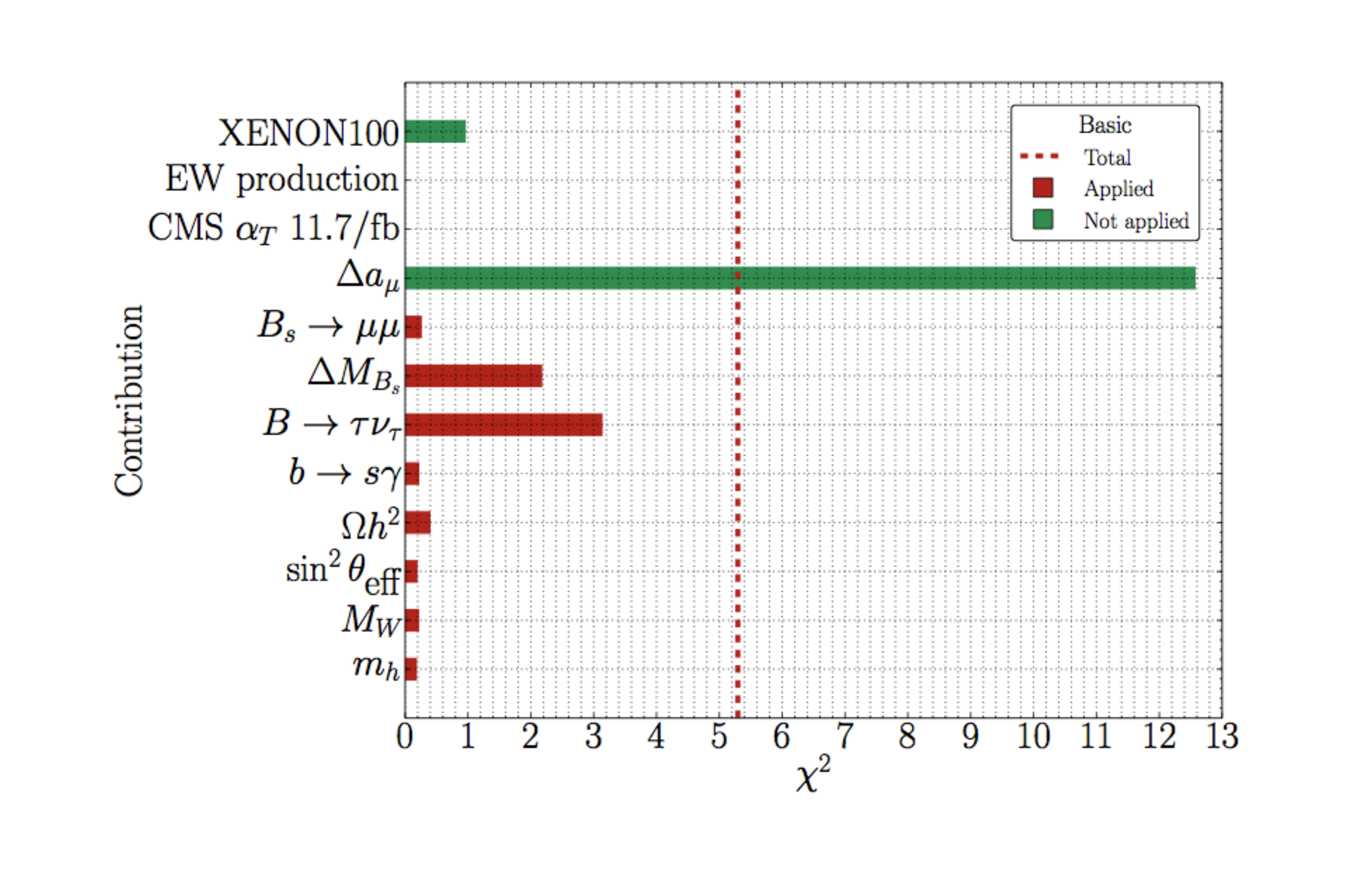}
}
\subfloat[][Basic likelihood and \gmtwo]{\label{fig:chisq:ii}
\includegraphics[width=0.49\linewidth, clip=true, trim=3cm 3cm 0cm 0cm]{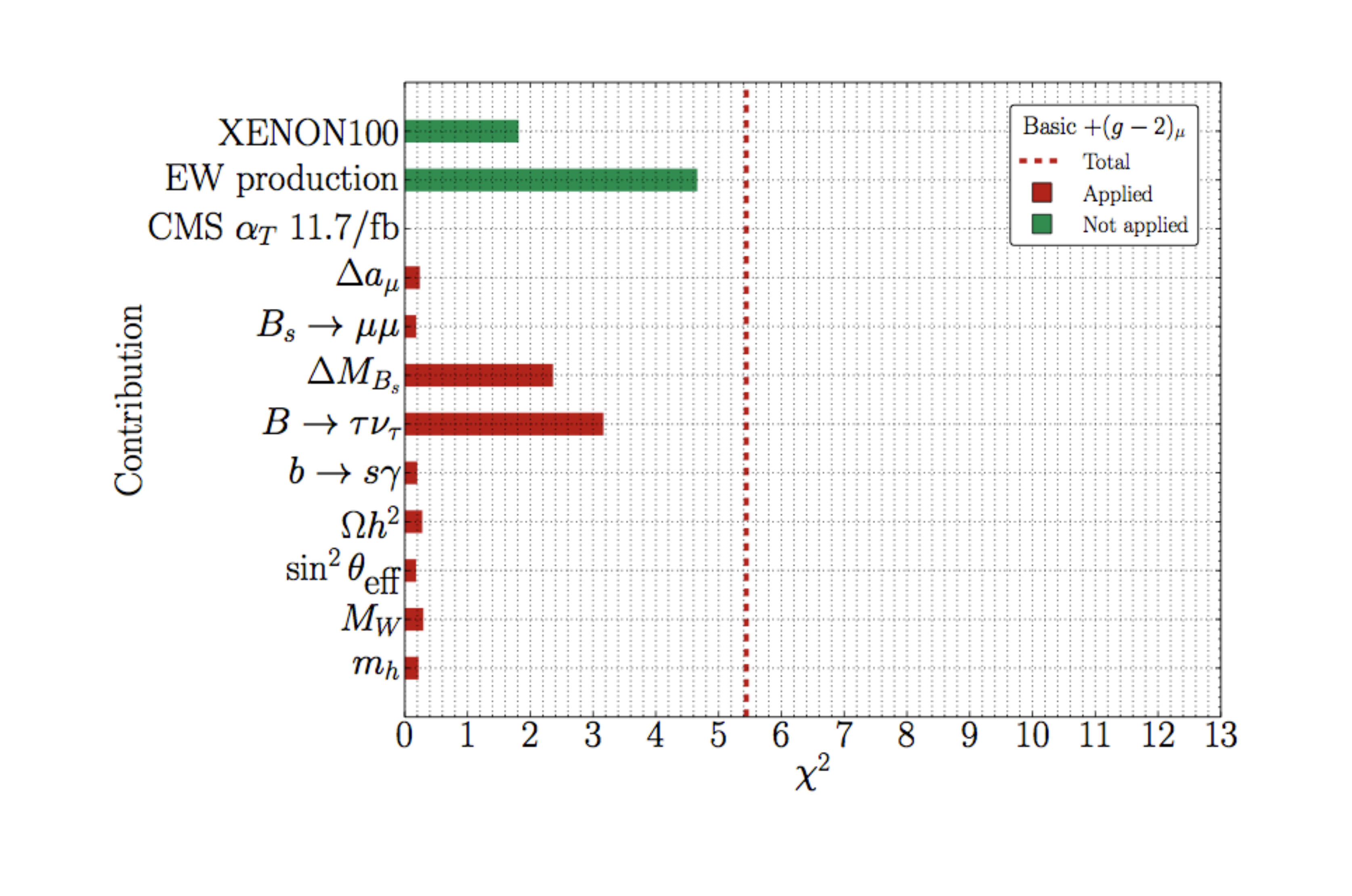}
}

\subfloat[][Basic likelihood and \xenon]{\label{fig:chisq:iii}
\includegraphics[width=0.49\linewidth, clip=true, trim=3cm 3cm 0cm 0cm]{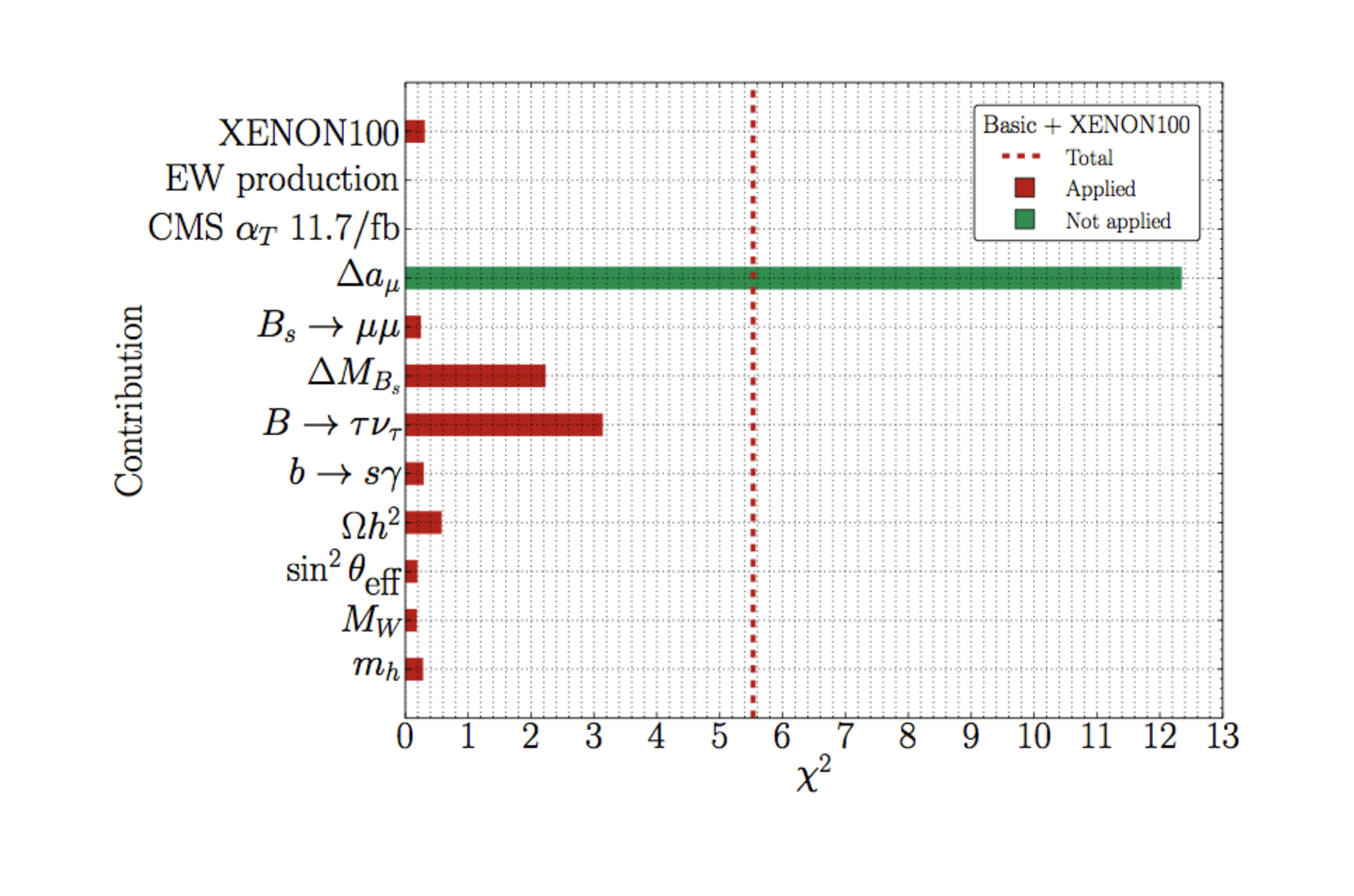}
}
\subfloat[][All likelihoods (basic, \gmtwo\ and \xenon)]{\label{fig:chisq:iv}
\includegraphics[width=0.49\linewidth, clip=true, trim=3cm 3cm 0cm 0cm]{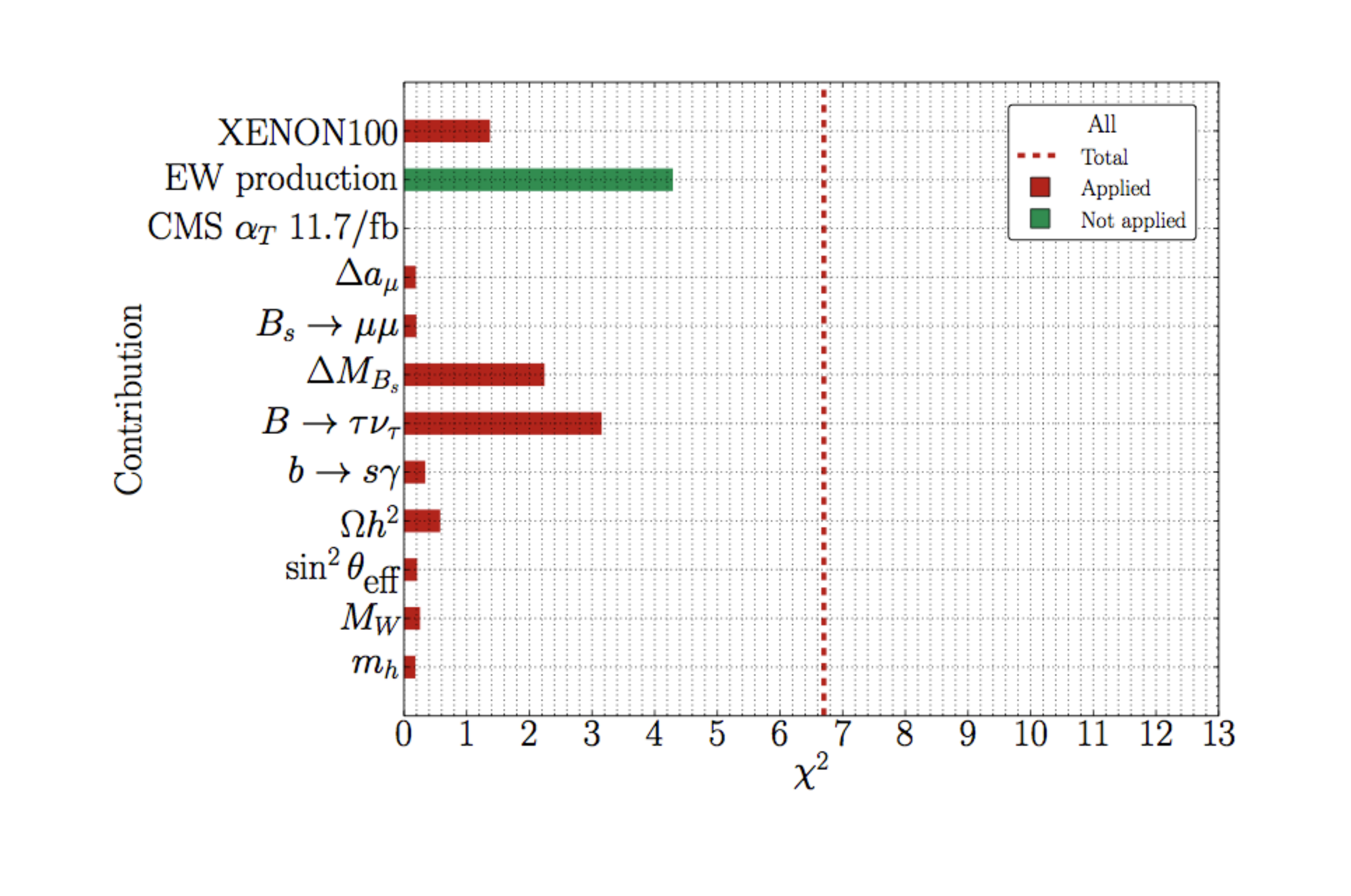}
}
\caption{
Breakdown of \chisq\ at the best-fit points in four likelihood combinations. 
Note that while in each plot all contributions are plotted, 
only specific contributions were minimized in each case. 
Contributions that were (were not) minimized are shown with red, darker (green, lighter) bars. 
The procedure for identifying these points is described in the text. The total \chisq\ 
in each case is marked with a vertical dashed red line. 
}
\label{fig:chisq}
\end{figure}

We identify the best-fit points (the points with the smallest total
\chisq) in four cases: the one due to the \textbf{basic} set of
constraints (hereafter called just \textbf{basic}); the one obtained
by adding the \gmtwo\ constraint to \textbf{basic}; the one obtained
by adding XENON100 to \textbf{basic}; and the one obtained by
considering all three sets together.  The relative breakdown of the
contributions from the different constraints is shown for each case in
Figs.~\ref{fig:chisq}\subref{fig:chisq:i},
\ref{fig:chisq}\subref{fig:chisq:ii},
\ref{fig:chisq}\subref{fig:chisq:iii}, and
\ref{fig:chisq}\subref{fig:chisq:iv}, respectively.
 
\begin{figure}[t]
\centering
\subfloat[][Basic likelihood only]{\label{fig:a}
\includegraphics[width=0.49\linewidth]{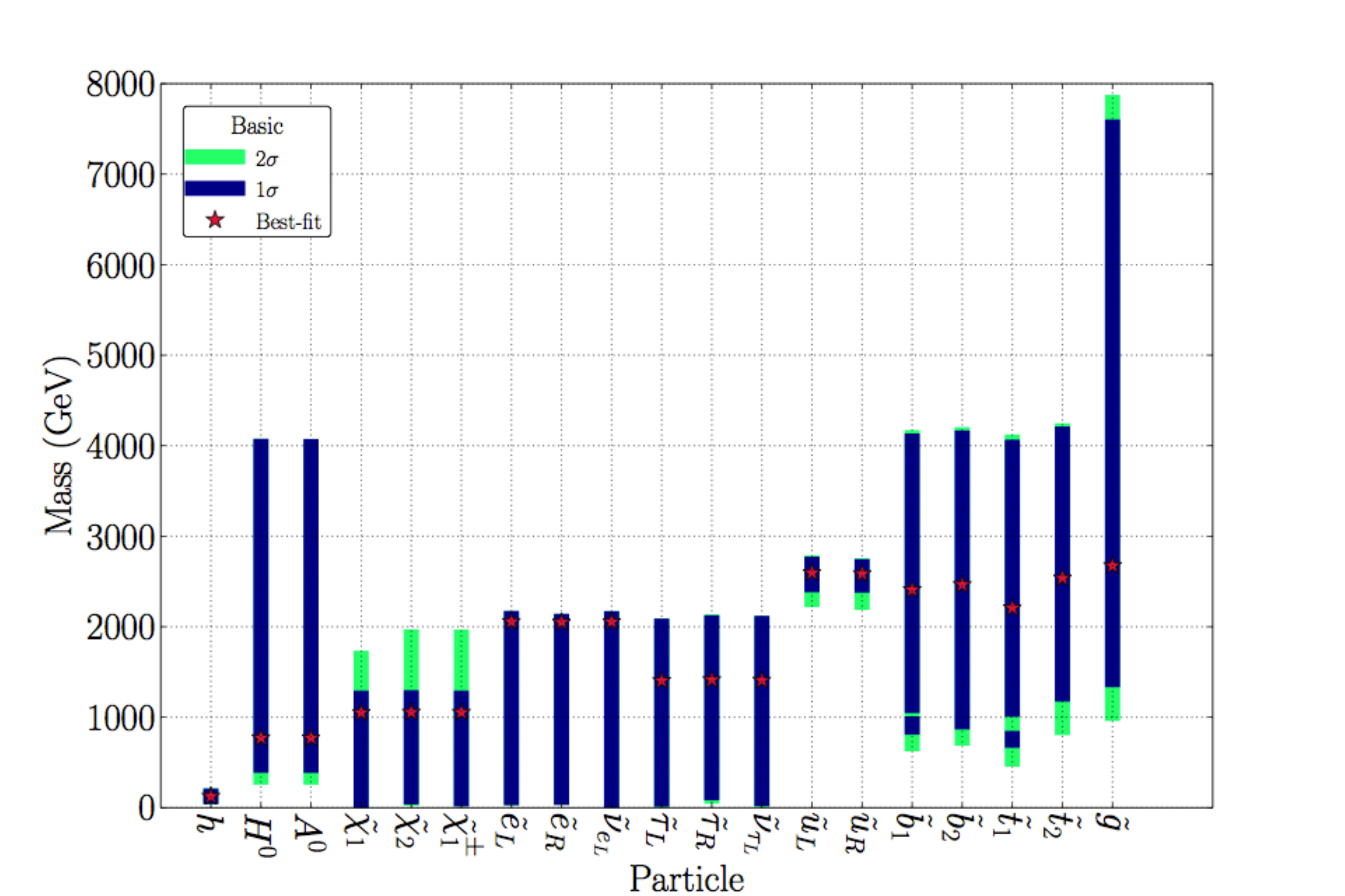}
}
\subfloat[][Basic likelihood and \gmtwo]{\label{fig:b}
\includegraphics[width=0.49\linewidth]{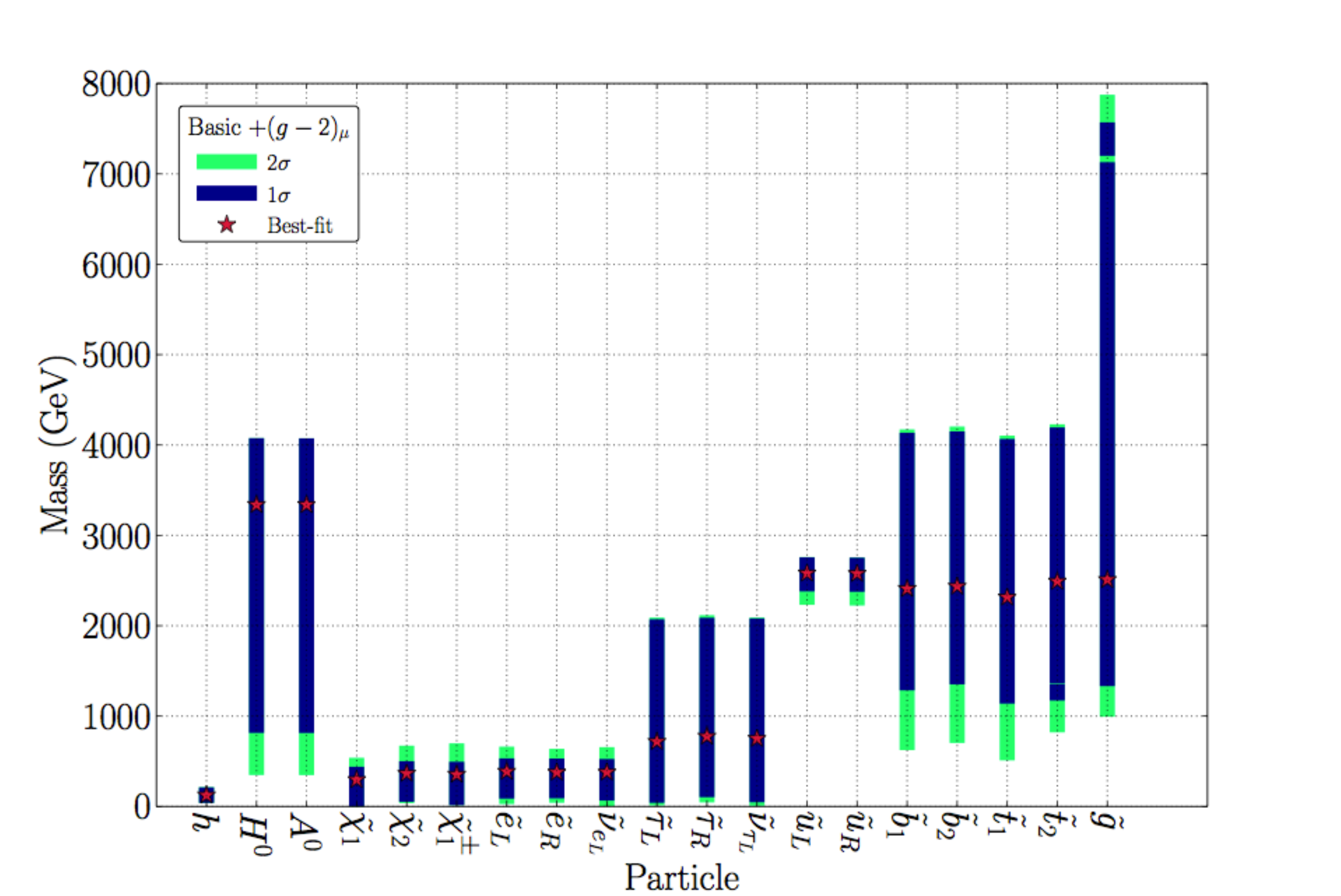}
}

\subfloat[][Basic likelihood and \xenon]{\label{fig:c}
\includegraphics[width=0.49\linewidth]{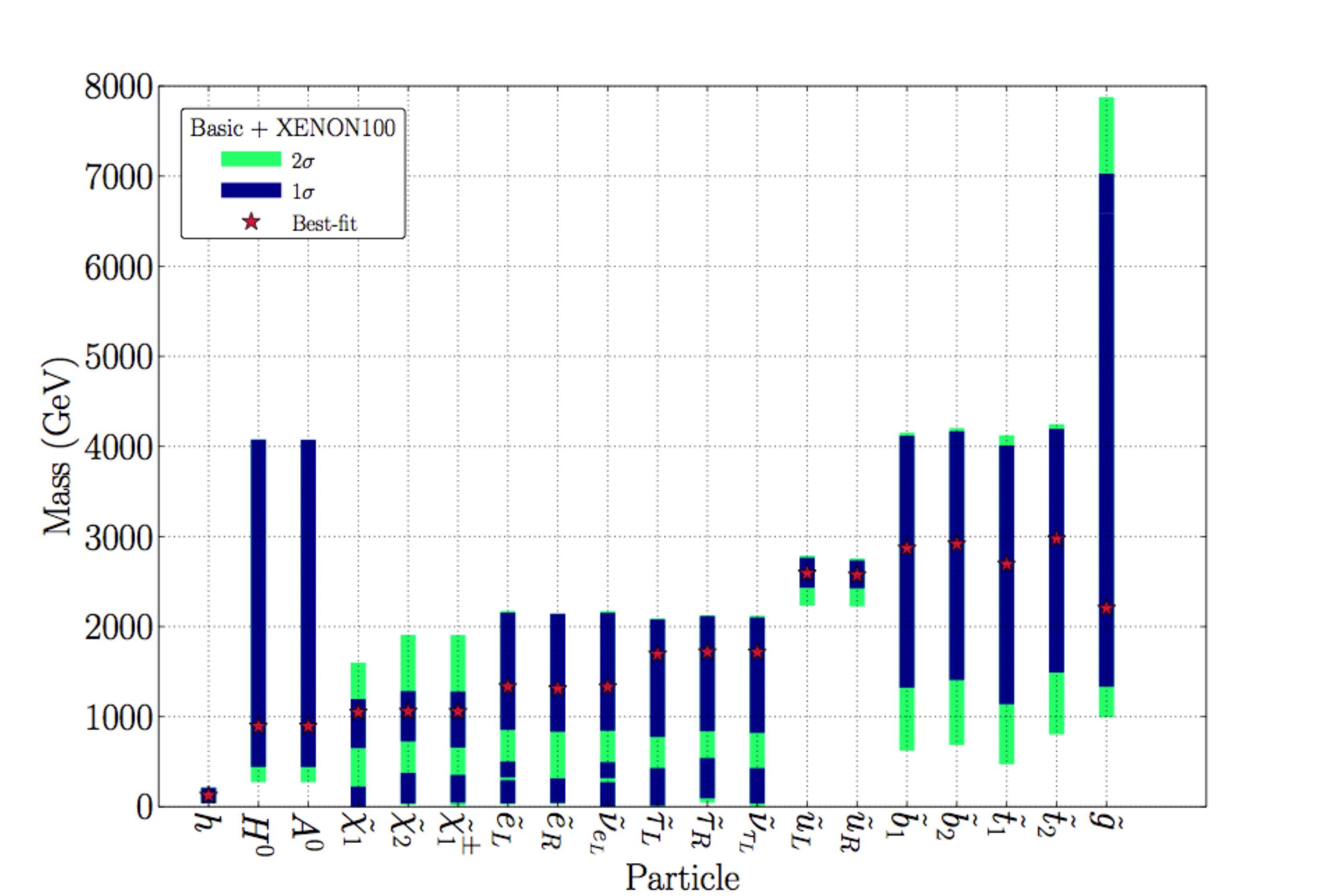}
}
\subfloat[][All likelihoods (basic, \gmtwo\ and \xenon)]{\label{fig:d}
\includegraphics[width=0.49\linewidth]{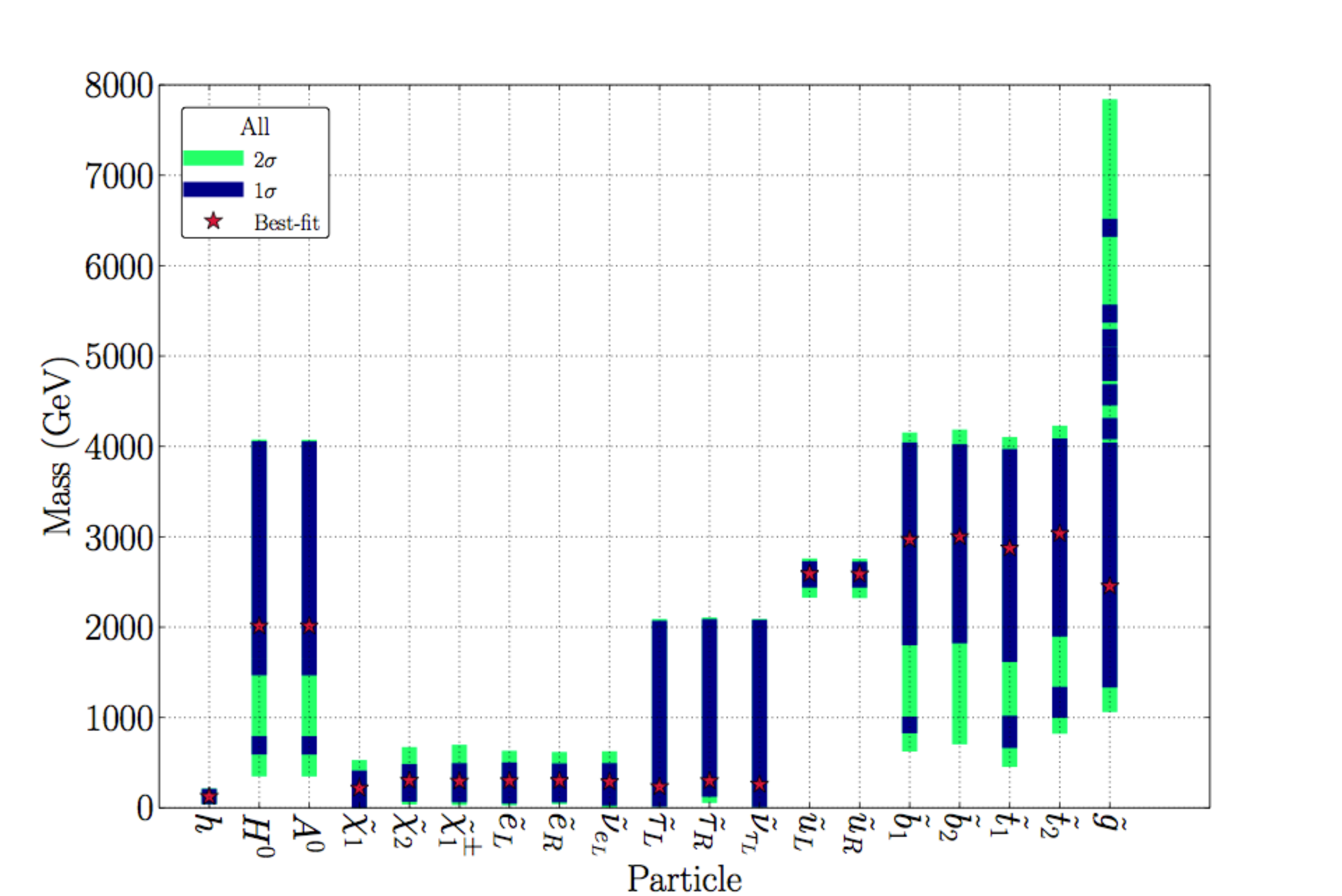}
}
\caption{Best-fitting mass spectra of the p9MSSM in four likelihood combinations (red stars), 
and the associated 1$\sigma$ (blue, darker bars) and 2$\sigma$ (green, lighter bars) confidence intervals.}
\label{fig:mass}
\end{figure}

The corresponding mass spectra for the best-fit points, together with
their 68\% and 95\% confidence regions for the cases in question, are
shown in Figs.~\ref{fig:mass}\subref{fig:a},
\ref{fig:mass}\subref{fig:b}, \ref{fig:mass}\subref{fig:c}, and
\ref{fig:mass}\subref{fig:d}.

For each best-fit point, we also show in \reffig{fig:chisq} the contribution 
due to the EW-production likelihood, 
although we remind the reader that it is not included in the 
total \chisq\ computation, but applied separately, for the reasons explained in~\refsec{subsec:LHC}.
As a consequence, it always appears in the figures in green (see caption).
In the cases where the \gmtwo\ constraint is included,
which tend to favor light sleptons, charginos and neutralinos   
(see Figs.~\ref{fig:mass}\subref{fig:b} and \ref{fig:mass}\subref{fig:d}), 
we have selected the points with the lower \chisq\ that are not 
excluded at the 95\%~C.L. by direct electroweakino searches at the LHC,
i.e., points with $\delchisq_{\textrm{EW}}<5.99$.
In particular, our selected points have $\delchisq_{\textrm{EW}}\simeq4.5$ and $\delchisq_{\textrm{EW}}\simeq4$, respectively.
Note also that when a particular experiment is not constraining for the parameter space its contribution to the \chisq\ is approximately zero
and the corresponding bar does not show in \reffig{fig:chisq}. This is particularly the case 
of the \alphaT\ constraint for the best-fit points since, as can be seen in all four panels of \reffig{fig:mass},
the squark and gluino masses are favored to be well above the present LHC sensitivity.

The four cases in \reffig{fig:chisq} exhibit similar behavior; 
only the contributions from \gmtwo, \xenon\ and EW
production differ between scenarios. The contributions from EW precision observables, 
Higgs mass, CMS direct search and relic density are negligible in the total \chisq\ at the best-fit points, 
while contributions from $B$-physics observables \butaunu\ and \delmbs\ are substantial. 
The discrepancy between the SM value $\brbutaunu_{\textrm{SM}}=8.014\times 10^{-5}$ and the measured value (Table~\ref{tab:exp_constraints}) is
known in the literature\cite{Hurth:2012vp}. 
\xenon's \chisq\ is moderate when not constrained by the likelihood (\reffig{fig:chisq}\subref{fig:chisq:i} 
and \reffig{fig:chisq}\subref{fig:chisq:ii}), but when it is included in the likelihood 
its \chisq\ is reduced without damaging other predictions 
(compare \reffig{fig:chisq}\subref{fig:chisq:i} with \reffig{fig:chisq}\subref{fig:chisq:iii}). 

The \chisq\ from \gmtwo\ is substantial when not constrained in the
likelihood (\reffig{fig:chisq}\subref{fig:chisq:i} and
\reffig{fig:chisq}\subref{fig:chisq:iii}) and even when \gmtwo\ is
constrained in the likelihood, its \chisq\ can only be reduced at the
expense of an increase in the EW-production \chisq\ and, to a lesser
extent, \xenon\ (\reffig{fig:chisq}\subref{fig:chisq:ii} and
\reffig{fig:chisq}\subref{fig:chisq:iv}), thus highlighting the
tension between these constraints.

As was shown in \refsec{sec:g2}, however, we identified a fairly broad
region of the parameter
space in good agreement with all the constraints included here. 

\section{\label{Summary}Summary and conclusions}

In this paper we have performed a global statistical analysis
of a parametrization of the MSSM with 9 free parameters defined at the SUSY scale,
the p9MSSM.
The parameters have been selected as a minimum set that allows investigation 
of the dark matter sector assuming that the lightest neutralino is the
only source of dark matter, and that its composition is either 
bino-like, or higgsino-like, or a mixture of the two.
We analyzed a sample of $1.8\times10^6$ points, which were subject to constraints 
from the relic density as measured by PLANCK, the Higgs boson mass 
measurement at the LHC, the recent evidence of 
$\brbsmumu=3.2^{+1.5}_{-1.2}\times 10^9$ at LHCb, 
the measurement of \deltagmtwomu, direct SUSY limits 
from the LHC at $\sqrt{s}=8\tev$, and the upper bound on \sigsip\ from XENON100.

The statistical analysis was performed by constructing a global likelihood function
and we presented 95\% confidence regions in 2D projections of the parameter space
according to the profile-likelihood method.
The global likelihood function was calculated for each scanned point, 
and it includes an approximate but reasonably accurate implementation 
of the inclusive search for squarks and gluinos 
with the $\alpha_T$ variable at CMS, with $11.7/\textrm{fb}$ of data at $\sqrt{s}=8\tev$.
It also includes an approximate but accurate implementation of the XENON100 bound. 
For the first time we included in the XENON100 likelihood a recent 
determination\cite{Stahov:2012ca} from CHAOS data of the $\Sigma_{\pi N}$ term, 
which parametrizes the 
effect of the theoretical uncertainties due to nuclear physics on \sigsip.
We quantified the effect of the new $\Sigma_{\pi N}$ determination on the parameter space
and we showed that, when compared to the case where the theoretical uncertainty is neglected,
or to cases where the theoretical uncertainty is calculated on the basis 
of alternative determinations of the same matrix element, the impact of the XENON100 bound is greatly reduced.

In this study, we also analyzed the compatibility of the \gmtwo\ constraint 
with the strongest lower bounds on the neutralino, chargino, and slepton 
masses from direct EW-production 
searches at the LHC. We did so by 
selecting a uniformly distributed sample of $\sim 2\%$ of the total points,
and by constructing an approximate but accurate likelihood that simulated 
the CMS $3l+E_T^{\textrm{miss}}$ search at $\sqrt{s}=8\tev$, and $9.2/\textrm{fb}$. 
We found that, when all constraints are taken into account,
there is at present a small window in the neutralino mass,
$200\gev\lesssim\mchi\lesssim500\gev$, and similarly in the sleptons masses, which is not excluded by any one of the constraints,
including LHC direct searches and \deltagmtwomu.
The LHC run at $\sqrt{s}=14\tev$ should, however, definitely probe this region.

In the last part of the paper, we calculated the impact on the parameter space of the 
constraints on the annihilation cross section $\sigma v$ from 
Fermi-LAT data on $\gamma$-ray fluxes from dSphs and from the Galactic Center.
We confirm the findings of other groups that, in both cases, the present data is constraining for the p9MSSM
only when including a phenomenological boost factor. 
The same conclusion pertains also to the analysis of data 
from the positron flux at AMS02.

We also considered sensitivities for direct and indirect detection of dark matter 
for those experiments which will run or collect data in the near future.
We showed the projected sensitivities for direct detection at LUX and XENON1T,
and for indirect detection we showed the 5-year, 95\%~C.L. sensitivity at IceCube-86.

Finally, we showed the best-fit points and relative spectra for different combinations of our constraints:
\textbf{basic} (for a definition of this box see \refsec{sec:expcons}), \textbf{basic}+\gmtwo, \textbf{basic}+XENON100,
and all combined.
The best-fit points show in all four cases squarks an gluinos around 2--3\tev,
as required by the Higgs mass measurement and LHC direct searches,
with the exception of small regions where the squarks of the third generation
are favored around 1\tev, for large stop mixing.
The basic set favors sleptons in the ballpark of 1--2\tev, 
and neutralinos and charginos almost degenerate and around
1\tev, as required in the higgsino region of good relic density.
When including the \gmtwo\ constraints the best-fit point shows 
sleptons, charginos and neutralinos lighter than 500--600\gev,
which is possible for bino, or mixed bino/higgsino LSPs.
When including XENON100, but neglecting \gmtwo, the situation is not dissimilar
from the \textbf{basic} case, due the reduced impact of the direct detection constraint
with the new determination of $\Sigma_{\pi N}$.

The best-fit \chisq\ for the case where all contraints are included together 
is slightly larger than in both the case with \textbf{basic}+\gmtwo\ and 
the case with \textbf{basic}+XENON100 constraints.
This points to a slight tension between the present constraints from direct 
detection and the measurement of \deltagmtwomu.
 
\bigskip
\begin{center}
\textbf{ACKNOWLEDGMENTS}
\end{center}

  We would like to thank Shoaib Munir for helpful discussions 
  through different stages of this work and useful input in the revision of the manuscript.
  This work has been funded in part by the Welcome Programme
  of the Foundation for Polish Science. A.J.F. is
funded by the Science Technology and Facilities Council (STFC).
  K.K. is supported by the EU and MSHE Grant No. POIG.02.03.00-00-013/09.
  L.R. is also supported in part by the Polish National Science Centre Grant No. N202 167440, a STFC
  consortium grant of Lancaster, Manchester, and Sheffield Universities,
  and by the EC 6th Framework Programme No. MRTN-CT-2006-035505. The use
  of the CIS computer cluster at the National Centre for Nuclear Research is gratefully acknowledged.

\newpage
\appendix
\section{Abbreviations used in the paper}

The abbreviation acronyms appearing in this paper are summarized in Table~\ref{table:Acronyms}.

\begin{table}[t]
\begin{center}
\begin{tabular}{|c|c|}
\hline
\hline
Abbreviation & Full text \\
\hline\hline
AF & $A$-funnel \\
(C)MSSM & (Constrained) Minimal Supersymmetric Standard Model \\
DD &  Direct detection \\
DM & Dark matter \\
dSphs & Dwarf spheroidal satellite galaxies\\
EW & Electroweak \\
FP/HB & Focus point/Hyperbolic branch\\
GC & Galactic Center \\
GUT & Grand Unified Theory \\
HR & Higgs-resonance \\
ID & Indirect detection \\
LSP & Lightest supersymmetric particle \\
NUHM & Non-universal Higgs Model \\
p$n$MSSM & Phenomenological MSSM with $n$ free parameters \\
SC & Slepton coannihilation \\
SI & Spin-independent \\
SM & Standard Model \\
SMS & Simplified model spectrum \\
SUSY & Supersymmetry \\
vev & Vacuum expectation value \\
1TH & 1 TeV higgsino \\
2D & Two-dimensional \\
\hline
\hline
\end{tabular}
\caption{
List of abbreviation acronyms appearing in the text.}
\label{table:Acronyms}
\end{center}

\end{table}

\section{Relation of $\sigma v$ with the relic density}
We dedicate this appendix to reviewing the relationship between 
the velocity-averaged cross section for annihilation and coannihilation of neutralinos in the early Universe,
which enters the calculation of the relic density,
and the cross section times velocity in the limit of small momenta $\sigma v$, 
shown in \reffig{fig:gf}.

Let us start with $\sigma v$, which is the relevant quantity in
indirect DM searches. Since the relative velocities
of DM particles are at present very small relative to those in the
early Universe at the time of freeze-out,
$\sigma v$ is the neutralino pair-annihilation cross section 
$\sigma v^{\textrm{ann}}_{\chi\chi} (p)$ in the limit $p\to 0$. 
More precisely, its value is determined by averaging over
the distribution of small DM velocities in the halo at the present time, 
and we can thus denote it as $\langle\sigma v\rangle^{\rm{ann}}_{\rm{ID}}$.

On the other hand, the value of the relic density, \abundchi, is calculated by
taking the thermally averaged $\langle\sigma
v\rangle^{\rm{ann+co}}_{\rm{abund}}$ that results from convolving
the total $\sigma v^{\rm{ann+co}}(p)$ due to both annihilation and coannihilation mechanisms,
with a Boltzmann distribution $\kappa(p,\mchi,T)$ over the neutralinos' (and coannihilating particles')
3-momenta of magnitude $p$ in the early Universe,
\begin{equation}
\langle\sigma v\rangle^{\rm{ann+co}}_{\rm{abund}}\sim\int \sigma
v^{\rm{ann+co}}(p)\kappa(p,\mchi,T) dp\,.\label{sigmav}
\end{equation}
The explicit form of $\kappa(p,\mchi,T)$ is not particularly illuminating
for the discussion,
and can be found for instance in Eq.~(3.9) and Fig.~1 of
Ref.\cite{Edsjo:2003us}.

The correct value of the relic density observed in the Universe,
$\abundchi\simeq 0.1$, corresponds 
to $\langle\sigma v\rangle^{\rm{ann+co}}_{\rm{abund}} \simeq 3\times
10^{-26}\cmeter^3 \text{s}^{-1}$. Note that, on the other hand, $\langle\sigma v\rangle^{\rm{ann}}_{\rm{ID}}$ 
is not bound to assume the same value as $\langle\sigma
v\rangle^{\rm{ann+co}}_{\rm{abund}}$. In fact, one can notice in \reffig{fig:gf} that for some points $\langle\sigma
v\rangle^{\rm{ann}}_{\rm{ID}}$ is significantly smaller than $3\times
10^{-26}\cmeter^3 \text{s}^{-1}$, while for others it is larger. 
Still,
the points in the plot have been selected on the basis of having the
correct \abundchi.

This is easy to explain for the regions of the parameter space where
early-Universe coannihilation with some particle other than the
neutralino is involved, like the SC region or part of the 1TH region,
characterized by $\chi_{1}\chi^\pm$ coannihilation and/or
$\chi_{1}\chi_{2}$ coannihilation.  There, the annihilation cross
section does not have to be $\sim 3\times10^{-26}\cmeter^3
\text{s}^{-1}$, because the deficit to fulfil the relic density
constraint is made up by the coannihilation cross section.

However, the discrepancy can also be explained for the regions 
where only neutralino pair-annihilation is involved, $\langle\sigma v\rangle^{\rm{ann+co}}_{\rm{abund}}\approx
\langle\sigma v\rangle^{\rm{ann}}_{\rm{abund}}$,
due to the effects of thermal
averaging. In what follows, we 
show the effects of Eq.~(\ref{sigmav}) for some benchmark points.

\begin{figure}[ht!]
\centering
\subfloat[]{%
\label{fig:a}%
\includegraphics[width=0.47\textwidth]{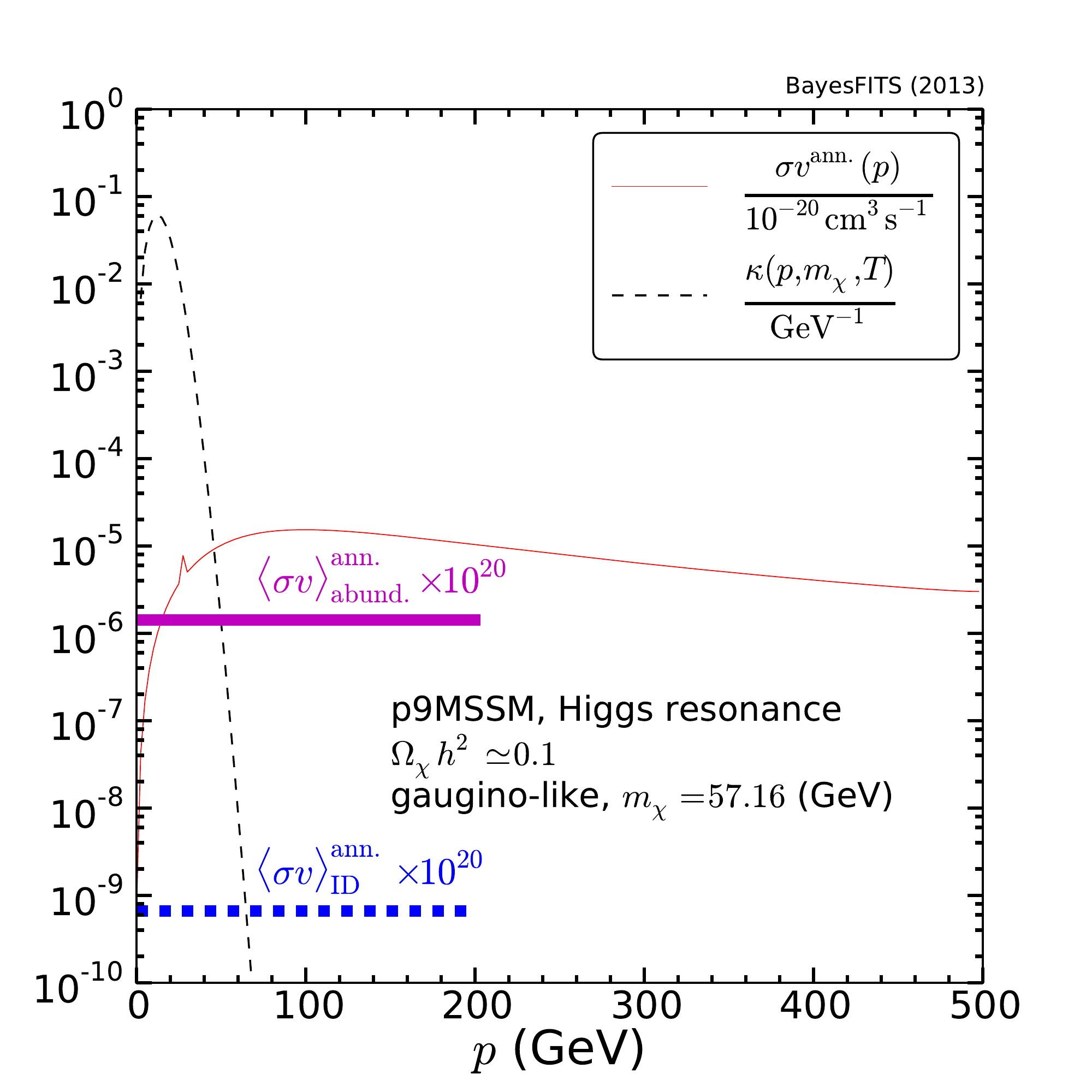}
}%
\subfloat[]{%
\label{fig:b}%
\includegraphics[width=0.47\textwidth]{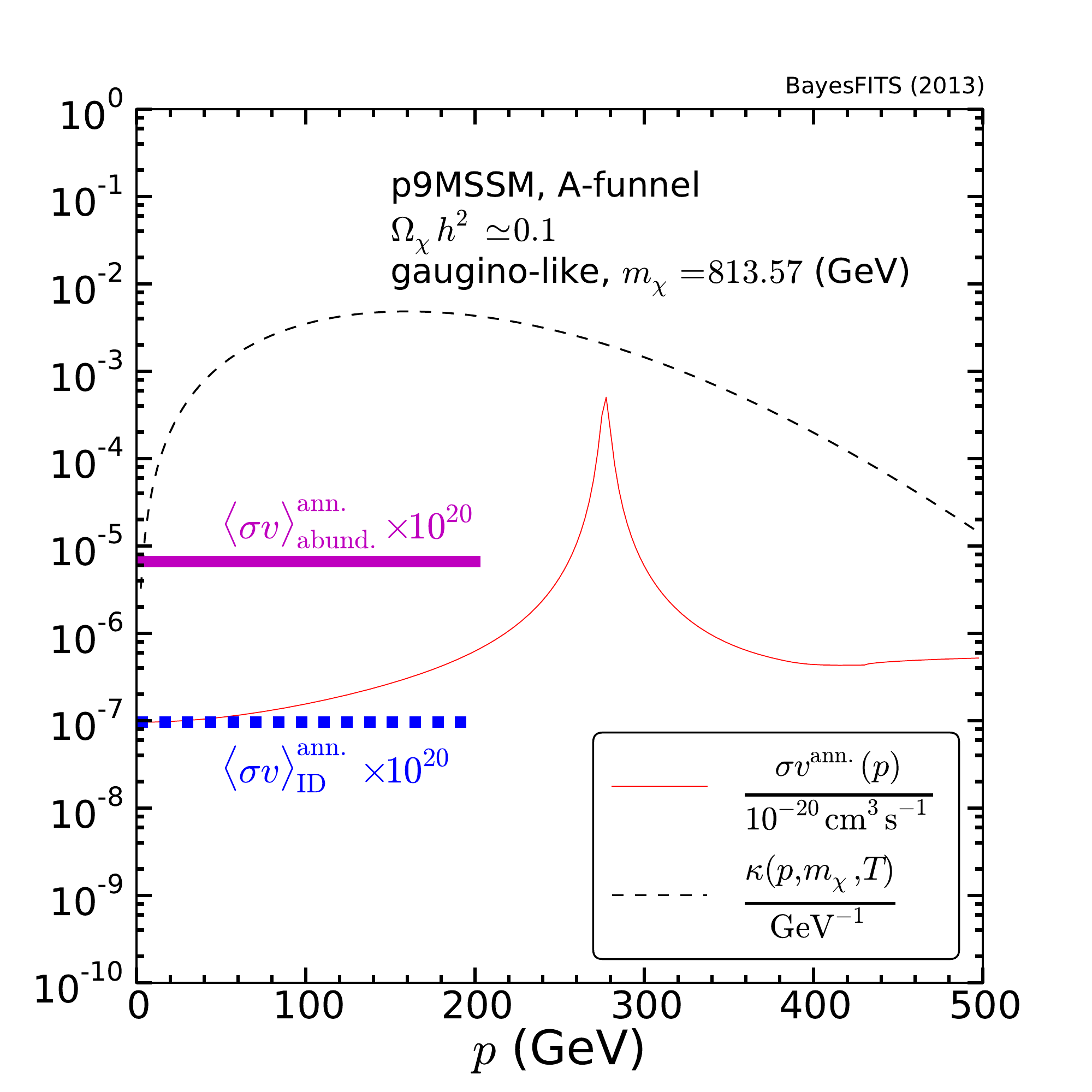}
}\\
\subfloat[]{%
\label{fig:c}%
\includegraphics[width=0.47\textwidth]{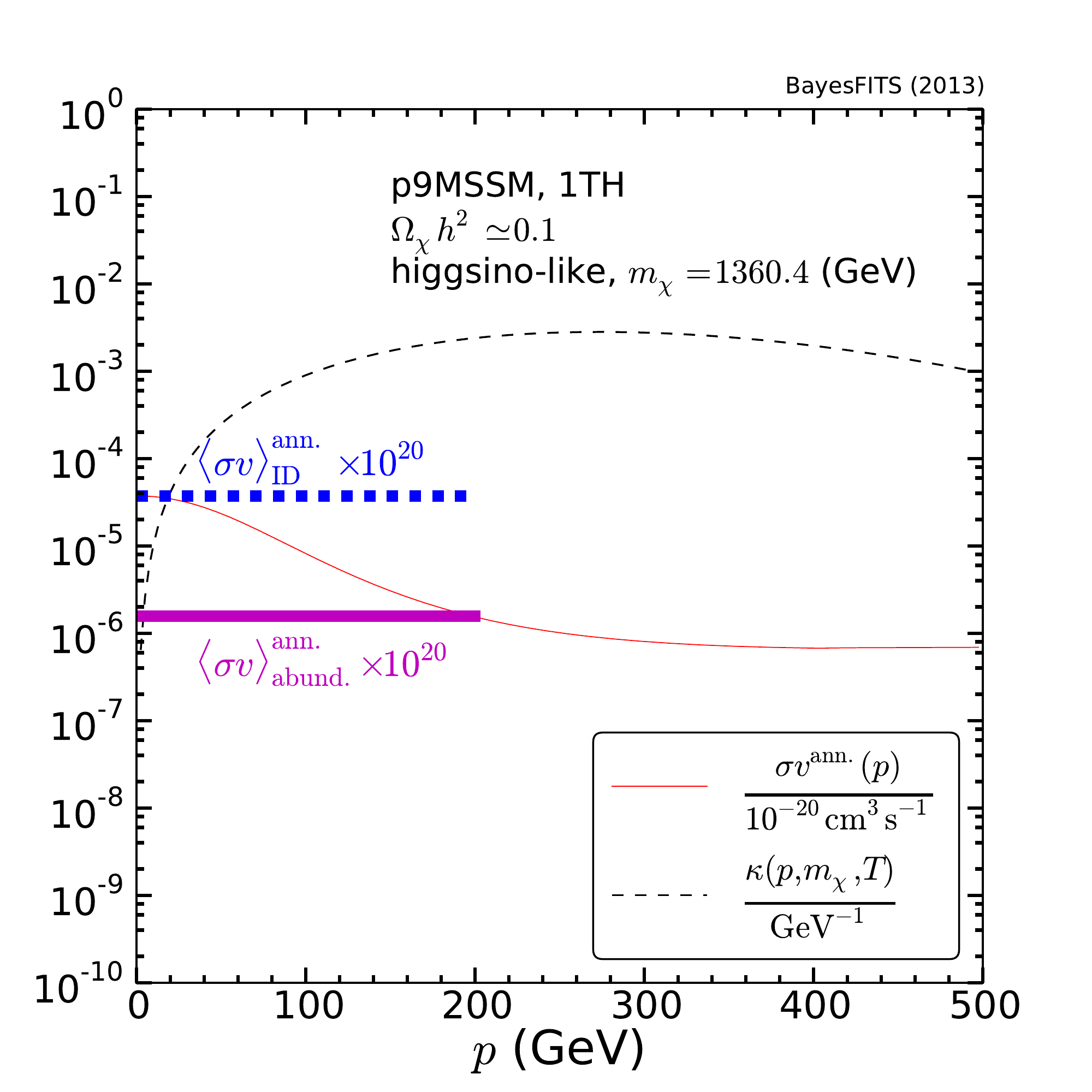}
}%
\subfloat[]{%
\label{fig:d}%
\includegraphics[width=0.47\textwidth]{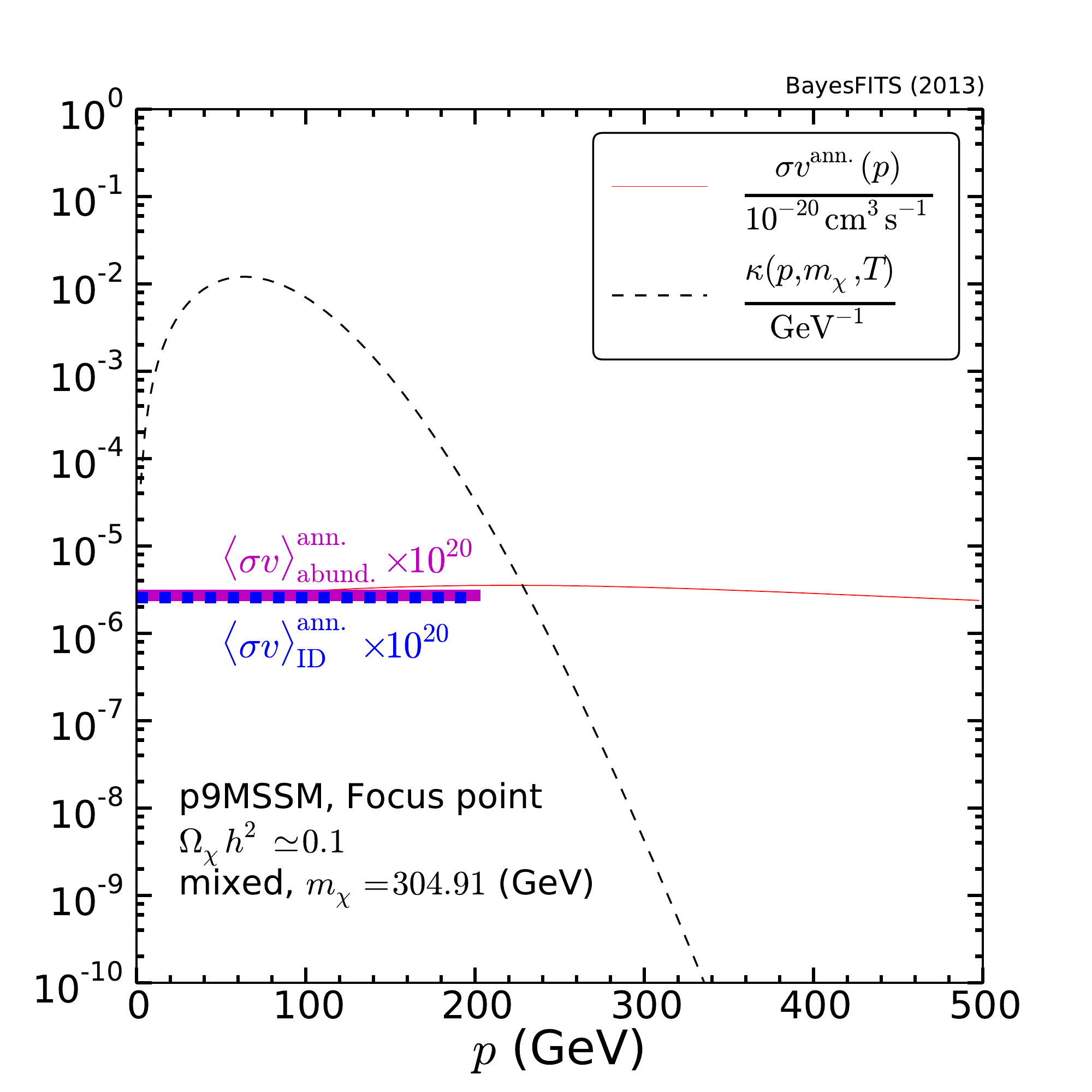}
}
\caption[]{ The Boltzmann factor, $\kappa$ (dashed thin in black), and the
  annihilation cross section, $\sigma v$ (solid thin in red), versus effective
  neutralino momentum ($p=|\vec{p}_2|=|\vec{p_1}|$ for annihilating
  neutralinos labelled $\chi_1\chi_2$) for four p9MSSM points for
  which the neutralino relic density is achieved via distinct
  mechanisms: \subref{fig:a} Higgs resonance, \subref{fig:b}
  $A$-funnel, \subref{fig:c} $\sim 1\tev$ higgsino, and \subref{fig:d}
  the focus point with mixed-composition neutralinos.  The thermally
  averaged cross section, which is $\sigma v$ weighted by $\kappa$,
  $\langle \sigma v \rangle^{\textrm{ann.}}_{\textrm{abund.}}$
  (applicable to neutralino annihilation in the early Universe), is
  shown with a solid thick magenta line, while that over low momenta,
  $\langle \sigma v \rangle^{\textrm{ann.}}_{\textrm{ID}}$ (applicable
  to indirect detection), is shown with a dashed thick blue line.  Note that while
  we display only annihilation cross sections, throughout our analysis
  the relic densities include coannihilation effects, as well.  }
\label{fig:app1}
\end{figure}

In \reffig{fig:app1}\subref{fig:a} we show with a solid thin red line the rescaled
annihilation cross section,
$\sigma v^{\rm{ann}}_{\chi\chi}(p)/(10^{-20}\cmeter^3 \textrm{s}^{-1})$ 
(we neglect the suffix $\chi\chi$ in what follows),
and we show with a dashed thin black line the rescaled Boltzmann factor, $\kappa(p,\mchi,T)/\gev^{-1}$, for
a point representative of the HR region.
The corresponding $\langle\sigma
v\rangle^{\rm{ann}}_{\rm{ID}}$ is shown with a dashed thick blue line,
and $\langle\sigma v\rangle^{\rm{ann}}_{\rm{abund}}$ with a solid thick magenta line.
One can see that the distribution of $\sigma
v^{\rm{ann}}(p)$ drops quickly for $p\to 0$. When the neutralino
annihilates at rest the cross section is suppressed by the
off-shell propagator.
At $p\simeq 25-40\gev$ the resonance with
$\mhl\simeq125\gev$ is met (a small peak in $\sigma v^{\rm{ann}}(p)$ is observed)
and for larger $p$ the cross section keeps increasing thanks to the opening $WW$ pair-production channels.   
Since the Boltzmann distribution $\kappa(p,\mchi,T)/\gev^{-1}$ has a sharp maximum
for $p>0$ (typical of a light neutralino mass) the thermal
averaging yields $\langle\sigma
v\rangle^{\rm{ann}}_{\rm{abund}}\simeq 10^{-26}\cmeter^3 \text{s}^{-1}$.

In \reffig{fig:app1}\subref{fig:b} we show $\sigma
v^{\rm{ann}}(p)$ and $\kappa(p,\mchi,T)$
for a point of the AF region, for which $\langle\sigma
v\rangle^{\rm{ann}}_{\rm{ID}}<\langle\sigma v\rangle^{\rm{ann}}_{\rm{abund}}$. 
In this region the
correct value of relic density is obtained through the exchange of the pseudoscalar $A$
in the $s$ channel, which requires $s\approx\ma^2$. Thus, one can see that $\sigma v^{\rm{ann}}(p)$ shows a relatively broad
peak at $p\simeq 300\gev$. 
The Boltzmann distribution is fairly flat given the large value
of the neutralino mass, so that the convolution can
produce the correct value of $\langle\sigma v\rangle^{\rm{ann}}_{\rm{abund}}$.  

In \reffig{fig:app1}\subref{fig:c} we consider a point of the 1TH region, for which $\langle\sigma
v\rangle^{\rm{ann}}_{\rm{ID}}>\langle\sigma v\rangle^{\rm{ann}}_{\rm{abund}}$.
This point is characterized by $\ma\approx 2\mchi$, so that the correct value of the relic density 
can be obtained, partially at least, again through $A$-resonance
in the $s$ channel. With increasing $p$, the
resonance region is increasingly left behind, so that $\sigma
v^{\rm{ann}}(p)$ decreases. 
The large value of \mchi\ makes $\kappa(p,\mchi,T)$ fairly flat over almost all of the momentum
range, with the exception of $p\rightarrow 0$, where it is suppressed by 
$p^2$ in the Jacobian. This is enough to bring the convolution down to the correct   
$\langle\sigma v\rangle^{\rm{ann}}_{\rm{abund}}$.

Finally, in \reffig{fig:app1}\subref{fig:d} we show 
the results for a point of the FP/HB region,
for which $\langle\sigma v\rangle^{\rm{ann}}_{\rm{abund}}$
and $\langle\sigma v\rangle^{\rm{ann}}_{\rm{ID}}$
are comparable. As one can see, $\sigma v^{\rm{ann}}(p)$ is
almost flat, since the rapidly falling cross section 
for $h$ resonance is enhanced at large $p$ by processes with $t$-channel
exchange of stops and/or charginos. 
The convolution with the Boltzmann factor will thus not change significantly the 
value of the cross section.

\bibliographystyle{utphysmcite}	

\bibliography{myref}



\end{document}